%% file: ms.tex
\documentclass[11pt,onecolumn,letterpaper]{article}
\usepackage[letterpaper,margin=1in]{geometry}
\usepackage{times}

\input{preamble}

\input{typing-rules}

\title{Compositional Security for Reentrant Applications}

\author{Ethan Cecchetti \and Siqiu Yao \and Haobin Ni \and Andrew C. Myers
  \end{tabular} \\[0.25em]
  {\normalsize %
    Department of Computer Science \\
    Cornell University \\
    \{ethan,yaosiqiu,haobin,andru\}@cs.cornell.edu}
  \begin{tabular}[t]{c}}
\date{}

\begin{document}
  \input{body}

\end{document}

%% file: preamble.tex
\usepackage{graphicx}
\usepackage{color}
\usepackage[dvipsnames]{xcolor}
\usepackage{eso-pic}
\usepackage{amsmath,amssymb,amsthm}
\usepackage{stmaryrd}
\usepackage{xspace}
\usepackage[hyphens]{url}
\usepackage{microtype}
\usepackage{listings}
\usepackage{tikz}
\usepackage{pgf-umlsd}
\usepackage{boxedminipage}
\usepackage{suffix}
\usepackage{mathpartir}
\usepackage[shortcuts]{extdash}
\usepackage{cancel}
\usepackage{calc}

\usepackage[T1]{fontenc}
\usepackage{inconsolata}

\usepackage{enumitem}
\setlist{
    topsep=0.25em,
    partopsep=0pt,
    parsep=0pt,
    itemsep=0.25em
}

\usepackage{subcaption}
\definecolor{scifblue}{rgb}{0.13,0.13,1}
\definecolor{scifgreen}{rgb}{0,0.5,0}
\definecolor{scifred}{rgb}{0.9,0,0}
\definecolor{scifgrey}{rgb}{0.46,0.45,0.48}

\usepackage[hidelinks]{hyperref}
\usepackage{cleveref}

\usepackage{tikz}
\usetikzlibrary{positioning,calc,fit,patterns}
\definecolor{secureRegion}{rgb}{0.75,0.93,1}
\definecolor{secureCall}{rgb}{0,0.4,0.75}
\definecolor{insecureColor}{rgb}{0.9, 0.2, 0.1}
\tikzset{%
  secure fill/.style={fill=secureRegion},
  contract/.style={solid,draw=black,rectangle,rounded corners=0.25em},
  security domain/.style={solid,thick,secure fill,rectangle,rounded corners=0.25em},
  call point/.style={circle,draw=black,fill=black!20!white,inner sep=0pt,minimum width=4pt},
  call/.style={thick,draw=black,solid,-latex},
  reentrant call/.style={call,line width=1.25pt,color=insecureColor,dashed},
}
\pgfdeclarelayer{background}
\pgfsetlayers{background,main}

\usepackage[numbers,sort&compress]{natbib}
\makeatletter
\def\NAT@spacechar{~}
\makeatother

\definecolor{codeBlue}{rgb}{0,0,0.85}
\definecolor{orangeGold}{rgb}{0.72, 0.4, 0}

\lstdefinelanguage{solidity} {
  keywords=[1]{
    contract, function, constructor, modifier, mapping, event, indexed,
    external, public, internal, private, returns,
    constant, view, pure, receive, payable, fallback,
    new, var,
    bool, int, int8, int256, uint, uint8, uint256, fixed, ufixed, address,
  },
  keywordstyle=[1]\color{codeBlue},
  keywords=[2]{
    if, else, do, while, for,
    return, assert, throw,
  },
  keywordstyle=[2]\color{orangeGold},
  keywords=[3]{
    msg, this, call, balance, send,
    sender,
    bot,
  },
  keywordstyle=[3]\color{Plum},
  sensitive=true,
  morecomment=[l]{//}, 
  morecomment=[s]{/*}{*/}, 
  string=[b]", 
  commentstyle=\upshape\color{ForestGreen},
  stringstyle=\upshape\color{Red},
  escapeinside={(*}{*)},
}
\lstdefinelanguage{scif}{
  keywords=[1]{
    class, extends, this,
    bool, int, unit,
  },
  keywordstyle=[1]\color{codeBlue},
  keywords=[2]{
    if, then, else,
    let, in,
    assert,
  },
  keywordstyle=[2]\color{orangeGold},
  keywords=[3]{
    endorse, lock,
    new, ref,
    bot, top,
  },
  keywordstyle=[3]\color{Plum},
  sensitive=true,
  morecomment=[l]{//}, 
  morecomment=[s]{/*}{*/}, 
  string=[b]", 
  commentstyle=\upshape\color{ForestGreen},
  stringstyle=\upshape\color{Red},
  escapeinside={(*}{*)},
  breaklines=true,
  postbreak=\mbox{\textcolor{red}{$\hookrightarrow$}\space}
}
\newcommand{\newretheorem}[2]{%
  \newtheorem{inner#1}{#2}
  \newenvironment{#1}[1]%
    {\expandafter\renewcommand\csname theinner#1\endcsname{\ref{##1}}%
      \csname inner#1\endcsname}
    {\csname endinner#1\endcsname}%
}

\theoremstyle{plain}
\newtheorem{theorem}{Theorem}
\newtheorem{lemma}{Lemma}
\newtheorem{prop}{Proposition}
\newtheorem{corollary}{Corollary}
\newtheorem*{claim}{Claim}

\newretheorem{retheorem}{Theorem}

\theoremstyle{definition}
\newtheorem{definition}{Definition}

\newcommand{\langname}{SeRIF\xspace}
\newcommand{\langnameLong}{Secure-Reentrancy Information Flow Calculus~(\langname)\xspace}

\newcommand{\programfont}[1]{\ensuremath{\mathsf{#1}}\xspace}

\makeatletter
\renewcommand{\paragraph}[1]{%
  \@startsection{paragraph}{4}{\z@}%
                {1ex plus 0.25ex minus 0.25ex}{-1.5ex plus -0.25ex minus -0.5ex}%
                {\normalfont\upshape\normalsize\bfseries}*{#1.}%
}
\makeatother

\newcommand{\rulefiguresize}{\small}
\newenvironment{ruleset}[1][\columnwidth]{\par\noindent\begin{minipage}{#1}\rulefiguresize\begin{mathpar}}{\end{mathpar}\end{minipage}\par}

\newcommand{\RuleTemplate}[5]{
  \ensuremath{%
    \if\relax\detokenize{#1}\relax
      \infer{#3}{#4}
    \else
      {\textsc{\rulefiguresize[#1]}}~\infer{#3}{#4}
    \fi%
  }%
  \if\relax\detokenize{#2}\relax\else\label{#2}\fi%
  \if\relax\detokenize{#5}\relax\else~\text{#5}\fi
}

\newcommand{\Rule}[4][]{%
  \if\relax\detokenize{#2}\relax
    \RuleTemplate{}{}{#3}{#4}{#1}
  \else
    \RuleTemplate{#2}{rule:#2}{#3}{#4}{#1}
  \fi}
\newcommand{\NlRule}[4][]{\RuleTemplate{#2}{}{#3}{#4}{#1}}

\WithSuffix\newcommand\RuleTemplate*[5]{
  \if\relax\detokenize{#1}\relax\else\textsc{[#1]}\fi%
  \if\relax\detokenize{#2}\relax\else\label{#2}\fi%
  \hfill\ensuremath{\infer{#3}{{#4}}}%
  \if\relax\detokenize{#5}\relax\else~\text{#5}\fi\hfill{}
}
\WithSuffix\newcommand\Rule*[4][]{%
  \if\relax\detokenize{#2}\relax
    \RuleTemplate*{}{}{#3}{#4}{#1}
  \else
    \RuleTemplate*{#2}{rule:#2}{#3}{#4}{#1}
  \fi}
\WithSuffix\newcommand\NlRule*[4][]{\RuleTemplate*{#2}{}{#3}{#4}{#1}}

\makeatletter
\newcommand\ruleref[1]{%
  \@ifundefined{r@rule:#1}{%
    \mbox{\sc#1} ({\bf ??})%
  }{%
    \mbox{\hyperref[rule:#1]{\sc#1}}%
  }%
}
\makeatother

\newcommand{\adv}{\mathcal{A}}

\newcommand{\T}{\mathcal{T}}

\newcommand{\Request}{\texttt{request}}
\newcommand{\Cancel}{\texttt{cancel}}
\newcommand{\Deliver}{\texttt{deliver}}

\newcommand{\dom}{\operatorname{dom}}
\newcommand{\pc}{\mathit{pc}}
\newcommand{\proves}{\vdash}
\newcommand{\nproves}{\mathrel{\cancel{\proves}}}
\newcommand{\join}{\mathbin{\vee}}
\newcommand{\meet}{\mathbin{\wedge}}
\newcommand{\bigmeet}{\bigwedge}
\newcommand{\bigjoin}{\bigvee}
\newcommand{\actsfor}{\mathrel{\Rightarrow}}
\newcommand{\nactsfor}{\not\actsfor}
\newcommand{\EndrsSymb}{\mathchoice%
    {\mkern2mu\raisebox{0.125ex}{\ensuremath{\scriptstyle\gg}}\mkern2mu}
    {\mkern2mu\raisebox{0.125ex}{\ensuremath{\scriptstyle\gg}}\mkern2mu}
    {{\scriptscriptstyle\gg}\mkern1mu}
    {{\scriptscriptstyle\gg}\mkern1mu}%
}
\newcommand{\ty}{\mkern2mu{:}\mkern2mu}
\newcommand{\seq}{\mathrel{;}}
\newcommand{\subtyp}{\mathrel{<:}}
\newcommand{\prot}{\triangleleft}
\newcommand{\alt}{\mkern3mu\mid\mkern3mu}
\newcommand{\getLab}{\operatorname{label}}

\newcommand{\lockLabBase}{\lambda}
\newcommand{\inLock}{\lockLabBase_{\textsc{I}}}
\newcommand{\outLock}{\lockLabBase_{\textsc{o}}}
\newcommand{\ellHigh}{\ell_t}
\WithSuffix\newcommand\ellHigh*{\mathchoice%
  {\ellHigh}
  {\ellHigh}
  {\makebox[5pt][l]{\ensuremath{\scriptstyle\ellHigh}}}
  {\makebox[4pt][l]{\ensuremath{\scriptscriptstyle\ellHigh}}}%
}
\newcommand{\ellAdv}{\ell_{\adv}}
\newcommand{\lequiv}[1][\ell]{\approx_{#1}}
\newcommand{\locIso}{\simeq}
\newcommand{\lequivLocIso}[1][\ell]{\locIso_{#1}}
\newcommand{\lerase}[2][\ellHigh]{{#2}|_{#1}}

\newcommand{\geqErased}{\geq_\bullet}

\newcommand{\subst}[3]{{#1}[{#2} \mapsto {#3}]}

\newcommand{\satPQTemplate}[6]{\mathrel{{#1} #2_{#3}} \{#4\} \mathbin{#5} \{#6\}}
\newcommand{\satPQ}[5][\ell]{\satPQTemplate{#2}{\vDash}{#1}{#3}{#4}{#5}}
\WithSuffix\newcommand\satPQ*[2][\Sigma]{\satPQ{#1}{P}{#2}{Q}}
\newcommand{\singleEntSatPQ}[5][\ell]{\satPQTemplate{#2}{\vDash^1}{#1}{#3}{#4}{#5}}
\WithSuffix\newcommand\singleEntSatPQ*[2][\Sigma]{\singleEntSatPQ{#1}{P}{#2}{Q}}

\newcommand{\Labs}{\ensuremath{\mathcal{L}}\xspace}
\newcommand{\CT}{\ensuremath{\mathit{CT}}\xspace}

\newcommand{\nat}{\programfont{nat}}
\newcommand{\unit}{\programfont{unit}}
\newcommand{\bool}{\programfont{bool}}
\newcommand{\RefType}[1]{\programfont{ref}~{#1}}

\newcommand{\mtypeExpr}[5]{{#1} \xrightarrow{{#2}\EndrsSymb{#3};{#4}} {#5}}
\WithSuffix\newcommand\mtypeExpr*{\mtypeExpr{\overline{\tau_a}}{\pc_1}{\pc_2}{\outLock}{\tau}}

\newcommand{\provesWt}[1]{\proves {#1}~\mathsf{wt}}

\newcommand{\True}{\programfont{true}}
\newcommand{\False}{\programfont{false}}
\newcommand{\loc}{\iota}
\newcommand{\Null}{\programfont{null}}

\newcommand{\super}{\programfont{super}}
\newcommand{\this}{\programfont{this}}
\newcommand{\new}{\programfont{new}}

\newcommand{\Lock}{\programfont{lock}}
\newcommand{\Let}{\programfont{let}}
\newcommand{\In}{\programfont{in}}

\newcommand{\RefOp}[2]{\programfont{ref}~{#1}~{#2}}
\newcommand{\deref}[1]{\programfont{!}{#1}}
\newcommand{\assign}{\mathrel{\programfont{:=}}}
\newcommand{\EndorseName}{\programfont{endorse}}
\newcommand{\Endorse}[3]{\EndorseName~{#1}~\programfont{from}~{#2}~\programfont{to}~{#3}}
\newcommand{\lockIn}[2]{\Lock~{#1}~\In~{#2}}
\newcommand{\If}{\programfont{if}}
\newcommand{\IfThenElse}[4][\pc]{\If\{{#1}\}~{#2}~\programfont{then}~{#3}~\programfont{else}~{#4}}
\newcommand{\letIn}[3][x]{\Let~{#1}={#2}~\In~{#3}}

\newcommand{\atpcName}{\programfont{at\text{-}pc}}
\newcommand{\atpc}[2][\pc]{{#2}~\atpcName~{#1}}
\newcommand{\withLock}[2][\ell]{{#2}~\programfont{with\text{-}lock}~{#1}}
\newcommand{\funend}[1]{\programfont{return}_{#1}}

\newcommand{\ignoreLocksName}{\programfont{ignore\text{-}locks\text{-}in}}
\newcommand{\ignoreLocks}[1]{\ignoreLocksName~{#1}}

\newcommand{\cdefTemplate}[4]{\programfont{class}~{#1}[{#2}]~\programfont{extends}~{#3}~\{{#4}\}}
\newcommand{\cdef}[6]{\cdefTemplate{#1}{#2}{#3}{{#4} \seq {#5} \seq {#6}}}
\WithSuffix\newcommand\cdef*{\cdef{C}{\ell_C}{D}{\overline{f}\ty\overline{\tau_f}}{K}{\overline{M}}}
\newcommand{\cdefNoBody}[3]{\cdefTemplate{#1}{#2}{#3}{\cdots}}
\WithSuffix\newcommand\cdefNoBody*{\cdefNoBody{C}{\ell_C}{D}}

\newcommand{\mdef}[7]{{#1}~{#2}\{{#3}\EndrsSymb{#4};{#5}\}({#6})~\{{#7}\}}
\WithSuffix\newcommand\mdef*{\mdef{\tau}{m}{\pc_1}{\pc_2}{\outLock}{\overline{x}\ty\overline{\tau_a}}{e}}

\newcommand{\erased}{\bullet}

\newcommand{\fields}{\mathit{fields}}
\newcommand{\mtype}{\mathit{mtype}}
\newcommand{\mbody}{\mathit{mbody}}
\newcommand{\override}{\mathit{can\text{-}override}}

\newcommand{\mbodyExpr}[7]{\left({#1}, {#2}, {#3}, {#4}\EndrsSymb{#5}, {#6}, {#7}\right)}
\WithSuffix\newcommand\mbodyExpr*[1][\mlabel]{\mbodyExpr{#1}{\overline{x}}{\overline{\tau_a}}{\pc_1}{\pc_2}{e}{\tau}}

\newcommand{\getLocks}{\mathit{getLocks}}
\newcommand{\innerPc}{\mathit{innerPc}}

\newcommand{\config}{\mathcal{C}}
\newcommand{\lockList}{L}
\newcommand{\mlabel}{\ell_m}
\newcommand{\mlabList}{\mathcal{M}}

\newcommand{\confTuple}[4][\CT]{({#1}, {#2}, {#3}, {#4})}
\WithSuffix\newcommand\confTuple*{\confTuple{\sigma}{\mlabList}{\lockList}}

\newcommand{\conf}[2]{\langle {#1} \mid {#2} \rangle}
\newcommand{\confExpr}[2][]{\conf{#2}{\config\if\relax\detokenize{#1}\relax\else{[#1]}\fi}}

\newcommand{\stepsone}{\longrightarrow}
\newcommand{\stepsmany}{\stepsone^*}
\newcommand{\bulletstepsone}{\mathrel{\rlap{\hspace{0.55em}$\bullet$}{\longrightarrow}}}
\newcommand{\bulletstepsmany}{\bulletstepsone^*}

\newcommand{\provesCwl}[3][\mkern-1mu\raisebox{-1pt}{$\scriptstyle\T$}]{{#2} \mathrel{\proves_{\mkern-4mu\raisebox{-1pt}{$\scriptstyle#1$}}} {#3}~\mathsf{cwl}}

\newcommand{\UpEvent}{\programfont{up}}
\newcommand{\DownEvent}{\programfont{down}}
\newcommand{\SetEvent}{\programfont{set}}
\newcommand{\RetEvent}{\programfont{ret}}
\newcommand{\ev}{\mathit{ev}}

%% file: typing-rules.tex
\newcommand\newRuleFull[5][]{
  \if\relax\detokenize{#3}\relax
    \expandafter\newcommand\csname#2\endcsname{\NlRule[#1]{#3}{#4}{#5}}
    \WithSuffix\expandafter\newcommand\csname#2\endcsname*{\NlRule*[#1]{#3}{#4}{#5}}
  \else
    \newcounter{#2@used}
    \setcounter{#2@used}{0}
    \expandafter\newcommand\csname#2\endcsname{%
      \ifnum\value{#2@used}=0
        \Rule[#1]{#3}{#4}{#5}%
        \setcounter{#2@used}{1}%
      \else
        \NlRule[#1]{#3}{#4}{#5}%
      \fi%
    }
    \WithSuffix\expandafter\newcommand\csname#2\endcsname*{%
      \ifnum\value{#2@used}=0
        \Rule*[#1]{#3}{#4}{#5}%
        \setcounter{#2@used}{1}%
      \else
        \NlRule*[#1]{#3}{#4}{#5}%
      \fi%
    }
  \fi
}
\newcommand{\typingRule}[4][]{
  \newRuleFull[#1]{#2Rule}{#2}{#3}{#4}
}
\newcommand{\semanticRule}[4][]{
  \newRuleFull[#1]{E#2Rule}{E-#2}{#3}{#4}
}

\typingRule{Var}{\Gamma(x) = \tau}{\Sigma; \Gamma \proves x : \tau}
\typingRule{Unit}{ }{\Sigma; \Gamma \proves () : \unit^\ell}
\typingRule{True}{ }{\Sigma; \Gamma \proves \True : \bool^\ell}
\typingRule{False}{ }{\Sigma; \Gamma \proves \False : \bool^\ell}
\typingRule{New}{
  \fields(C) = \overline{f}\ty\overline{\tau} \\\\
  \Sigma; \Gamma \proves \overline{v} : \overline{\tau}
}{\Sigma; \Gamma \proves \new~C(\overline{v}) : C^\ell}
\typingRule{Loc}{\Sigma(\loc) = \tau}{\Sigma; \Gamma \proves \loc : (\RefType{\tau})^\ell}
\typingRule{Null}{ }{\Sigma; \Gamma \proves \Null : (\RefType{\tau})^\ell}
\typingRule{SubtypeV}{
  \Sigma; \Gamma \proves v : \tau' \\
  \tau' \subtyp \tau
}{\Sigma; \Gamma \proves v : \tau}

\typingRule{Val}{\Sigma; \Gamma \proves v : \tau}{\Sigma; \Gamma; \pc; \inLock \proves v : \tau \dashv \outLock}
\typingRule{Variance}{
  \Sigma; \Gamma; \pc'; \inLock' \proves e : \tau' \dashv \outLock' \\\\
  \tau' \subtyp \tau \\
  \pc \actsfor \pc' \\\\
  \inLock' \actsfor \inLock \\
  \outLock' \actsfor \outLock
}{\Sigma; \Gamma; \pc; \inLock \proves e : \tau \dashv \outLock}
\typingRule{If}{
  \Sigma; \Gamma \proves v : \bool^\ell \\
  \ell \actsfor \pc \\
  \ell \prot \tau \\\\
  \Sigma; \Gamma; \pc; \inLock \proves e_1 : \tau \dashv \outLock \\
  \Sigma; \Gamma; \pc; \inLock \proves e_2 : \tau \dashv \outLock
}{\Sigma; \Gamma; \pc; \inLock \proves \IfThenElse{v}{e_1}{e_2} : \tau \dashv \outLock}
\typingRule{Cast}{
  \Sigma; \Gamma \proves v : D^\ell
}{\Sigma; \Gamma; \pc; \inLock \proves (C)v : C^\ell \dashv \outLock}
\typingRule{Field}{
  \Sigma; \Gamma \proves v : C^\ell \\\\
  \fields(C) = \overline{f}\ty\overline{\tau} \\\\
  \tau_i \subtyp \tau \\
  \ell \prot \tau
}{\Sigma; \Gamma; \pc; \inLock \proves v.f_i : \tau \dashv \outLock}
\typingRule{Call}{
  \mtype(C, m) = \mtypeExpr{\overline{\tau_a}}{\pc_1}{\pc_2}{\outLock}{\tau_0} \\
  \Sigma; \Gamma \proves v : C^\ell \\
  \Sigma; \Gamma \proves \overline{v_a} : \overline{\tau_a} \\\\
  \ell \actsfor \pc_1 \\
  \pc_1 \actsfor \pc_2 \join \inLock \\
  \tau_0 \subtyp \tau \\
  \pc_2 \join \ell \prot \tau
}{\Sigma; \Gamma; \pc_1; \inLock \proves v.m(\overline{v_a}) : \tau \dashv \outLock \join \pc_2}
\typingRule{Ref}{
  \Sigma; \Gamma \proves v : \tau \\
  \pc \prot \tau
}{\Sigma; \Gamma; \pc; \inLock \proves \RefOp{v}{\tau} : (\RefType{\tau})^\ell \dashv \outLock}
\typingRule{Deref}{
  \Sigma; \Gamma \proves v : (\RefType{\tau'})^\ell \\\\
  \tau' \subtyp \tau \\
  \ell \prot \tau
}{\Sigma; \Gamma; \pc; \inLock \proves \deref{v} : \tau \dashv \outLock}
\typingRule{Assign}{
  \Sigma; \Gamma \proves v_1 : (\RefType{\tau})^\ell \\\\
  \Sigma; \Gamma \proves v_2 : \tau \\
  \ell \prot \tau
}{\Sigma; \Gamma; \ell; \inLock \proves v_1 := v_2 : \unit^{\ell'} \dashv \outLock}
\typingRule{Endorse}{
  \Sigma; \Gamma \proves v : t^{\ell'}
}{\Sigma; \Gamma; \ell; \inLock \proves \Endorse{v}{\ell'}{\ell} : t^\ell \dashv \outLock}
\typingRule{Lock}{
  \Sigma; \Gamma; \pc; \inLock' \proves e : \tau \dashv \outLock' \\\\
  \inLock' \meet \ell \actsfor \inLock \\
  \outLock' \meet \ell \actsfor \outLock
}{\Sigma; \Gamma; \pc; \inLock \proves \lockIn{\ell}{e} : \tau \dashv \outLock}
\typingRule{Let}{
  \Sigma; \Gamma; \pc; \inLock \proves e_1 : \tau_1 \dashv \outLock' \\
  \outLock' \actsfor \inLock \\\\
  \Sigma; \Gamma, x\ty\tau_1; \pc; \inLock \proves e_2 : \tau_2 \dashv \outLock
}{\Sigma; \Gamma; \pc; \inLock \proves \letIn{e_1}{e_2} : \tau_2 \dashv \outLock}

\typingRule{AtPc}{
  \Sigma; \Gamma; \pc; \inLock \proves s : \tau \dashv \outLock
}{\Sigma; \Gamma; \pc'; \inLock \proves \atpc{s} : \tau \dashv \outLock}
\typingRule{WithLock}{
  \Sigma; \Gamma; \pc; \inLock' \proves s : \tau \dashv \outLock' \\\\
  \inLock' \meet \ell \actsfor \inLock \\
  \outLock' \meet \ell \actsfor \outLock
}{\Sigma; \Gamma; \pc; \inLock \proves \withLock{s} : \tau \dashv \outLock}
\typingRule{Return}{
  \Sigma; \cdot; \pc; \inLock' \proves s : \tau \dashv \outLock' \\\\
  \inLock' \join \outLock' \actsfor \outLock
}{\Sigma; \Gamma; \pc; \inLock \proves \funend{\tau}~s : \tau \dashv \outLock}

\typingRule{IgnoreLocks}{
  \Sigma; \Gamma; \pc; \inLock' \proves e : \tau \dashv \outLock'
}{\Sigma; \Gamma; \pc; \inLock \proves \ignoreLocks{e} : \tau \dashv \outLock}

\newRuleFull{MethodOkRule}{Method-Ok}{
  \inLock \actsfor \pc_2 \\
  \ell_C \actsfor \pc_2 \\
  \inLock \join \outLock' \actsfor \outLock \\
  \pc_1 \prot \overline{\tau_a} \\\\
  \raisebox{0pt}[1em][0.5em]{$\Sigma; \overline{x}\ty\overline{\tau_a}, \this\ty C^{\pc_2}; \pc_2; \inLock \proves e : \tau \dashv \outLock'$} \\\\
  \CT(C) = \cdefNoBody* \\
  \override(D, m, \mtypeExpr*)
}{\Sigma \proves \mdef*~\mathsf{ok~in}~C}
\newRuleFull{ClassOkRule}{Class-Ok}{
  \fields(D) = \overline{g}\ty\overline{\tau_g} \\\\
  K = C(\overline{g}\ty\overline{\tau_g} \seq \overline{f}\ty\overline{\tau_f})~\{ \super(\overline{g}) \seq \this.\overline{f} = \overline{f} \} \\\\
  \Sigma \proves \overline{M}~\mathsf{ok~in}~C
}{\Sigma \proves \cdef*~\mathsf{ok}}
\newRuleFull{CtOkRule}{CT-Ok}{
  C~\text{referenced in any type} ~\Longrightarrow~ C \in \dom(\CT) \\\\
  \forall C \in \dom(\CT).\, \Sigma \proves \CT(C)~\mathsf{ok}
}{\Sigma \proves \CT~\mathsf{ok}}

\newRuleFull{FieldListRule}{}{
  \CT(C) = \cdef* \\\\
  \fields(D) = \overline{g}\ty\overline{\tau_g}
}{\fields(C) = \overline{g}\ty\overline{\tau_g} \seq \overline{f}\ty\overline{\tau_f}}

\newRuleFull{DefinedMethodRule}{}{
  \CT(C) = \cdef* \\\\
  \mdef* \in \overline{M}
}{
  \mtype(C, m) = \mtypeExpr* \\\\
  \mbody(C, m) = \mbodyExpr*[\ell_C]
}
\newRuleFull{InheritedMethodRule}{}{
  \CT(C) = \cdef* \\\\
  m~\text{not defined in}~\overline{M}
}{
  \mtype(C, m) = \mtype(D, m) \\\\
  \mbody(C, m) = \mbody(D, m)
}
\newRuleFull{CanOverrideRule}{}{
  (D, m) \in \dom(\mtype) ~\Longrightarrow~ \mtype(D, m) = \mtypeExpr*
}{\override(D, m, \mtypeExpr*)}

\semanticRule{Eval}{
  \confExpr{s} \stepsone \conf{s'}{\config'}
}{\confExpr{E[s]} \stepsone \conf{E[s']}{\config'}}

\semanticRule{Let}{ }{\confExpr{\letIn{v}{e}} \stepsone \confExpr{\subst{e}{x}{v}}}

\semanticRule{IfT}{ }{\confExpr{\IfThenElse{\True}{e_1}{e_2}} \stepsone \confExpr{\atpc{e_1}}}

\semanticRule{IfF}{ }{\confExpr{\IfThenElse{\False}{e_1}{e_2}} \stepsone \confExpr{\atpc{e_2}}}

\semanticRule{AtPc}{ }{\confExpr{\atpc{v}} \stepsone \confExpr{v}}

\semanticRule{Ref}{
  \loc \notin \dom(\sigma) \\
  \Sigma_\sigma \proves v : \tau \\
  \mlabList = \mlabList', \mlabel \\
  \mlabel \prot \tau
}{\confExpr{\RefOp{v}{\tau}} \stepsone \confExpr[\subst{\sigma}{\loc}{(v, \tau)}/\sigma]{\loc}}

\semanticRule{Deref}{\sigma(\loc) = (v, \tau)}{\confExpr{\deref{\loc}} \stepsone \confExpr{v}}

\semanticRule{Assign}{
  \Sigma_\sigma(\loc) = \tau \\
  \Sigma_\sigma \proves v : \tau \\
  \mlabList = \mlabList', \mlabel \\
  \mlabel \prot \tau
}{\confExpr{\loc := v} \stepsone \confExpr[\subst{\sigma}{\loc}{(v, \tau)}/\sigma]{()}}

\semanticRule{Cast}{
  D \subtyp C
}{\confExpr{(C)(\new~D(\overline{v}))} \stepsone \confExpr{\new~D(\overline{v})}}

\semanticRule{Field}{ }{\confExpr{\new~C(\overline{v}).f_i} \stepsone \confExpr{v_i}}

\semanticRule{Call}{
  \mbody(C, m) = \mbodyExpr* \\\\
  \mlabList = \mlabList', \mlabel' \\
  \mlabel' \actsfor \pc_1 \\
  \bigwedge_{\ell \in \lockList} (\pc_1 \actsfor \pc_2 \join \ell) \\\\
  \Sigma_\sigma \proves \overline{w} : \overline{\tau_a} \\
  e' = e[\overline{x} \mapsto \overline{w}, \this \mapsto \new~C(\overline{v})]
}{\confExpr{\new~C(\overline{v}).m(\overline{w})} \stepsone \confExpr[\mlabList,\mlabel/\mlabList]{\funend{\tau}~(\atpc[\pc_2]{e'})}}

\semanticRule{CallAtk}{
  \mbody(C, m) = \mbodyExpr* \\\\
  \mlabList = \mlabList', \mlabel' \\
  \mlabel' \actsfor \pc_1 \\
  \ellAdv \actsfor \pc_2 \\\\
  \Sigma_\sigma \proves \overline{w} : \overline{\tau_a} \\
  e' = e[\overline{x} \mapsto \overline{w}, \this \mapsto \new~C(\overline{v})]
}{\confExpr{\new~C(\overline{v}).m(\overline{w})} \stepsone \confExpr[\mlabList,\mlabel/\mlabList]{\funend{\tau}~(\atpc[\pc_2]{e'})}}

\semanticRule{Return}{
  \Sigma_\sigma \proves v : \tau \\
  \mlabList = \mlabList', \mlabel
}{\confExpr{\funend{\tau}~v} \stepsone \confExpr[\mlabList'/\mlabList]{v}}

\semanticRule{Lock}{ }{\confExpr{\lockIn{\ell}{e}} \stepsone \confExpr[\lockList, \ell/\lockList]{\withLock{e}}}

\semanticRule{Unlock}{
  \lockList = \lockList', \ell
}{\confExpr{\withLock{v}} \stepsone \confExpr[\lockList'/\lockList]{v}}

\semanticRule{Endorse}{ }{\confExpr{\Endorse{v}{\ell'}{\ell}} \stepsone \confExpr{v}}

\semanticRule{IgnoreLocks}{ }{\confExpr{\ignoreLocks{v}} \stepsone \confExpr{v}}

%% file: body.tex
\maketitle

\begin{abstract}
  \input{abstract}
\end{abstract}

\section{Introduction}
\label{sec:intro}

Compositional security remains a fundamental concern for software security.
Code might appear secure, yet expose vulnerabilities when it interacts with other code.
Blockchain smart contracts offer multiple prominent recent examples of this problem~\citep{dao-hack-nyt,uniswap-heist,parity-freeze},
but other instances exist.
JavaScript code is difficult to secure when running on the same web page as code
from a different source~\citep{cmjl09, conscript, hs12}.
Web browsers themselves have fallen victim to attacks when executing code on web pages~\citep{cve-2014-1772,cve-2018-8174}.
In these settings, securing code in isolation is not sufficient.
Reasoning about the behavior of a combination of interacting systems, however, is notoriously difficult.
This work therefore aims for a way to build software with
\emph{compositional} security guarantees, meaning
the security of an entire system follows from the security of its
components.

Complex control flow, and in particular reentrant executions,
pose a fundamental challenge for compositional security.
Developers are increasingly building applications from separate communicating services
that may belong to different trust domains~\citep{dragoni2017microservices,opengroup-soa-standards}.
In such architectures, one service waiting for another to respond
must be prepared to handle separate incoming requests.
These \emph{reentrant calls} effectively interrupt the
execution of the application and, if the developer is
not careful, can catch it in an inconsistent state, creating
security vulnerabilities~\citep{cwe-1265}.

Reentrancy security has received much more attention since July 2016, when
the Decentralized Autonomous Organization~(DAO)---an Ethereum smart contract intended to function as a distributed venture capital fund---
lost \$50 million in tokens to such an attack, making global news~\citep{dao-hack-nyt}.
Since then, a variety of methods have emerged to analyze or eliminate
reentrancy attacks~\citep{obsidian-toplas,grossman2017online,luu2016making,das2019resource,albert2020taming},
but vulnerabilities continue to appear.
For example, a January 2019 audit uncovered a reentrancy vulnerability in the Uniswap decentralized exchange~\citep{uniswap-audit}.
The attack leveraged a subtle interaction between two contracts that
were secure in isolation, and a third malicious contract.
The first contract implicitly assumed the second would not call the malicious contract.
Because the interface could not specify this expectation, developers used the exchange for a token standard that allowed for such calls.
This choice led to the theft of \$25~million worth of tokens in April 2020~\citep{uniswap-heist}, over a year after the original vulnerability disclosure.

We follow our previous suggestion~\citep{cecchetti-fab20}
and use a general language-based technique to obtain compositional security even in the presence of reentrant executions.
We define and enforce security using a semantic specification of trust in the form of information flow labels.
Information flow control~(IFC) has long been an appealing technique for obtaining compositional security
and has proven useful in practice~\citep{ifappstore}.
IFC type systems can guide software development with compile-time checking
and provably enforce strong security guarantees such as noninterference.
But while IFC is a good starting point for compositional security,
existing approaches break down in the presence of reentrancy.
Standard IFC rules either reject useful, secure applications by blocking
requests from untrusted sources, or they allow insecure applications
that are vulnerable to reentrancy attacks.
We extend standard IFC rules to define a secure type system that efficiently and provably prevents attacks,
yet is expressive enough to build interesting applications.

This approach addresses fundamental shortcomings of existing solutions.
Current stand-alone reentrancy analyses~\citep{luu2016making,grossman2017online,albert2020taming} are non-compositional.
That is, analyzing two pieces of code separately might not yield useful guarantees about their combination---%
the exact failing that led to the Uniswap attack.
These tools also focus specifically on blockchain smart contracts.
While smart contracts have provided notable recent examples of reentrancy vulnerabilities,
similar exploits appear elsewhere~\citep{cwe-1265,cve-2014-1772,cve-2018-8174} and there is no reason to limit solutions.
The focus on smart contracts and the absence of trust specifications
forces the tools to rely on contract boundaries---a syntactic construct---as a proxy for semantic security boundaries.
This choice leads to a reentrancy definition we call \emph{object reentrancy}
that can judge the security of two semantically equivalent implementations differently,
merely because the code has different structure.

There exist other language-based approaches that provide compositional guarantees and consider reentrancy,
but they are again smart-contract focused and use object-based reentrancy definitions.
Moreover, some limit expressiveness by outlawing reentrancy entirely~\citep{obsidian-toplas,das2019resource},
while others provide only heuristic reentrancy protection~\citep{sergey2019safer,move-lang,flint}.
In addition, they universally assume that all code is written in the same language.
This strong assumption clearly does not apply to open systems where anyone can submit code, like Ethereum contracts or JavaScript on web pages.
Even in closed systems with controlled environments and known code,
new code might need to interact with legacy applications that do not respect the language~rules.

We address these shortcomings by defining a new general-purpose security type
system that tracks the integrity of data and computation.
In addition to providing standard IFC data security guarantees,
the type system combines with a run-time mechanism to provably
eliminate dangerous reentrancy while allowing safe reentrancy.
The guarantees, moreover, continue to hold even when trusted code interacts
with untrusted code that does not obey the same restrictions.

The remainder of the paper is structured as follows:
\begin{itemize}[leftmargin=*]
  \item Examples in Section~\ref{sec:motivation} show the complexity of reentrancy.

  \item \Cref{sec:ifc} provides background on information flow control
  and exposes its failure to handle reentrancy.

  \item \Cref{sec:secure-reentrancy} presents a new definition of security in the presence of reentrancy.

  \item \Cref{sec:language} defines \langname, a core calculus that
        eliminates insecure reentrancy by combining a static IFC type system
        with a dynamic locking mechanism. 

  \item \Cref{sec:proof} shows formally that \langname
  enforces our formal, compositional security condition.

  \item \Cref{sec:implementation} describes a prototype type checker implementation
  and our experience using it on realistic programs.

  \item \Cref{sec:related} discusses related work in more detail and
  \Cref{sec:conclusion} concludes.
\end{itemize}

\section{Motivation}
\label{sec:motivation}

By their very nature, reentrancy vulnerabilities are often hard to spot.
For instance, the attack on Ethereum's Decentralized Autonomous Organization~(DAO)
was considered subtle at the time~\citep{dao-hack-analysis},
despite being one of the simplest examples of reentrancy.
To build intuition, we present three
running examples of applications with reentrancy.
Though we have distilled them to their core components,
the vulnerabilities have undermined security in real-world applications.

\subsection{Uniswap}
\label{sec:uniswap}

We begin with the Uniswap/Lendf.me reentrancy vulnerability first identified in January 2019~\citep{uniswap-audit}
and later exploited in April 2020~\citep{uniswap-heist}.
The vulnerability arises from the combination of two contracts.
Though each may be considered secure in isolation, they combine in unexpected ways,
demonstrating the need for \emph{compositional} reentrancy security.

\begin{figure}
  \centering
  \hspace*{6pt}\mbox{%
\begin{lstlisting}[language=solidity]
contract Uniswap {
  Token tX, tY;

  function sellXForY(uint xSold) returns uint {
    uint prod = tX.getBal(this) * tY.getBal(this);
    uint yKept = prod / (tX.getBal(this) + xSold);
    uint yBought = tY.getBal(this) - yKept;

    assert tX.transferTo(msg.sender, this, xSold);
    assert tY.transferTo(this, msg.sender, yBought);
    return yBought;
  }
}

contract Token {
  function transferTo(address from, address to,
        uint amount) returns bool {

    (*\raisebox{0.5\baselineskip}[0pt]{... {\color{ForestGreen}// check and update balances}}*)
    from.alertSend(to, amount);(*\label{lst:ln:token-alert-send}*)
    to.alertReceive(from, amount);(*\label{lst:ln:token-alert-receive}*)
    return true;
  }
}
\end{lstlisting}%
  }
  \caption{Distilled Solidity~\citep{solidity-0.7.5} code for the Uniswap bug.}
  \label{fig:uniswap}
\end{figure}

Uniswap is a smart contract platform where users can exchange one token for another.
Figure~\ref{fig:uniswap} shows a simplified portion of the Uniswap contract:
the exchange function "sellXForY" allows users to sell tokens of type $X$ for tokens of type $Y$.
Uniswap determines the exchange rate by the amount of $X$ and $Y$ it currently holds.
It holds the product of the two amounts constant,
allowing Uniswap to maintain the same total asset value as exchange rates fluctuate.
The tokens themselves are implemented by independent contracts.

To perform an exchange, Uniswap queries its balance with each token, computes how much of token~$Y$ the user bought,
and transfers tokens by calling \texttt{transferTo} on each token contract.
Tokens execute transfers by first checking and updating balances,
and then notifying the sender and recipient, allowing each in turn to execute arbitrary code.

Both contracts appear secure in isolation, following the best-practice
recommendation of modifying state before making external calls to
avoid reentrancy concerns~\citep{solidity-sec-consider}.
However, when combined, they expose a dangerous exploit.
Suppose the exchange begins with 6 units each of $X$ and $Y$.
\begin{enumerate}
  \item An attacker $\adv$ calls \texttt{sellXForY} selling 6 units of~$X$.
  \item Uniswap correctly computes $\texttt{prod} = \text{36}$ and $\texttt{yBought} = \text{3}$.
  \item Uniswap calls token $X$ to transfer 6 units from $\adv$.
  \item The token notifies $\adv$, giving it control of the execution.
  \item\label{atk:li:reenter-call} Before returning, $\adv$ calls \texttt{sellXForY} again to sell 6 more units of~$X$, reentering the Uniswap contract.
  \item Uniswap now has 12 units of~$X$, but still 6 units of~$Y$,
    so it computes $\texttt{prod} = \text{72}$, not 36, and $\texttt{yBought} = \text{2}$.
\end{enumerate}
When the dust settles, Uniswap has 18 units of~$X$ and only 1~unit of~$Y$,
having given $\adv$ an extra unit of~$Y$ and having broken the invariant that the product of the balances is 36.
If desired, $\adv$ can reclaim their original 12 units of~$X$ for only 2 units of~$Y$, keeping the other 3 as illicit profit.

The fundamental problem is a mismatch between Uniswap's notion of secure behavior and the token's.
The token correctly checks that all transfers are valid and authorized
and follows programming patterns that avoid (internal) reentrancy concerns.
No user can transfer more tokens than they have.
Uniswap, however, implicitly assumes that \texttt{transferTo}
transfers tokens and returns \emph{without allowing
an adversary to call Uniswap before it reestablishes the invariant that $\texttt{prod} = \textup{36}$.}

This insight suggests two approaches to fixing the bug:
(1)~token contracts could respect Uniswap's assumption by not calling unknown, untrusted code,
or (2)~Uniswap could stop relying on the assumption.
Current platforms provide no way to guarantee the first option.
Uniswap could state its assumption in documentation,
but there is no technical means of specifying or enforcing it.
Tokens that violate it could continue to freely interface with Uniswap, with disastrous results.
The exchange can, however, implement the second option by acquiring a run-time lock on entry to the contract.
It could then recognize the above attack and produce an error
at step~\ref{atk:li:reenter-call}.

Our approach detects this vulnerability and can specify and correctly analyze either proposed
solution.
Among existing tools, only Nomos~\citep{das2019resource} can express
the assumption of approach~(1),
which it mandates to statically eliminate all reentrancy.
Other tools either cannot properly secure the application~\citep{sergey2019safer,move-lang,flint}
or force the use of computationally expensive dynamic locks even when they are unnecessary~\citep{obsidian-toplas,albert2020taming}.

\subsection{Key--Value Store}
\label{sec:key-value-store}

\begin{figure}
  \centering
  \hspace*{6pt}\mbox{%
\begin{lstlisting}[language=solidity,morekeywords={[3]null}]
getOrCompute(key, computeFun) {
    i = _getIdx(key) // index of mapping if it exists
    if (mappings[i] == null) {
        mappings[i] = computeFun();
    }
    return mappings[i];
}
\end{lstlisting}%
  }
  \caption{The \texttt{getOrCompute} function of a key--value store.
    Here \texttt{mappings} is an array that the store resizes as mappings are added.}
  \label{fig:kv-store}
\end{figure}

Smart contracts have made reentrancy concerns highly visible,
but reentrancy is not unique to that domain.
It has led to multiple critical security vulnerabilities in Internet Explorer~\citep{cve-2014-1772,cve-2018-8174},
and is a known concern for any application executing user-provided code~\citep{cwe-1265}.

For example, key--value stores often compute
missing mappings with user-supplied functions~\citep{java-map-compute-if-absent,rust-entry-or-insert-with}.
A careless implementation of this functionality
can enable dangerous reentrancy.
Consider the code in Figure~\ref{fig:kv-store}, along with a \texttt{clear}
method that frees \texttt{mappings} and installs a new empty array.
An attacker can call \texttt{getOrCompute}, providing as arguments
an unmapped key and a malicious function
that calls \texttt{clear} and then returns a value.
First \texttt{getOrCompute} computes~\texttt{i}, then it calls the malicious function,
which calls \texttt{clear} and replaces the \texttt{mappings} array.
Finally \texttt{getOrCompute} attempts to write the attacker-provided value into index~\texttt{i} of the new array.

If \texttt{i} is large---which is likely if the store previously
contained many mappings---the write would be past the end of the new empty array.
In languages like C/C++ without array bounds checking, an
attacker-provided value would thus be written into an arbitrary memory location,
enabling remote code execution or other critical security vulnerabilities.
Even memory-safe languages like Java explicitly recommend developers
check for reentrant modifications and throw exceptions~\citep{java-map-compute-if-absent}.

Notably, while this attack appears very similar to
concurrent-modification attacks on key--value stores, it requires no concurrency.
Single-threaded applications or applications using simple thread-level locking are still vulnerable.

\subsection{Town Crier}
\label{sec:towncrier}

Banning all reentrancy might seem appealing, but this solution
would be overly restrictive.
Town Crier~(TC)~\citep{Zhang2016} is an example
where safe reentrancy enables important functionality.
TC provides authenticated data to smart contracts upon request.
Users place requests with a smart-contract front end,
and TC processes them asynchronously and delivers the data to user-specified callbacks when it is available.
TC also allows users to cancel pending requests for a refund.
Figure~\ref{fig:town-crier} shows simplified versions of TC's \texttt{deliver} and \texttt{cancel}~\mbox{methods}.

\begin{figure}
  \centering
  \hspace*{6pt}\mbox{%
\begin{lstlisting}[language=solidity]
contract TownCrier {
  address[] requesters, callbacks;

  function deliver(uint reqId, bytes data) {
    if (msg.sender == SERVICE_ADDR
        && requesters[reqId] != 0) {
      requesters[reqId] = 0;(*\label{lst:ln:tc:deliver-record}*)
      SERVICE_ADDR.call{value: FEE}("");
      callbacks[reqId].call(bytes);(*\label{lst:ln:tc:deliver-callback}*)
    }
  }

  function cancel(uint reqId) {
    if (msg.sender == requesters[reqId]) {(*\label{lst:ln:tc:cancel-cond}*)
      requesters[reqId] = 0;(*\label{lst:ln:tc:cancel-record}*)
      msg.sender.call{value: FEE}("");(*\label{lst:ln:tc:cancel-send}*)
    }
  }
}
\end{lstlisting}%
  }
  \caption{Solidity~\citep{solidity-0.7.5} code for simplified partial Town Crier contract.
    Here \texttt{SERVICE\_ADDR} is TC's trusted wallet address, and \texttt{FEE} is the request fee.}
  \label{fig:town-crier}
\end{figure}

Invoking a user-provided callback in \texttt{deliver} opens the possibility of reentrant calls.
Unlike in the previous examples, however, these calls are safe.
By ensuring that the request status is updated (lines~\ref{lst:ln:tc:deliver-record} and~\ref{lst:ln:tc:cancel-record})
before calling untrusted code (lines~\ref{lst:ln:tc:deliver-callback} and~\ref{lst:ln:tc:cancel-send}),
TC prevents attackers from receiving refunds for canceling requests that are mid-delivery or already canceled.
Honest users, however, can still respond to data they receive from one request by creating or canceling \emph{other} requests.

For instance, a user contract may ask TC to function as a real-world timer and alert it at a specific real-world time.
When woken up, the contract might determine that it needs to wait longer and request that TC send another alert, say, 2 hours later.
A different user could make multiple parallel requests to retrieve the same data, e.g., a stock price, from several sources.
Once enough responses have arrived, the user might wish to cancel the outstanding requests to reduce costs.
Both of these patterns require safe reentrant calls into TC.
This work aims to allow this \emph{secure} reentrancy while still eliminating the vulnerabilities described~\mbox{above}.

\section{Information Flow Control}
\label{sec:ifc}

To obtain compositional security, it is natural to build on top of
information flow control (IFC), a classic way to obtain compositional
security guarantees such as noninterference~\citep{GM82}.
Most IFC work has focused on data
confidentiality~\citep{sm-jsac,YangYS12}, but IFC can also
protect integrity~\citep{integrity,jif-split} and
availability~\citep{zm05}. As our
goal is to guard against attackers performing unexpected calls into
trustworthy code, we track only integrity.

IFC systems assign labels to computation and data
within a system. As information flows through the system, the label on
the destination of information is constrained to be no less restrictive
than the label on its source. Since our goal is to enforce integrity,
less trusted information should be prevented from influencing more trusted information.

Secure information flow is statically enforceable by a type system~\citep{sm-jsac}.
When linking separate code modules together, the security guarantees offered by the
type system are automatically compositional, as long as the linked
modules agree on types at interface boundaries and account for
the confidentiality and integrity of the code itself~\cite{oakland12}.
Of course, real-world systems often have to interact with user-provided code
or legacy applications that do not obey the rules of the type system.
As we show, such noncompliant code can only
violate the security guarantees of code that
expresses trust in it.

\subsection{Label model}
\label{sec:ifc-label-model}

We specify integrity using a set of integrity labels $\Labs$
and give each piece of data $x$ a label $\ell_x$ representing its trust level.
The labels have a reflexive, transitive relation $\ell_1 \actsfor \ell_2$, which we read ``$\ell_1$ \emph{acts for} $\ell_2$,'' to denote that $\ell_1$ is at least as trusted as $\ell_2$.
That is, anything that can influence data labeled~$\ell_1$ can also influence data labeled~$\ell_2$.
\footnote{Most IFC systems use \emph{flows-to}, denoted $\sqsubseteq$.
  We use acts-for as we find it intuitive, and the two mean the same thing when only tracking integrity.}
Data $x$ can thus safely influence data $y$ only when $\ell_x \actsfor \ell_y$.
Influence can be either \emph{explicit}---by assigning $x$ directly to $y$---or
\emph{implicit}---by conditioning on $x$ and assigning different values to $y$ in each branch.
For explicit flows, a simple check that $\ell_x \actsfor \ell_y$ at the point of assignment is sufficient.
To control implicit flows, a \emph{program counter label}, written~$\pc$,
tracks the integrity of the computation itself, as is standard~\citep{sm-jsac}.
Inside a branch conditioned on $x$, the value of $x$ has influenced
control flow, so we require the constraint $\ell_x\actsfor\pc$.
Assigning a variable~$y$ to some value then requires $\pc \actsfor \ell_y$,
ensuring transitively that $\ell_x \actsfor \ell_y$.

$\Labs$ must also have some additional structure.
Any pair of labels $\ell_1$ and~$\ell_2$ must have a join, denoted $\ell_1 \join \ell_2$, and a meet, denoted $\ell_1 \meet \ell_2$.
The join is the least upper bound and the meet is the greatest lower bound, so
\begin{eqnarray*}
  \ell_1 \join \ell_2 \actsfor \ell & \iff & \ell_1 \actsfor \ell \text{ and } \ell_2 \actsfor \ell \\
  \ell \actsfor \ell_1 \meet \ell_2 & \iff & \ell \actsfor \ell_1 \text{ and } \ell \actsfor \ell_2.
\end{eqnarray*}
We can then safely label information influenced by both $\ell_1$ and~$\ell_2$ with label $\ell_1 \join \ell_2$, for example.
Lastly, the join and meet operators must distribute: $\ell_1 \join (\ell_2 \meet \ell_3) = (\ell_1 \join \ell_2) \meet (\ell_1 \join \ell_3)$.
These properties collectively make $(\Labs, \actsfor)$ a \emph{distributive lattice}.%

This additional structure supports the precision and flexibility of
our approach to enforcing reentrancy security,
discussed in Section~\ref{sec:type-system}.
Luckily, existing label models are typically distributive lattices,
including two-point lattices, subset lattices of permissions~\citep{histar},
and free distributive lattices over a set of principals~\citep{ml-sp98,flam}.
In smart-contract systems, for example, it is natural to view contracts themselves as principals with different trust relationships among them.
We might then employ \emph{decentralized} information flow control~\citep{ml-tosem}
where labels are constructed from principals (e.g., contracts) that can influence data or computation.

\subsection{Endorsement}
\label{sec:ifc-endorsement}

Strictly enforcing IFC allows systems to enforce strong security properties like noninterference,
which forbids \emph{any} influence from untrusted information to trusted information.
Noninterference, however, is too restrictive to build real applications,
so practical IFC systems allow \emph{downgrading}.
Downgrading integrity, known as \emph{endorsement}~\citep{zcmz03},
treats information with a low-integrity label as being more trustworthy than its source would indicate.

From the IFC perspective, services like smart contracts endorse frequently, though implicitly.
They expose functions that accept calls from untrusted users,
yet modify trusted local state.
In other words, untrusted state affects trusted state,
which an IFC system should only allow via endorsement.

Existing IFC languages support these trusted functions, but make them explicit.
For example, the Jif language~\citep{jif-release35} supports
\emph{autoendorse} methods that can be called by an untrusted caller
and that boost the integrity of the $\pc$~label on entry.

Viewed from the perspective of $\pc$ integrity, reentrancy attacks
all exhibit a distinctive pattern: they involve trusted
(high-integrity) code calling lower-integrity code, which
then calls back into high-integrity code by exploiting endorsement.
However, existing endorsement mechanisms in Jif and other
systems~\citep{asbestos,histar,flume,jfabric} do not prevent this
potentially dangerous control-flow pattern. These IFC systems are 
thus vulnerable to reentrancy attacks.
Preventing reentrancy attacks requires new restrictions on endorsement.

\section{Reentrancy and Security}
\label{sec:secure-reentrancy}

The examples in Section~\ref{sec:motivation} show the need
across application domains to constrain reentrancy
without eliminating it entirely.
We build on our previous work~\citep{cecchetti-fab20} to provide
flexible definitions of reentrancy and security
based on information flow control.
This choice gives access to existing IFC tools and techniques
with their strong data security guarantees,
while making possible a precise, semantic specification of~\mbox{security}.

\subsection{Defining Reentrancy}
\label{sec:reentrancy-defn-informal}

Prior work~\citep{luu2016making, obsidian-toplas,
grossman2017online,albert2020taming} focuses on smart contracts and defines reentrancy in those terms:
if contract $A$ calls contract~$B$, which calls back into
contract $A$, the second call, and thus the entire execution, is
considered reentrant.
If no calls to $A$ occur before the call to~$B$ returns, the execution is non-reentrant.
We refer to this notion of reentrancy as \emph{object reentrancy}, viewing contracts as a form of object.

We avoid object reentrancy because it relies on object boundaries---a fundamentally syntactic construct---to define security.
Instead we define reentrancy with respect to the integrity level of computation.
As integrity levels are part of a semantic security specification,
using them to define a security-relevant property is sensible.
This view leads to the following informal definition.
\begin{definition}[$\ell$\=/Reentrancy (informal)]
  \label{defn:reentrancy-informal}
  If computation~$C_1$ calls computation~$C_2$, which then (possibly indirectly) calls~$C_3$,
  the execution is reentrant with respect to label~$\ell$, or \emph{$\ell$\=/reentrant},
  if $C_1$ and $C_3$ are trusted at~$\ell$, but $C_2$ is not.
\end{definition}
Note that $C_1$ and $C_3$ may be the same or different, as long as they are both trusted at~$\ell$.

\begin{figure}
  \centering
  \newlength{\secFigHeight}
  \newlength{\secFigGap}
  \newlength{\ctrctGap}
  \newlength{\ctrctWidth}
  \setlength{\secFigGap}{0.25em}
  \setlength{\ctrctGap}{0.33em}
  \setlength{\secFigHeight}{6em}%
  \setlength{\ctrctWidth}{4em}%
  \hspace*{\stretch{1}}%
  \begin{subfigure}[t]{0.3\columnwidth}
    \centering
    \begin{tikzpicture}[sec dom size/.style={minimum height=\secFigHeight, minimum width=\ctrctWidth + 2*\secFigGap}]
      \node[security domain, sec dom size] (sec-dom) {};

      \node[contract, minimum height=(\secFigHeight - 2*\secFigGap), minimum width=\ctrctWidth] (contract-a) at (sec-dom) {};
      \node[anchor=north west] at (contract-a.north west) {$A$};

      \node[contract, minimum height=(\secFigHeight - 2*\secFigGap), minimum width=\ctrctWidth,right=\ctrctGap of sec-dom] (contract-b) {};
      \node[anchor=north west] at (contract-b.north west) {$B$};

      \node[call point] (a-call) at ($(contract-a)+(1.25em,0.175*\secFigHeight)$) {};
      \node[call point] (a-receive) at ($(contract-a.south)+(-1.25em,0.175*\secFigHeight)$) {};
      \node[call point] (b-call) at (contract-b|-a-receive) {};

      \path (a-call) edge[call,bend left] (b-call);
      \path (b-call) edge[reentrant call,bend right] (a-receive);
    \end{tikzpicture}
    \caption{Object reentrancy and $\ell$\=/reentrancy are the same when object and trust boundaries match.}
    \label{subfig:reentrancy:classic}
    \end{subfigure}%
  \hspace*{\stretch{1}}%
  \begin{subfigure}[t]{0.3\columnwidth}
    \centering
    \begin{tikzpicture}[sec dom size/.style={minimum height=0.3*\secFigHeight, minimum width=\ctrctWidth - 2*\secFigGap}]
      \node[security domain, sec dom size] (sec-dom) {};

      \node[contract, minimum height=(\secFigHeight - 2*\secFigGap), minimum width=\ctrctWidth, anchor=north] (contract-a) at ($(sec-dom.north)+(0,\secFigGap+1em)$) {};
      \node[anchor=north west] at (contract-a.north west) {$A$};

      \node[contract, minimum height=(\secFigHeight - 2*\secFigGap), minimum width=\ctrctWidth, right=2*\secFigGap of contract-a] (contract-b) {};
      \node[anchor=north west] at (contract-b.north west) {$B$};

      \node[call point] (trusted-a) at (sec-dom) {};
      \node[call point] (untrusted-a) at ($(contract-a)-(0,0.25*\secFigHeight - 0.375*\secFigGap)$) {};
      \node[call point] (untrusted-b) at (contract-b|-untrusted-a) {};

      \path (trusted-a) edge[call,bend left=35] (untrusted-b);
      \path (untrusted-b) edge[reentrant call,dotted,color=secureCall,bend right] (untrusted-a);

      \begin{pgfonlayer}{background}
        \node[fill=white, draw=none, minimum height=\secFigHeight, minimum width=1pt] (spacer) at (contract-a) {};
      \end{pgfonlayer}
    \end{tikzpicture}
    \caption{Partially-trusted objects can create object reentrancy that is not $\ell$\=/reentrancy.}
    \label{subfig:reentrancy:partial-trust}
  \end{subfigure}%
  \hspace*{\stretch{1}}%
  \begin{subfigure}[t]{0.3\columnwidth}
    \centering
    \begin{tikzpicture}[sec dom size/.style={minimum height=\secFigHeight, minimum width=\ctrctWidth + 2*\secFigGap}]
      \node[security domain, sec dom size] (sec-dom) {};

      \node[contract, minimum height=0.5*\secFigHeight - 1.5*\secFigGap, minimum width=\ctrctWidth, anchor=north] (contract-a) at ($(sec-dom.north)-(0,\secFigGap)$) {};
      \node[anchor=north west] at (contract-a.north west) {$A$};

      \node[contract, minimum height=0.5*\secFigHeight - 1.5*\secFigGap, minimum width=\ctrctWidth, anchor=south] (contract-c) at ($(sec-dom.south)+(0,\secFigGap)$) {};
      \node[anchor=north west] at (contract-c.north west) {$C$};

      \node[contract, minimum height=(\secFigHeight - 2*\secFigGap), minimum width=\ctrctWidth, right=\secFigGap of sec-dom] (contract-b) {};
      \node[anchor=north west] at (contract-b.north west) {$B$};

      \node[call point] (trusted-a) at (contract-a) {};
      \node[call point] (trusted-c) at (contract-c) {};
      \node[call point] (untrusted-b) at (contract-b|-trusted-c) {};

      \path (trusted-a) edge[call,bend left=40] (untrusted-b);
      \path (untrusted-b) edge[reentrant call,bend right] (trusted-c);
    \end{tikzpicture}
    \caption{Mutually-trusting objects can create $\ell$\=/reentrancy that is not object reentrancy.}
    \label{subfig:reentrancy:mutual-trust}
  \end{subfigure}%
  \hspace*{\stretch{1}}
  \caption{Comparing $\ell$-reentrancy to object reentrancy.
    Boxes represent objects, the blue shaded region is high-integrity code, and arrows represent calls.}
  \label{fig:reentrancy-pic}
\end{figure}
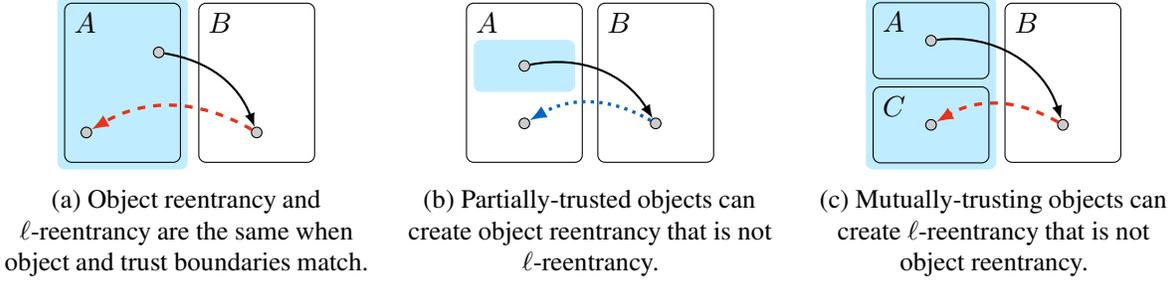

Figure~\ref{fig:reentrancy-pic} depicts how $\ell$\=/reentrancy relates to object reentrancy.
If an entire object is trusted at~$\ell$ and nothing else is (Figure~\ref{subfig:reentrancy:classic}),
$\ell$-reentrancy and object reentrancy align.
However, object and trust boundaries may differ, leading to different definitions.
If a trusted operation in $A$ calls untrusted $B$,
a call to an \emph{untrusted} portion of $A$ (Figure~\ref{subfig:reentrancy:partial-trust}),
would be considered reentrant in an object-based definition but not $\ell$\=/reentrancy.
Such a call could correspond to a Town Crier user updating a request callback during data delivery
or a web app accessing untrusted user profile data while modifying a trusted billing key--value store.
These operations are never dangerous, as low-integrity operations cannot damage high-integrity data.
By contrast, one application may be split across multiple mutually trusting objects.
For example, such a split in Ethereum's Parity Wallet led to two famous attacks~\citep{parity-extract,parity-freeze}.
For an application split across $A$~and~$C$, if $A$~calls~$B$,
then a call from $B$ into $C$ (Figure~\ref{subfig:reentrancy:mutual-trust}) is a reentrant call into the application.
By relying on trust levels, $\ell$\=/reentrancy properly identifies this pattern as reentrancy,
while object reentrancy does not.

To employ $\ell$\=/reentrancy, each operation needs an integrity level.
Conveniently, the $\pc$~label used to control implicit information flows
(Section~\ref{sec:ifc-label-model}) provides such a label.
It combines the integrity of the code and the integrity of data influencing the control flow
to specify how trusted an operation is to execute when it does,
making it ideal to define a property of trusted and untrusted operations calling each other.

\subsection{Reentrancy Security}
\label{sec:security-informal}

While $\ell$\=/reentrancy defines reentrancy based on integrity patterns of the control flow,
it does not tell us when it is secure.
An option taken by some work~\citep{obsidian-toplas,das2019resource} is to declare
all reentrancy (according to their definition) dangerous and to outlaw it entirely.
With an appropriate definition of reentrancy, this would eliminate vulnerabilities,
but safe reentrancy has legitimate uses, as illustrated by
the Town Crier example.

To eliminate the need for difficult manual reentrancy analysis,
we define ``secure reentrancy''
as reentrancy that programmers can ignore when analyzing correctness.
In general, a safe way to accomplish this goal is to ensure that
reentrancy cannot enable program behaviors that would not exist without it.
These behaviors could be program invariants, such as
Uniswap holding the product of its asset quantities constant
or the key--value store never writing to unallocated memory;
they could be statements about how state changes, like Town Crier's request ID monotonically increasing;
or they could be more complex properties like noninterference.

\begin{figure}
  \tikzset{%
    behavior/.style={circle},
    all behavior/.style={behavior,draw=black},
    safe behavior/.style={behavior,thick},
    danger fill/.style={pattern=north east lines,pattern color=insecureColor},
  }
  \newlength{\secPicSize}
  \newlength{\safeBehaveSize}
  \setlength{\secPicSize}{5.5em}
  \setlength{\safeBehaveSize}{4em}
  \centering
  \begin{tikzpicture}[
      text node/.style={font=\footnotesize},
      caption/.style={text node,inner sep=0pt},
    ]
    \node[all behavior,danger fill,thick,minimum size=\secPicSize] (danger-all) {};
    \node[safe behavior,draw=secureRegion!50!black,secure fill,minimum size=\safeBehaveSize] (danger-safe) at ($(danger-all)-(0.25em,0.25em)$) {};
    \node[caption,anchor=south] (caption-a) at ($(danger-all.south)-(0,\baselineskip)$) {(a) Vulnerable system};

    \node[all behavior,secure fill,line width=2pt,minimum size=\safeBehaveSize,right=10em of danger-safe] (safe-all) {};
    \node[safe behavior,dashed,fill=none,draw=secureRegion,minimum size=\safeBehaveSize] (safe-safe) at (safe-all) {};
    \node[caption] (caption-b) at (safe-all|-caption-a) {(b) Secure system};

    \node[draw=none] (anchor) at ($(danger-safe-|danger-all.east)!0.5!(safe-all.west)$) {};
    \node[text node,anchor=south] (all-behavior) at ($(anchor)+(0,0.5em)$) {All behavior};
    \node[text node,below=0.67em of all-behavior] (safe-behavior) {\parbox{6em}{\centering Non-reentrant behavior}};

    \draw[color=black,-latex] (all-behavior.west) -- (danger-all.east|-all-behavior.south);
    \draw[color=black,-latex] (all-behavior.east) -- ($(safe-all.west|-all-behavior.south)+(0.5pt,0)$);
    \draw[color=black,-latex] (safe-behavior.west) -- ($(danger-safe)+(0.5em,-1em)$);
    \draw[color=black,-latex] (safe-behavior.east) -- ($(safe-safe)-(0.5em,1em)$);
  \end{tikzpicture}
  \caption{The set of possible behaviors in a secure vs a vulnerable system.
    In a vulnerable system, reentrancy can introduce behaviors not possible without it.
    In a secure system, all behaviors are possible in non-reentrant executions.}
  \label{fig:security-picture}
\end{figure}
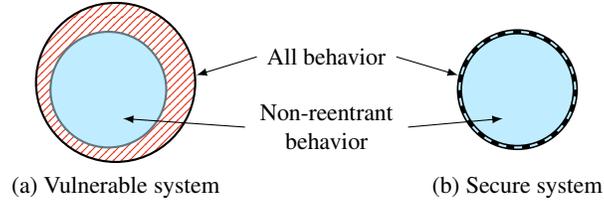

Programmers cannot hope to guarantee properties
that unknown or untrusted code can directly violate,
so our definition ignores such properties entirely.
Specifically, $\ell$\=/reentrancy security considers only
properties defined over state trusted at label~$\ell$.
We refer to these as \emph{$\ell$\=/integrity properties},
leading to the following security definition, depicted visually in Figure~\ref{fig:security-picture}.
\begin{definition}[Reentrancy Security (informal)]
  \label{defn:reentrancy-security-informal}
  A program is \emph{$\ell$\=/reentrancy-secure} if every $\ell$\=/integrity property,
  such as a program invariant, that holds
  for all non-$\ell$\=/reentrant executions holds for all executions.
\end{definition}

Definition~\ref{defn:reentrancy-security-informal} specifies a semantic notion of security and helps identify safe forms of reentrancy.
For instance, a high-integrity computation making a low-integrity call
as its last operation---in tail position---no longer needs high integrity.
That is, any reentrant call will have the same effect as making a second, non-reentrant call after the first computation returns.
We refer to this secure form of reentrancy as \emph{tail reentrancy}.
Tail reentrancy also provides a principled explanation for a common smart-contract programming best practice:
performing all state modifications before calling other contracts~\citep{solidity-sec-consider}.
Done properly, this design pattern ensures that all reentrant calls are tail-reentrant, and thus safe.

Definition~\ref{defn:reentrancy-security-informal} is also flexible.
For a specific application, we could refine it to require only that reentrancy
does not violate particular programmer-specified application properties.
To keep annotation burden low and to avoid the need to
specify detailed program properties,
our definition requires that $\ell$\=/reentrant executions maintain \emph{all} properties that hold without reentrancy.
However, the later formal definition (Definition~\ref{defn:reentrancy-security}) allows such refinement
simply by restricting a universal quantifier.

\subsection{Enforcing Reentrancy Security}
\label{sec:enforcing-security}

As described above, $\ell$-reentrancy occurs when high-integrity code calls low-integrity code that then calls back into high-integrity code before returning.
IFC only permits this pattern through the autoendorse mechanism described in Section~\ref{sec:ifc-endorsement}.
Many services, including the examples in Section~\ref{sec:motivation}, require untrusted users to make requests into trusted code,
making some version of autoendorse necessary.
We therefore allow it, but with additional restrictions.

In particular, endorsement of control flow is restricted by \emph{locking integrity}.
When a function endorses the integrity of the control flow to label~$\ell$,
integrity~$\ell$ is locked, preventing further endorsement up to~$\ell$ until the original call returns.
Locking
allows an honest user to invoke a service one or more times in sequence using a call-and-return pattern,
but prevents an adversary from reentering into high-integrity code.

The semantics of these locks is to prevent autoendorsement
from granting integrity that is locked.
A trusted operation is then always given the chance to reestablish any
high-integrity invariants or properties it may have temporarily invalidated
before an attacker can invoke another trusted operation.
To safely autoendorse from integrity~$\pc_1$ to integrity~$\pc_2$,
for any operation~$\pc_2$ is trusted to perform,
either $\pc_1$ must already be trusted at that level or the requisite integrity must be unlocked.
Formally, when integrity $\ell_L$ is locked, then for all labels~$\ell$,
if $\ell_L \actsfor \ell$ and $\pc_2 \actsfor \ell$, then $\pc_1 \actsfor \ell$.
The definition of lattice join quickly shows that this rule is equivalent to $\pc_1 \actsfor \pc_2 \join \ell_L$.

We could track and enforce locks statically, as part of the type system, or dynamically in the runtime.
Static locking---proving that a dynamic lock would never prevent execution---%
imposes no overhead and avoids unexpected errors at run time.
Unfortunately, purely static locks interact poorly with code that may not enforce the same guarantees.
If some unknown code might call autoendorse functions---violating a static lock,
meaning a dynamic lock would halt execution---a sound type system must assume the worst
and prevent all calls to that code when integrity may be locked.
This highly restrictive outcome would violate a core design goal of this work:
providing compositional security even when interacting with unknown code.
Dynamic locks avoid this constraining over-approximation
at the expense of run-time cost.

We therefore take a hybrid approach and separate locked integrity into a static component and a dynamic one.
The type system automatically adds endorsed control flow to the static component,
but programmers can explicitly move integrity from the static component to the dynamic one.
This approach achieves the run-time efficiency and predictability of static mechanisms when security
can be proved statically, while still supporting safe interaction with
unknown or untrusted code through more expressive dynamic locks.

The calculus does not specify how to implement dynamic locks.
They could be built into the runtime, tracked by a security monitor, or even implemented as a library.
So long as all code trusted at level~$\ell$ is well-typed and agrees on \emph{some} protocol to enforce the dynamic portion of the locks,
the system will preserve $\ell$\=/reentrancy security.
There is no requirement that untrusted check integrity locks statically or dynamically.

\section{A Core Calculus for Secure Reentrancy}
\label{sec:language}
\label{sec:lang-syntax}

We present the \langnameLong, an object-oriented core calculus that
models how a programming language can implement the above ideas.
Figure~\ref{fig:lang-syntax} gives the syntax for \langname.
It extends Featherweight Java (FJ)~\citep{ipw01}
with information flow labels and, to support mutation, also
reference cells~\citep[Chapter 13]{tapl}.

\begin{figure}
  \small\centering
  $\displaystyle
    \begin{array}{rcl}
      f,m,x & \in & \mathcal{V} \quad \text{(variable, method, and field names)} \\[0.25em]
      \ell, \pc & \in & \Labs \quad \text{(integrity labels)} \\[0.25em]
      t & ::= & \unit \alt \bool \alt \RefType{\tau} \alt C \\[0.25em]
      \tau & ::= & t^\ell \\[0.25em]
      \mathit{CL} & ::= & \cdef{C}{\ell}{C}{\overline{f}\ty\overline{\tau}}{K}{\overline{M}} \\[0.25em]
      K & ::= & C(\overline{f}\ty\overline{\tau})~\{\super(\overline{f}) \seq \this.\overline{f} = \overline{f}\} \\[0.25em]
      M & ::= & \mdef{\tau}{m}{\pc}{\pc}{\ell}{\overline{x}\ty\overline{\tau}}{e} \\[0.25em]
      v & ::= & x \alt () \alt \True \alt \False \alt \loc \alt \Null \alt \new~C(\overline{v}) \\[0.25em]
      e & ::= & v \alt \IfThenElse{v}{e}{e} \\
      & | & \RefOp{v}{\tau} \alt \deref{v} \alt v \assign v \\
      & | & (C)v \alt v.f \alt v.m(\overline{v}) \\
      & | & \Endorse{v}{\ell}{\ell} \alt \lockIn{\ell}{e} \\
      & | & \letIn{e}{e} \\
    \end{array}$
  \caption{Syntax for \langname}
  \label{fig:lang-syntax}
\end{figure}

\langname employs fine-grained IFC, so each type $\tau$ consists
of a base type $t$ and an integrity label $\ell$.
For simplicity, we limit base types to $\unit$, $\bool$, references, and object types.
To simplify proofs, null references are allowed.

Class and method definitions extend
those in FJ with integrity labels.
To model distributed systems, we consider code a form of data that may come from multiple sources,
so each class definition $\mathit{CL}$ includes a label $\ell_C$ for the integrity of the code.

A method definition $M$ contains labels $\pc_1 \EndrsSymb \pc_2;\ell$.
Most IFC systems give functions a single $\pc$ label, but \langname has two:
$\pc_1$~specifies the minimum integrity required to call~$m$,
while~$\pc_2$ specifies the integrity at which~$m$ operates.
Separating these labels supports autoendorsement as described in Section~\ref{sec:ifc-endorsement}.
If $\pc_1 \nactsfor \pc_2$, then $m$ is an autoendorse function.
Both $\pc$ labels are bounded by $\ell_C$, so code may only perform
operations that $\ell_C$ is trusted to perform.
The label $\ell$ specifies the locks method $m$ promises not to violate.

The $\If$ syntax includes the $\pc$ label used for the branches.
We make this label explicit only to simplify the operational semantics.
In practice, it is easy to infer automatically.

The $\EndorseName$ expression endorses data as in other IFC systems with downgrading.
The term $\lockIn{\ell}{e}$ converts static locks to dynamic ones.
In the operational semantics, $e$~executes with~$\ell$ dynamically locked,
so the type system can safely release any static lock on~$\ell$ when type-checking $e$.

Expression subterms consist mostly of (open) values, not arbitrary expressions.
In particular, $\Let$ statements are
the only way to sequentially compose computation.

Because \langname is object-oriented, it can
model interacting services and reentrancy concerns.
An application or contract implementation is a class,
and a contract or instance of that application is an object of that class type,
allowing easy interaction between different services.
Moreover, inheritance allows applications that share common features to inherit form a common parent.
For instance, a blockchain smart contract system can be modeled by having all contracts inherit
from a \programfont{Contract} class that implements tracking of~currency.

\subsection{\langname Operational Semantics}
\label{sec:semantics}

\langname has a small-step substitution-based semantics.
Most rules are standard for an object-oriented language with mutable references~\citep{ipw01,tapl},
with a few additions for security.

Because expressions are built mostly out of values, evaluation contexts are simple.
Indeed, $\Let$ expressions are the only surface syntax to serve as evaluation contexts.
We introduce three new syntactic forms as evaluation contexts to
enable precise tracking
of function boundaries, execution integrity, and dynamic locks.
These \emph{statements} are denoted by $s$.
$$\begin{array}{rcl}
  E & ::= & [\cdot] \alt \letIn{E}{e} \alt \funend{\tau}~E \alt \atpc{E} \alt \withLock{E} \\[0.25em]
  s & ::= & E[e] \\
\end{array}$$

Semantic steps are defined on a pair of a statement~$s$ and a \emph{semantic configuration}:
a four-tuple\linebreak
$\config = \confTuple*$.
Unlike in FJ, the class table $\CT$ is explicit,
as the security definitions in Section~\ref{sec:proof} quantify over possible class tables.
A heap $\sigma$ maps locations to value--type pairs,
and $\Sigma_\sigma$ denotes the location-to-type mapping induced by~$\sigma$.
That is, $\Sigma_\sigma(\loc) = \tau$ if and only if $\sigma(\loc) = (v, \tau)$ for some~$v$.
The final two elements, $\mlabList$ and $\lockList$ are both lists of integrity labels.
$\mlabList$ tracks the integrity of executing code,
and $\lockList$ tracks the dynamic portion of the currently-locked integrity.
For notational ease, we reference the components of $\config$ freely when only one group is in scope
and we write $\config[X/\lockList]$ to denote $\confTuple{\sigma}{\mlabList}{X}$, and similarly for $\sigma$~and~$\mlabList$.

\begin{figure}[tb]
  \centering
  \begin{ruleset}[0.8\textwidth]
    \EIfTRule* \\ \EAtPcRule*
    \\
    \ERefRule* \\ \EAssignRule*
    \\
    \ECallRule* \\ \EReturnRule*
    \\
    \ELockRule* \\ \EUnlockRule*
  \end{ruleset}
  \caption{Selected small-step semantic rules for \langname.}
  \label{fig:semantics}
\end{figure}

Figure~\ref{fig:semantics} presents selected semantic rules.
The complete semantics is in Figure~\ref{fig:full-semantics} (Appendix~\ref{sec:full-lang}).
In the semantic rules, $v$ refers to a closed value, not a variable.
In addition to many standard rules,
the rules \ruleref{E-Lock} and \ruleref{E-Unlock}
dynamically lock and unlock labels.
The semantics abstracts out the many possible lock implementations,
merely tracking the set of locked labels and defining where to check them.
The rules for conditionals (\ruleref{E-IfT} and \ruleref{E-IfF}) now include tracking terms.

The key rule is \ruleref{E-Call}.
It looks up the definition of a method with $\mbody$ (Appendix~\ref{sec:full-lang}) and performs several dynamic checks:
it verifies that the arguments all have the correct types,
that the caller has sufficient integrity to invoke the function,
and that calling the method does not violate any dynamically locked label~\mbox{$\ell \in \lockList$}.

\paragraph{Dynamic Security Checks}
Four rules---\ruleref{E-Ref}, \ruleref{E-Assign}, \ruleref{E-Call},
and \ruleref{E-Return}---contain dynamic checks for type safety and information security.
These checks prevent untrusted code from placing ill-typed values in the heap or passing them to trusted code.
They similarly prevent untrusted code from modifying trusted heap locations in any way.
Such checks are critical for trusted code to safely interact with ill-typed attacker code in any information flow system.
While we do not detail how to implement dynamic typing or label checks here, there is considerable research into both.
Gradually typed languages do run-time type checking~\citep{gradual-typing},
and distributed IFC systems include run-time label checks~\citep[e.g.,][]{jfabric,dstar,hails}.
Moreover, when all high-integrity code is well-typed,
it is sufficient to isolate memory between objects, as in Ethereum contracts~\citep{wood2014ethereum},
and to execute run-time checks when entering trusted code.

\subsection{Type System for \langname}
\label{sec:type-system}

The type system for \langname contains two different forms for typing judgments: one for values and one for expressions.
The typing judgment for values is straightforward for a stateful language.
It takes the form $\Sigma; \Gamma \proves v : \tau$
where $\Sigma$ is a heap type mapping references to types and $\Gamma$ is a typing environment mapping variables to types.
We write $\Sigma \proves v : \tau$ when $\Gamma$ is empty, as we did in Section~\ref{sec:semantics}.

Values specify no computation so they require no security reasoning.
Typing judgments for expressions are more complex,
including a standard $\pc$ label to track the integrity of the control flow.
To secure reentrancy with static locks when possible,
they also include a label~$\lockLabBase$ representing locked integrity.

Allowing tail reentrancy while eliminating other forms of $\ell$\=/reentrancy
requires treating calls in tail position differently from calls in other positions.
We accomplish this goal not by restricting when a given call can occur,
but instead by restricting what can occur \emph{after the call returns}.
Instead of one lock label, this strategy uses two:
an \emph{input lock}~$\inLock$ that an expression must maintain to safely execute outside tail position,
and an \emph{output lock}~$\outLock$ specifying the locks the expression \emph{actually} maintains.
The typing judgment now takes the form $\Sigma; \Gamma; \pc; \inLock \proves e : \tau \dashv \outLock$.

For an expression $e$ to type-check with input lock $\inLock$,
each subexpression of $e$ outside tail position must maintain $\inLock$.
As non-value expressions only appear outside of tail position in $\Let$ expressions,
the following typing rule enforces this restriction.
$$\LetRule$$
This rule is standard except that it requires $\outLock' \actsfor \inLock$,
capturing the intuition above:
$e_1$ must maintain at least lock~$\inLock$, as it is outside tail position.
Because $e_2$ is in tail position in this expression, there is no similar restriction on~$\outLock$.

\begin{figure*}
  \centering
  \begin{ruleset}[\textwidth]
    \IfRule \and \AssignRule
    \and
    \parbox[b]{27.3em}{$\CallRule$}
    \and \LockRule
    \and
    \Rule{Method-Ok}{
      \inLock \actsfor \pc_2 \\
      \ell_C \actsfor \pc_2 \\
      \inLock \join \outLock' \actsfor \outLock \\
      \pc_1 \prot \overline{\tau_a} \\\\
      \raisebox{0pt}[1em][0.5em]{$\Sigma; \overline{x}\ty\overline{\tau_a}, \this\ty C^{\pc_2}; \pc_2; \inLock \proves e : \tau \dashv \outLock'$} \\\\
      \CT(C) = \cdefNoBody* \\\\
      (D, m) \in \dom(\mtype) ~\Longrightarrow~ \mtype(D, m) = \mtypeExpr*
    }{\Sigma \proves \mdef*~\mathsf{ok~in}~C}
    \setcounter{MethodOkRule@used}{1}
  \end{ruleset}
  \caption{Selected typing rules for \langname}
  \label{fig:type-system-selected}
\end{figure*}

Figure~\ref{fig:type-system-selected} contains selected typing rules for \langname.
The notation $\ell \prot \tau$ indicates that data of type $\tau$ is
no more trusted than~$\ell$;
that is, $\ell \prot t^{\ell'}$ if and only if $\ell \actsfor \ell'$.
The rules also use auxiliary lookup functions $\fields$ and $\mtype$ and a subtyping relation $\subtyp$
that includes both standard object subtyping and safe relabeling---$t^\ell \subtyp t^{\ell'}$ if and only if $\ell \actsfor \ell'$.
The complete type system
is in Figures~\ref{fig:type-system-expr} and~\ref{fig:type-system-class} (Appendix~\ref{sec:full-lang}).

Most typing rules (e.g., \ruleref{If} and \ruleref{Assign}) are standard for an information flow calculus~\citep{sm-jsac}.
The only non-standard rules are those that directly reference or constrain static locks:
sequential composition (\ruleref{Let}), method calls (\ruleref{Call}), and dynamic locking~(\ruleref{Lock}).

Most premises of \ruleref{Call} are standard.
They check that the object and arguments have appropriate types and ensure information security of the return type and control flow of the call.
They also check that the call does not violate any static locks ($\pc_1 \actsfor \pc_2 \join \inLock$)
and that it attenuates trust in the output by the integrity of both the object and the method ($\pc_2 \join \ell \prot \tau$).

This rule has two notable features.
The first is not what it requires, but rather what it \emph{does not} require.
There is no relation between the static input locks $\inLock$ of the surrounding environment
and $\outLock$, the locks maintained by the method itself.
This lack of constraint is precisely what enables tail reentrancy.
A call in tail position need not maintain any locks, so it may result in reentrancy.
Outside tail position, however, the \ruleref{Let} rule requires that the output locks
of the call expression---bounded by the locks maintained by the method---must act for $\inLock$.
\ruleref{Call} and \ruleref{Let} therefore combine to enable
safe tail reentrancy while ruling out other potentially dangerous reentrancy.

The second feature is that \ruleref{Call} does not maintain locks $\outLock$---the
locks maintained by the method---but instead only $\outLock \join \pc_2$.
This adjustment enables safe interaction with untrusted code that
might not enforce the same guarantees as \langname.
Such code may claim to maintain locks, but fail to do so.
Our safeguard follows the principle of decentralized IFC~\citep{ml-tosem}:
you can only be hurt by an adversary you trust.
We therefore attenuate the claimed lock label~$\outLock$ by the integrity of the code.

Due to \langname's inheritance structure, however, there is no way to determine the exact integrity of the code.
The implementation of~$m$ may come from~$C$ or any of its superclasses or subclasses.
We instead need a bound on the implementation's integrity.
The class typing rule \ruleref{Method-Ok} requires that the code's integrity act for~$\pc_2$ to define or override a method with integrity~$\pc_2$.
As a result, $\pc_2$ is the most precise bound on the code's integrity available to the type system.

To understand the \ruleref{Lock} rule, recall that the $\Lock$ term is designed to convert static locks to dynamic ones.
The type system must ensure that $\inLock$, the previous input locks, remain locked in some manner,
but it can safely release the portion that is dynamically checked.
In particular, \ruleref{Lock} splits $\inLock$ into $\ell$ and some $\inLock'$ such that $\inLock' \meet \ell \actsfor \inLock$.
Now $\inLock$ will remain locked as long as $e$ type-checks with static input lock $\inLock'$.
Similarly, $\lockIn{\ell}{e}$ actually maintains locks on both $\outLock'$---the locks $e$ maintains---and~$\ell$.
It is thus safe to trust $\outLock$ up to $\outLock' \meet \ell \actsfor \outLock$.
Notably, allowing these arbitrary label divisions is only secure because the label lattice is distributive.
Otherwise, separately locking $\inLock'$ and~$\ell$ could be insufficient to lock~$\inLock$, and similarly for $\outLock'$ and~$\outLock$.

Finally, \ruleref{Method-Ok} defines when a method is well-typed.
This rule implements the idea that autoendorse methods statically lock integrity by default.
Specifically, it requires $\inLock \actsfor \pc_2$,
so any expression outside tail position must
respect locks on the new, higher integrity of control flow.
The integrity of the code must also act for the integrity with which the function executes ($\ell_C \actsfor \pc_2$),
ensuring code cannot do anything its source is not trusted to do.
Next, the locks the method claims to enforce~($\outLock$) must be maintained both initially~($\inLock$) and throughout~($\outLock'$).
The last information-security check ($\pc_1 \prot \overline{\tau_a}$)
guarantees that any code trusted to call the method is also trusted to provide its arguments.

\subsection{Modeling Application Operation}

We aim to model applications that, like smart contracts,
service user requests and may persist state across requests.
We represent the current state of the world by a set of class definitions in a class table $\CT$ and a state map $\sigma$.
A single user interaction, which we term an \emph{invocation} $I$,
is a label specifying the user's integrity and a call to a single method of an object stored in $\sigma$.

Execution of an invocation $I = (\loc, m(\overline{v}), \ell)$ with state $\sigma$
starts from a semantic configuration
with the expression,
integrity $\ell$, and no locks, and step it to completion.
The notation $(I, \CT, \sigma) \Downarrow \sigma'$ signifies that it terminates
in updated state~$\sigma'$.
The following rule formalizes this idea, using
$\deref{\loc}.m(\overline{v})$ as shorthand for $\letIn[o]{\deref{\loc}}{o.m(\overline{v})}$.
\begin{mathpar}
  \Rule{E-Invoke}{
    \conf{\deref{\loc}.m(\overline{v})}{(\CT, \sigma, \ell, \cdot)} \stepsmany \conf{w}{(\CT, \sigma', \ell, \cdot)}
  }{(I, \CT, \sigma) \Downarrow \sigma'}
\end{mathpar}

The same notation denotes running a list of invocations $\overline{I}$ in sequence,
using the output state from one as the input state from the next.
That is, if $\overline{I} = I_1, \dotsc, I_n$ and $(I_i, \CT, \sigma_{i-1}) \Downarrow \sigma_i$ for each $1 \leq i \leq n$,
then we write $(\overline{I}, \CT, \sigma_0) \Downarrow \sigma_n$.

To type-check an invocation,
the expression used in the evaluation must be well-typed in the evaluation environment:
\begin{mathpar}
  \Rule{Invoke}{
    \Sigma; \cdot; \ell; \inLock \proves \deref{\loc}.m(\overline{v}) : \tau \dashv \outLock
  }{\Sigma \proves (\loc, m(\overline{v}), \ell)}
\end{mathpar}

\subsection{Examples Revisited}
\label{sec:examples-revisited}

We now revisit the examples from Section~\ref{sec:motivation}
to see how \langname detects application vulnerabilities while permitting secure implementations.

\paragraph{Uniswap}
The vulnerability (Section~\ref{sec:uniswap}) stems from an unexpected interaction between
an exchange, tokens, and a malicious user.
While they may all have different integrity, for simplicity,
we give the exchange and the tokens the same trusted label~$T$
and the user an untrusted label~$U$ with $U \nactsfor T$.

Anyone can call \texttt{sellXForY}, but it computes how much of asset $Y$ to move
and transfers tokens, so it must have label $U \EndrsSymb T; \outLock$ for some~$\outLock$.
Similarly, the token's \texttt{transferTo} method modifies high-integrity records,
so it needs label $\pc \EndrsSymb T; \outLock'$ for some labels~$\pc$ and~$\outLock'$.

The \ruleref{Method-Ok} rule requires \texttt{sellXForY} to type-check with some $\inLock$ where $\inLock \actsfor T$.
Because we sequence two calls to \texttt{transferTo},
\ruleref{Let} requires either $\outLock' \actsfor \inLock \actsfor T$,
or a dynamic lock on label~$T$ around (at least) the first transfer.
These options correspond precisely to the solutions suggested in Section~\ref{sec:uniswap}.
Requiring $\outLock' \actsfor T$ is a statement that Uniswap expects
the tokens not to call untrusted code.
A dynamic lock, by contrast, secures the exchange without assuming any particular token behavior
and correspondingly allows any value of~$\outLock'$.

Notably, \texttt{transferTo} can type-check with $\outLock' \actsfor T$ in either of two ways:
it can decline to call unknown code (i.e., remove lines~\ref{lst:ln:token-alert-send} and~\ref{lst:ln:token-alert-receive} in Figure~\ref{fig:uniswap}),
or the token itself could acquire a dynamic lock while making the calls.
The first option straightforwardly eliminates the vulnerability.
By locking $T$, the second option dynamically prevents reentrant calls
during a transfer to either the token or the exchange.

\paragraph{Key--value store}
We use the same labeling scheme: the key--value store application gets a trusted label~$T$
while the user gets an untrusted label~$U$.
Because anyone can call \texttt{getOrCompute} but it modifies trusted data, it must have label $U \EndrsSymb T; \outLock$ for some $\outLock$.
The user-provided computation function is not trusted, so it gets label $\pc \EndrsSymb U; \outLock'$ for some labels~$\pc$ and~$\outLock'$.

As in the Uniswap example above, \ruleref{Method-Ok} requires \texttt{getOrCompute} to type-check with some $\inLock \actsfor T$.
Because the user-provided fallback function executes in sequence before another trusted operation,
\ruleref{Let} and \ruleref{Call} combine to require either a dynamic lock or $\outLock' \join U \actsfor \inLock \actsfor T$.
This second option, however, is impossible because $U \nactsfor T$.

This forced reliance on a dynamic lock stems from the type system not trusting the user-provided callback to even type-check.
In a modified type system that separated trust in the code's execution from trust that it type-checks,
it would be sufficient to require that it type-check with high-integrity and some $\outLock' \actsfor T$.
This solution would correspond to a static guarantee that the user-provided callback does not invoke \texttt{clear}
or any other method modifying the store's internal state.

\paragraph{Town Crier}
As described in Section~\ref{sec:towncrier} and the original paper~\citep{Zhang2016},
Town Crier is secure despite using (object) reentrancy, and the type system can verify that.
Using the same labels again, we label Town Crier and the trusted service address~$T$ and the user~$U$.
We can give the functions the following signatures.
\begingroup
  \par\centering
  \begin{minipage}{23.2em}
\begin{lstlisting}[
      language=scif,
      morekeywords={[1]uint,void,address,bytes},
      numbers=none,
      basicstyle=\linespread{1.3}\small\ttfamily,
      framesep=0pt,
      frame=none
    ]
int request{(*$U \EndrsSymb T; T$*)}(params:(*$t^U$*), callback:address(*$^U$*))
void cancel{(*$U \EndrsSymb T; U$*)}(id:int(*$^U$*))
void deliver{(*$T \EndrsSymb T; U$*)}(id:int(*$^T$*), data:bytes(*$^T$*))
\end{lstlisting}%
  \end{minipage}%
  \par\noindent\ignorespacesafterend
\endgroup
The $\Request$ method---which just records the request parameters and updates a counter---type-checks simply.
The $\Cancel$ method type-checks with an endorsement on the condition on line~\ref{lst:ln:tc:cancel-cond} of Figure~\ref{fig:town-crier}.
Type-checking $\Deliver$ relies on TC trusting \texttt{SERVICE\_ADDR} not to call attackers when receiving money.
However, \texttt{SERVICE\_ADDR} is a hard-coded wallet address with no code that is already trusted to provide data to $\Deliver$,
so the operation sending it money can safely have the signature $T \EndrsSymb T; T$.
These labels allow $\Deliver$ to type-check as written.

\section{Formalizing Security Guarantees}
\label{sec:proof}

We now have the tools needed to formalize reentrancy and security from Section~\ref{sec:secure-reentrancy}.

\subsection{Attacker Model}
\label{sec:attacker-model}

Proving a security guarantee requires a well-defined attacker.
As $\ell$\=/reentrancy is parameterized on a label, we also parameterize attackers over what they compromise.
We assume that an attacker $\adv$ controls some collection of system components,
including anything that trusts any combination of those components.
For simplicity, we require a label~$\ellAdv$ representing the combined attacker power
and a label~$\ellHigh$ representing the minimum honest integrity,
where every label is either attacker-control or honest.
That is, for all $\ell \in \Labs$,
either $\ellAdv \actsfor \ell$ or $\ell \actsfor \ellHigh$, but not both.
\footnote{
  Our results hold for any partition of $\Labs$ into a downward-closed sublattice~$\mathcal{T}$ and an upward-closed sublattice~$\adv$,
  letting~$\ell$ be ``trusted'' if $\ell \in \mathcal{T}$.
  If $\mathcal{T}$ and $\adv$ are complete,
  this formulation is equivalent with $\ellHigh = \bigjoin \mathcal{T}$ and $\ellAdv = \bigmeet \adv$.
}
We prove that, for \emph{any} such~$\ellHigh$ and~$\ellAdv$, if all code trusted at~$\ellHigh$
abides by the static and dynamic locking requirements,
the system is $\ell$\=/reentrancy secure whenever $\ell \actsfor \ellHigh$.
This parameterization of the attacker ensures that only someone you trust can damage your security.

Notably, the requiring $\ellHigh$~and~$\ellAdv$ to exist means that,
to guaranteeing security at $\ell_1 \meet \ell_2$,
one or both of $\ell_1$ and $\ell_2$ must act for~$\ellHigh$, and therefore be honest.
In other words, trusting the combined power of two labels
is a statement that you believe at least one of those labels is honest, though you may not know which.
Combined with trust in $\ell_1 \join \ell_2$ expressing trust in both $\ell_1$ and $\ell_2$,
this idea supports modeling complex assumptions like ``at least $k$ of $n$ nodes are honest.''

Because reentrancy attacks stem from attacker code performing unexpected operations, we grant attackers considerable power.
Specifically, attackers can modify or replace any code that executes with low integrity---that is, any code where $\ellAdv \actsfor \pc$.
Allowing attackers to modify high-integrity code executing with a low-integrity~$\pc$ may seem unrealistic,
but experience has shown that code bases contain ``gadgets'' that
attackers can combine to achieve arbitrary functionality~\citep{rop,rop07}.
This expansive power conservatively models
the ability to exploit such gadgets without modeling the gadgets explicitly.

To model the attacker's ability to sidestep static security features,
we introduce a new term to ignore static lock labels.
\begin{mathpar}
  \small
  \begin{array}{rcl}
    e & ::= & \cdots \alt \ignoreLocks{e} \\[0.25em]
    E & ::= & \cdots \alt \ignoreLocks{E}
  \end{array}
  \\
  \EIgnoreLocksRule
  \\
  \IgnoreLocksRule
\end{mathpar}

Reasoning explicitly about ill-typed code is challenging,
so the formal model requires all code to type-check, but allows low-integrity code to use this new term.
Using $\ignoreLocksName$ may not appear to grant the full power of ignoring the type system.
After all, the type system limits the location of method calls and state modifications
based on the $\pc$~label, which attackers cannot modify.
However, low-integrity code can only interact with high-integrity code in three ways:
calling high-integrity methods, returning values to high-integrity contexts,
or writing to memory that high-integrity code will later read.
In each case, the operational semantics includes dynamic checks
to ensure memory safety and to ensure that method calls and state modifications
are only performed by sufficiently trusted code---exactly what the type system asks.

Indeed, the only constraint the type system imposes that these dynamic checks do not enforce
is the static locking that $\ignoreLocksName$ is designed to avoid.
Modeling well-typed high-integrity code and unknown attacker code is therefore
as simple as demanding that all code type-checks and high-integrity code does not use $\ignoreLocksName$,
formalized as follows.
\begin{samepage}
\begin{definition}[Lock Compliance]
  A class table $\CT$ \emph{complies with locks in $\ellHigh$\=/code}
  if, whenever
  \[
    \CT(C) = \cdef*
  \]
  and $\ell_C \actsfor \ellHigh$,
  then \ignoreLocksName does not appear syntactically in the body of any method $m \in \overline{M}$.
\end{definition}
\end{samepage}

Strong object-level memory isolation, like that in Ethereum, reduces the information
security checks of the semantics to type-checking high-integrity code.
Forcing dynamic lock checks, however, requires direct support in the system runtime.
As such features are uncommon, we model a system where attackers can freely ignore dynamic locks.
Specifically, we extend the operational semantics with a second rule for function calls, \ruleref{E-CallAtk},
which enables calls to attacker-controlled code without checking dynamic label locks.
\[
  \rulefiguresize
  \ECallAtkRule
\]
This rule is identical to \ruleref{E-Call},
except instead of checking dynamic locks, it checks that $\pc_2$ is untrusted ($\ellAdv \actsfor \pc_2$).

Interestingly, in systems that require even untrusted calls to check dynamic locks---admitting only \ruleref{E-Call} and not \ruleref{E-CallAtk}---%
trust of $\ell_1 \meet \ell_2$ can be safe even when neither $\ell_1$ nor~$\ell_2$ is honest.
Such systems enforce $\ellHigh$\=/reentrancy security whenever $\CT$ complies with locks in $\ellHigh$\=/code.
There can even exist labels~$\ell_1$ and~$\ell_2$ where $\CT$ does not comply with locks in $\ell_1$\=/code or $\ell_2$\=/code,
but $\ell_1 \meet \ell_2 \actsfor \ellHigh$,
meaning $\ellAdv$ cannot be a well-defined label.
The proofs in Appendix~\ref{sec:security-proofs-full} consider both system and attacker models.

\paragraph{Attacker-provided code}
In addition to having ill-typed code, attackers can tailor their attacks to the specific application.
We therefore define security with respect to \emph{any} system with the same high-integrity code.
Specifically, we employ a notion of $\ellHigh$\=/equivalent code that allows an attacker to add, remove, or replace code whenever $\pc \nactsfor \ellHigh$.

We formalize the equivalence using erasure on the code in a class table $\CT$.
Let $\lerase{\CT}$ denote $\CT$, but erasing
any class $C$ with low-integrity code ($\ell_C \nactsfor \ellHigh$),
any method $m$ that executes with low integrity ($\pc_2 \nactsfor \ellHigh$),
and the branches of $\If$ statements executing with low integrity ($\pc \nactsfor \ellHigh$).
Two class tables are then $\ellHigh$\=/equivalent if they erase to the same thing.
$$\CT \lequiv \CT' \overset{\triangle}{\iff} \lerase{\CT} = \lerase{\CT'}$$

Attackers can also freely modify low-integrity locations in the heap,
so we define $\ellHigh$\=/equivalent heaps using similar erasure.
As a heap $\sigma$ is a partial function from locations to value--type pairs,
memory is erased to $\lerase{\sigma}$ simply by erasing mappings with low-integrity types.
Formally, $\lerase{\sigma}(\loc) = \sigma(\loc)$ if $\sigma(\loc) = (v, t^\ell)$ with $\ell \actsfor \ellHigh$,
and it is undefined otherwise.
As with code, the equivalence follows directly from this erasure:
$$\sigma \lequiv \sigma' \overset{\triangle}{\iff} \lerase{\sigma} = \lerase{\sigma'}.$$

\subsection{Noninterference}
\label{sec:noninterference}

A typical goal for security in IFC systems, including our core
calculus, is \emph{noninterference}~\citep{GM82}, which for integrity
means untrusted data should not influence trusted data at all.
As we argued in Section~\ref{sec:ifc-endorsement}, noninterference is too restrictive,
and indeed, endorsement exists to violate it.
However, explicit endorsement should be the \emph{only} way to violate noninterference.

To state this,
we first need a notion of a class table $\CT$ being \emph{endorsement-free} for a label $\ell$.
\begin{definition}[Endorsement-Free]
  \label{defn:endorse-free}
  $\CT$ is \emph{$\ell$\=/endorsement-free} if,
  for all classes $C$ and methods $m$ such that
  \begin{align*}
    \cdef* & \in \CT \\
    \mdef* & \in \overline{M}
  \end{align*}
  the following two properties hold.
  (1) Either $\pc_1 \actsfor \ell$ or $\pc_2 \nactsfor \ell$,
  and (2) for any subexpression of $e$ of the form $\Endorse{v}{\ell_1}{\ell_2}$,
  similarly, either $\ell_1 \actsfor \ell$ or $\ell_2 \nactsfor \ell$.
\end{definition}
Intuitively, this definition says that $\CT$ is $\ell$\=/endorsement-free if
$\CT$ contains no means of endorsing either control flow or data from a label that $\ell$ does not trust to one that it does.

This condition is sufficient to prove a strong notion of noninterference at $\ell$.
Because the \langname semantics are nondeterministic with respect to selection of location names (\ruleref{E-Ref}),
we use a modified equivalence~$\lequivLocIso$ that allows renaming locations in addition to erasing low-integrity state.
See Appendix~\ref{sec:loc-name-iso} for the formal definition of this equivalence.

For partial functions $f$ and $f'$, we write $f \subseteq f'$ to mean $\dom(f) \subseteq \dom(f')$ and $f(x) = f'(x)$ wherever $f$ is defined.
\begin{theorem}[Noninterference]
  \label{thm:ni}
  Let $\CT$ be a class table where $\Sigma \proves \CT~\mathsf{ok}$ is $\ell$\=/endorsement-free.
  For any well-typed heaps $\sigma_1$ and $\sigma_2$ such that $\Sigma \subseteq \Sigma_{\sigma_i}$
  and any invocation $I$ such that $\Sigma \proves I$ and $(I, \CT, \sigma_i) \Downarrow \sigma_i'$,
  if $\sigma_1 \lequivLocIso \sigma_2$, then $\sigma_1' \lequivLocIso \sigma_2'$.
\end{theorem}

Theorem~\ref{thm:ni} follows by a complicated induction on the operational semantics,
erasing untrusted values in the heap.
See Appendix~\ref{sec:ni-proof} for details.

Note also that the theorem says nothing about lock compliance,
only endorsement freedom.
Indeed, reentrancy locks are unnecessary to enforce noninterference.

\subsection{Formalizing Reentrancy}
\label{sec:reentrancy-defn-formal}

Definition~\ref{defn:reentrancy-informal} in Section~\ref{sec:reentrancy-defn-informal} informally defines $\ell$\=/reentrancy
as a trusted computation calling an untrusted one, which then calls a trusted computation before returning.
We also noted that the $\pc$~label specifies the integrity of the control flow and is therefore ideal for defining reentrancy.

Because \langname's semantics has no explicit call stack,
it must insert $\atpcName$ tracking terms in the only places where the $\pc$~label of the currently-executing code can change:
conditionals and method calls.
The terms surround the body of the condition or method and remain until execution returns to the previous $\pc$~label.
Nested tracking terms appear precisely when code in one conditional or method body calls a second before returning.
We therefore formalize $\ell$\=/reentrancy as three nested $\atpcName$ terms
where~$\ell$ trusts the label of the first and third, but not the second.
As each condition or call may still have pending computation, we allow arbitrary evaluation contexts at each integrity level.
\begin{definition}[$\ell$\=/Reentrancy]
  \label{defn:reentrancy}
  A statement $s$ is \emph{$\ell$\=/reentrant} if, for some
  evaluation contexts $E_0$, $E_1$, $E_2$,
  \[
    s = E_0\Big[\atpc[\pc_1]{E_1\big[\atpc[\pc_2]{E_2[\atpc[\pc_3]{s'}]}\big]}\Big]
  \]
  where $\pc_1, \pc_3 \actsfor \ell$ but $\pc_2 \nactsfor \ell$.

  We say an invocation $I = (\loc, m(\overline{v}), \ell')$ is \emph{$\ell$-reentrant in $\sigma$}
  if $\conf{\deref{\loc}.m(\overline{v})}{\confTuple{\sigma}{\ell'}{\cdot}} \stepsmany \confExpr{s}$ where $s$ is $\ell$-reentrant.
\end{definition}

With a definition of reentrancy and a formal attacker model, we can formalize
the notion of security described in Section~\ref{sec:security-informal}.
Recall that ``secure reentrancy'' meant that
any program behavior possible with reentrancy is also possible without reentrancy.
Equivalently, state changes made by reentrant executions
must be possible using non-reentrant ones.

We describe the properties a program \emph{maintains}
using a modified Hoare logic~\citep{hoare-pofc}.
Because high-integrity code may interact with arbitrary attacker code,
we consider all possible invocations with $\ell$\=/equivalent code.
Specifically, the high-integrity component of $\CT$ maintains a property defined by a predicate pair~$(P, Q)$ if,
whenever $P$ holds on the input state, $Q$~must hold on the output state.

\begin{definition}[Predicate Satisfaction]
  \label{defn:pred-satisfy}
  Given a class table $\CT$, a heap type $\Sigma$, and state predicates $P$ and $Q$,
  we say that $\CT$ \emph{satisfies $(P,Q)$ at $\ell$ in $\Sigma$}, denoted $\satPQ*{\CT}$,
  if, for any $\CT'$ such that $\CT \lequiv \CT'$,
  any well-typed state~$\sigma_1$ where $\Sigma \subseteq \Sigma_{\sigma_1}$,
  and any invocation sequence $\overline{I}$ such that $\Sigma_{\sigma_1} \proves \overline{I}$ and $(\overline{I}, \CT', \sigma_1) \Downarrow \sigma_2$,
  then $P(\sigma_1)$ implies~$Q(\sigma_2)$.
\end{definition}

To simplify proofs, the definition requires invocations to be well-typed.
The requirement does not, however, weaken the security guarantee.
In a system like Ethereum without a strong type system, a
high-integrity contract would need to examine its arguments to ensure
they are well-typed. We assume this facility is built into the runtime.

The predicates $P$ and $Q$ can capture a variety of program properties.
A simple example is program invariants---such as Uniswap's invariant
on the product of the token balances---in which case $P$ and $Q$ would be the same.
Quantifying over a potentially infinite set of predicates, as the security definition does below, allows for arbitrarily complex properties.
For example, requiring a specific high-integrity output state for each possible high-integrity input state would enforce noninterference.
A demonstration of interference would demonstrate that one such predicate pair is not satisfied.

Our goal, however, is not to guarantee any specific properties,
but to formalize the idea that reentrancy should not introduce new behavior.
Definition~\ref{defn:pred-satisfy} says nothing about reentrancy.
It captures the \emph{entire} set of possible behaviors, including the reentrant ones.
Saying that a complete set of behaviors is equivalent to the non-reentrant behaviors
requires a definition of non-reentrant behaviors.
For that, we simply restrict our previous definition to executions that are not $\ell$\=/reentrant.

\begin{definition}[Single-Entry Predicate Satisfaction]
  \label{defn:single-entry-pred-satisfy}
  Given a class table $\CT$, a heap type $\Sigma$, and state predicates $P$ and $Q$,
  we say that $\CT$ \emph{single-entry satisfies $(P,Q)$ at $\ell$ in $\Sigma$}, denoted $\singleEntSatPQ*{\CT}$,
  if $\CT$ satisfies $(P,Q)$ at $\ell$ in $\Sigma$ when restricted to invocation sequences $\overline{I}$ that are \emph{not} $\ell$\=/reentrant.
\end{definition}

These two definitions combine to specify the difference between non-reentrant program behavior and all program behavior.
To compare them, note that a program satisfies predicate pair~$(P, Q)$ precisely when no behavior violates it.
Therefore, if reentrancy can exhibit new behaviors---the program is insecure---there should be a predicate pair
that is single-entry satisfied, but not satisfied in general.

Because attackers can arbitrarily modify low-integrity state,
\emph{any} changes to low-integrity state are possible without $\ell$\=/reentrancy.
We correspondingly restrict our security notion to predicates that are unaffected by low-integrity state.
\begin{definition}[$\ell$\=/integrity Predicate]
  \label{defn:high-integ-predicate}
  We say a predicate $P$ is \emph{$\ell$\=/integrity} if,
  for all pairs of states $\sigma_1$ and $\sigma_2$,
  $$\sigma_1 \lequiv \sigma_2 ~\Longrightarrow~ P(\sigma_1) \Leftrightarrow P(\sigma_2).$$
\end{definition}

\noindent
We now define $\ell$\=/reentrancy security formally.
\begin{definition}[Reentrancy Security (formal)]
  \label{defn:reentrancy-security}
  We say a class table $\CT$ is \emph{$\ell$\=/reentrancy secure in $\Sigma$}
  if for all pairs $(P,Q)$ of $\ell$\=/integrity predicates, $\singleEntSatPQ*{\CT}$ implies \mbox{$\satPQ*{\CT}$}.
\end{definition}

\noindent
Definition~\ref{defn:reentrancy-security} is the core security definition \langname enforces.
\begin{theorem}
  \label{thm:security}
  For any label~$\ell$, class table~$\CT$, and heap type~$\Sigma$,
  if $\ell \actsfor \ellHigh$
  and $\Sigma \proves \CT~\mathsf{ok}$ complies with locks in $\ellHigh$\=/code,
  then $\CT$ is $\ell$\=/reentrancy secure in $\Sigma$.
\end{theorem}

Theorem~\ref{thm:security} follows from two core results.
First, all reentrancy allowed by \langname is tail reentrancy.
That is, if an invocation passes through an $\ell$\=/reentrant state,
then the outer high-integrity call ($\atpc[\pc_1]{E_1}$ in Definition~\ref{defn:reentrancy}) must be in tail position.

\begin{theorem}
  \label{thm:all-tail-reenter}
  For a label~$\ell$, class table~$\CT$, and well-typed heap~$\sigma_1$,
  if $\ell \actsfor \ellHigh$ and
  $\Sigma_{\sigma_1} \proves \CT~\mathsf{ok}$ complies with locks in $\ellHigh$\=/code,
  then for any invocation~$I$ and heap $\sigma_2$ where $\Sigma_{\sigma_1} \proves I$ and $(I, \CT, \sigma_1) \Downarrow \sigma_2$,
  all $\ell$\=/reentrant states in the execution are $\ell$\=/tail-reentrant.
\end{theorem}

\begin{proof}[Proof Sketch]
  The theorem follows from two facts.
  First, if a statement $s$ steps to a call to a method that grants integrity~$\ell$, then $s$ cannot maintain a lock on~$\ell$.
  Second, any statement executing with integrity~$\ell$ must maintain a lock on~$\ell$ (either statically or dynamically) unless it is in tail position.
  We provide a complete proof in Appendix~\ref{sec:only-tail-reenter-proof}.
\end{proof}

Once we know that all reentrant executions are tail-reentrant,
we need only show that tail reentrancy is secure.
The following theorem formalizes this idea by proving that,
if all $\ell$\=/reentrant states are $\ell$\=/tail-reentrant,
then single-entry predicate satisfaction translates to predicate satisfaction.

\begin{theorem}
  \label{thm:tail-reenter-safe}
  Let $\CT$ be a class table, $\sigma_1$ and $\sigma_2$ be well-typed heaps, and $I$ be an invocation
  such that \mbox{$(I, \CT, \sigma_1) \Downarrow \sigma_2$}
  where all $\ell$\=/reentrant states are $\ell$\=/tail-reentrant.
  For any $\ell$\=/integrity predicates $P$ and $Q$,
  if $\singleEntSatPQ*[\Sigma_{\sigma_1}]{\CT}$ and $P(\sigma_1)$, then $Q(\sigma_2)$.
\end{theorem}

\begin{proof}[Proof Sketch]
  Examine the execution of $I$ and build a $\CT'$ and $\overline{I}$ that
  produce a $\ell$\=/equivalent final state with no reentrancy.
  Whenever a high-integrity environment transitions to a low-integrity one in $\CT$,
  replace the low-integrity code in $\CT'$ with code that returns the same value as a hard-coded constant
  and makes no calls to high-integrity code.
  For each call from a low-integrity environment to a high-integrity method,
  add an invocation to $\overline{I}$ that makes the same call with the same arguments.
  Add additional invocations between each high-integrity call to update the low-integrity state
  to match the low-integrity state in the original execution when the call occurred.
  The result is clearly a non-reentrant set of executions.
  Because all $\ell$\=/reentrant states are $\ell$\=/tail-reentrant in the original execution,
  placing a reentrant call sequentially after the call it was originally inside produces the same result.

  Since the start and end states $\sigma_1'$ and $\sigma_2'$ of this new execution are $\ell$\=/equivalent to $\sigma_1$ and $\sigma_2$
  and $\singleEntSatPQ*[\Sigma_{\sigma_1}]{\CT}$,
  $$P(\sigma_1) \iff P(\sigma_1') ~\Longrightarrow~ Q(\sigma_2') \iff Q(\sigma_2).$$
  See Appendix~\ref{sec:tail-reenter-secure-proof} for details.
\end{proof}

From here, we have enough to prove our desired result.

\begin{proof}[Proof of Theorem~\ref{thm:security}]
  For a class table $\CT'$, invocation $I$, and heaps $\sigma_1$ and $\sigma_2$
  such that $\CT \lequiv \CT'$ and $(I, \CT', \sigma_1) \Downarrow \sigma_2$,
  Theorem~\ref{thm:all-tail-reenter} says all $\ell$\=/reentrant states are $\ell$\=/tail-reentrant.
  For $\ell$\=/integrity predicates $P$ and $Q$
  such that $\singleEntSatPQ*[\Sigma_{\sigma_1}]{\CT}$,
  Theorem~\ref{thm:tail-reenter-safe} says that
  if $P(\sigma_1)$ then $Q(\sigma_2)$, which is
  precisely the definition of $\satPQ*[\Sigma_{\sigma_1}]{\CT}$.
\end{proof}

\section{Implementation}
\label{sec:implementation}

We implemented a type checker for \langname
in 4,200 lines of Java, using JFlex~\citep{jflex} and CUP~\citep{cup-0.11b}.
We employ the SHErrLoc constraint solver~\citep{sherrloc} to analyze
information flow constraints, infer missing integrity labels,
and identify likely error locations.

We ran the type checker on four examples:
the three from Section~\ref{sec:motivation}, but without simplifying Town Crier, and one we call Multi-DAO.
Multi-DAO is a multi-contract version of the vulnerable portion of Ethereum's DAO contract~\citep{dao-hack-nyt}.
It is one application split across multiple contracts that synchronize on each transaction.
This structure allows for the DAO's original reentrancy vulnerability,
as well as a second attack where the attacker reenters the application
by leaving one contract and entering another before they synchronize.
By definition, this attack is not object reentrancy,
but as long as the Multi-DAO contracts trust each other, it \emph{is} $\ell$\=/reentrancy.
As with the original DAO, the exploits can be patched
either with dynamic locks or by performing local state changes \emph{and inter-contract synchronization operations}
before external calls.

For each example, the type checker correctly identified vulnerabilities
in the initial versions presented in Section~\ref{sec:motivation}.
It also accepted as secure patched implementations following the suggested fixes,
both with and without dynamic locks.

\paragraph{Developer Overhead}
Table~\ref{table:evaluation} presents several metrics for developer overhead.
As each example application is designed to distill complex security logic into minimal code,
the examples are all relatively short---ranging from 35 to 133 lines of code.
On these examples, the type checker is able to run in a few seconds
on a consumer desktop from 2015 with an Intel i7-4790 CPU.
Because the type system and the associated guarantees are
compositional, modules can be checked independently, so
running time should scale well as the code grows.

\begin{table}
  \centering
  \begin{tabular}{c|ccc}
    Application
    & LoC & \parbox{\widthof{type-check}}{\centering type-check time (s)\vspace{2pt}}
    & \parbox{\widthof{annotations}}{\centering necessary annotations\vspace{2pt}}
    \\
    \hline
    Uniswap 1         & 57  & 4.1  & 11  \\
    Uniswap 2         & 49  & 4.0  & 9   \\
    Uniswap 3$^*$     & 53  & 4.3  & 9   \\
    Town Crier 1      & 133 & 6.3  & 17  \\
    Town Crier 2$^*$  & 133 & 6.5  & 17  \\
    Town Crier 3$^*$  & 133 & 6.4  & 17  \\
    KV Store 1        & 38  & 2.1  & 10  \\
    KV Store 2$^*$    & 35  & 2.0  & 9   \\
    Multi-DAO 1       & 38  & 3.5  & 8   \\
    Multi-DAO 2       & 36  & 3.3  & 7   \\
    Multi-DAO 3$^*$   & 36  & 3.3  & 7   \\
  \end{tabular}
  \caption{Evaluation of \langname type checker. Asterisks indicate vulnerable implementations.}
  \label{table:evaluation}
\end{table}

Another important practical concern is the annotation burden of adding information flow labels to the code.
Labels on classes, fields, methods, and data endorsements
are necessary to define the security of a program.
Though \langname requires explicit labels elsewhere to ease formal reasoning,
many of these---such as the $\pc$ labels on $\If$ statements---are simple to infer.
Considering only the labels with no obvious inference mechanism,
we found that 13\% of the lines required explicit labels in Town Crier.
The other examples required more annotations per line
as their distilled nature led to more function declarations and explicit endorsements.
As even Town Crier is a short application with complex security concerns,
we expect many applications would have lower annotation burdens.

Finally, SHErrLoc is capable of localizing errors, helping guide development.
To see its utility, we look at the Uniswap example in more detail.
As in Section~\ref{sec:examples-revisited}, we use two labels: $U$~and~$T$.
Recall that the exchange must either utilize a lock or state its assumption that
the token will not call untrusted code.
The following signature for the token's \texttt{transferTo} method makes the assumption explicit,
where \texttt{H} is a token holder class.
\begingroup
  \par\centering
  \begin{minipage}{25.4em}
\begin{lstlisting}[language=scif, numbers=none]
bool(*$^T$*) transferTo{(*$T \EndrsSymb T; T$*)}(from:H(*$^T$*), to:H(*$^T$*), amount:int(*$^T$*))
\end{lstlisting}%
  \end{minipage}
  \par\noindent\ignorespacesafterend
\endgroup
To model the alert functions in \texttt{H} being unknown code from unknown sources,
the interface can state the following entirely-untrusted signatures.
\begingroup
  \par\centering
  \begin{minipage}{23em}
\begin{lstlisting}[language=scif, morekeywords={[1]void}, numbers=none]
 void alertSend{(*$U \EndrsSymb U; U$*)}(to:H(*$^U$*), amount:int(*$^U$*))
 void alertReceive{(*$U \EndrsSymb U; U$*)}(from:H(*$^U$*), amount:int(*$^U$*))
\end{lstlisting}
  \end{minipage}
  \par\noindent\ignorespacesafterend
\endgroup
With these signatures, the calls to the alert functions in \texttt{transferTo}
on lines~\ref{lst:ln:token-alert-send} and~\ref{lst:ln:token-alert-receive} of Figure~\ref{fig:uniswap}
cannot type-check without a dynamic lock.
SHErrLoc helpfully identifies line~\ref{lst:ln:token-alert-receive} as the most likely error.
The type checker correctly identifies the program as secure
if we either wrap both alerts in a dynamic lock or remove them entirely.

\section{Related Work}
\label{sec:related}

We now discuss other work on reentrancy security, secure smart contracts, and information flow control.

\paragraph{Formal Reentrancy Security}

\citet{grossman2017online} define Effectively Callback-Free~(ECF) executions,
the only other formal definition of reentrancy security of which we are aware.
An ECF execution is one where the operations can be reordered to produce the same result without callbacks (reentrancy).
Their definition is object-based, which
we have argued fails to separate the security specification from the program design,
and they focus on dynamic analysis of individual executions.

\citet{albert2020taming} present a static analysis tool to check if code produces only ECF executions.
The authors advertise the tool as providing modular guarantees, but define ``modular'' to mean that
a contract remains secure against any possible outside code.
Our approach provides the same guarantees when applied to a single program with
no assumptions on others, but also enables developers to safely compose
independently-checked modules by stating assumptions on each other's behavior.
Furthermore, \citeauthor{albert2020taming}'s analysis relies on an SMT solver,
limiting its scalability.
In comparison, \langname only relies on checking acts-for relationships of information flow labels.

We previously proposed the intuition of using information flow control
with a mix of static and dynamic locks to enforce $\ell$\=/reentrancy~\citep{cecchetti-fab20}.
In this work we add a core calculus with static and dynamic semantics,
formal definitions, proofs, and an evaluation.

\paragraph{Reentrancy-aware Languages}

Several languages---all smart-contract oriented---attempt to guard against reentrancy using a variety of techniques.

Scilla~\citep{sergey2019safer} constrains programming style by removing the call-and-return model of contract interaction.
Instead, it queues requests and executes them when the caller completes.
While this structure makes object-level reentrancy difficult,
it prevents contracts from using the return values from remote calls.
Moreover, by allowing multiple unconstrained requests,
it fails to detect or eliminate bugs like Uniswap (see Section~\ref{sec:uniswap}).

Obsidian~\citep{obsidian-toplas} and Flint~\citep{flint} ease reasoning about contract behavior using typestate.
Obsidian includes a dynamic check that prevents (object) reentrancy entirely, while Flint has no such check.
Both languages and Move~\citep{move-lang} have a notion of linear assets that cannot be created or destroyed.
Asset linearity prevents attacks like the DAO,
but fails to address the challenges of Uniswap.
The errant send in Uniswap does not create or destroy tokens; it merely sends the wrong number because it the invariant it relies on is broken.

Nomos~\citep{das2019resource} enforces security using resource-aware session types.
Since linearity of session types is insufficient to eliminate reentrancy,
it uses the resources tracked by the session types to
prevent attackers from acquiring permission to call an in-use
contract---again, eliminating all (object) reentrancy.

\paragraph{Smart Contract Analysis Tools}

There are many static analysis tools for blockchain smart contracts.
Some tools operate as unsound best-effort bug finding tools.
\textsc{Oyente}~\citep{luu2016making} searches for anti-patterns in code,
\textsc{teEther}~\citep{teether} automatically generates exploits based on commonly-exploitable operations,
and Ethainter~\citep{ethainter} uses information flow taint analysis to attempt to locate a predefined set of security concerns,
such as tainted owner variables and access to self-destruct.

Other tools use formal analysis techniques to soundly analyze contracts.
\citet{bhargavan2016formal} prove functional correctness through translation to F$^\star$.
\textsc{Maian}~\citep{nikolic2018finding} and \textsc{EthBMC}~\citep{ethbmc}
prove security against specific classes of vulnerabilities using symbolic execution and bounded model checking, respectively.
EtherTrust~\citep{ethertrust} allows developers to specify program properties as Horn clauses
and verify them using a formal semantics for EVM~\citep{grishchenko2018semantic}.
\textsc{Solythesis}~\citep{solythesis} combines static and dynamic mechanisms
It statically determines what checks are necessary for correctness and compiles them into run-time checks.

These tools are valuable for securing smart contracts, but they all
analyze individual contracts, and their analyses often fail to compose.
As a result, they are unable to verify security of applications like Uniswap
that span multiple contracts.

\paragraph{Information Flow Control}

Several distributed and decentralized systems enforce security using IFC.
Fabric~\citep{jfabric} is a system and language for building distributed
systems that allows secure data and code sharing between nodes despite mutual distrust.
DStar~\citep{dstar} uses run-time tracking at the OS level to control information flow in a distributed system.
These previous systems have the same limitation
of information flow systems that is described
in Section~\ref{sec:intro}: they do not defend against
reentrancy attacks.
The IFC-based instruction set of \citet{zsm19} restricts endorsement of $\pc$~labels
using a purely dynamic mechanism that appears to prevent all $\ell$\=/reentrancy.
However, this property is neither stated nor proved.

\section{Conclusion}
\label{sec:conclusion}

Despite decades of work on techniques for making software more secure and
trustworthy, recent smart contract bugs have vividly shown that
avoiding critical security vulnerabilities can be difficult even in very short programs.
The essential challenge is composition of code
with complex control flow across trust boundaries.
Prior static information flow analyses provide compositional guarantees,
but are missing a key ingredient: security against reentrant executions.
Smart contracts have produced the most salient reentrancy vulnerabilities to date
due to their structure of interacting service in different trust domains.
As more applications adopt distributed service-oriented architectures mirroring this design,
we expect reentrancy to become more of a concern elsewhere.

This paper provides a flexible general-purpose security definition that permits secure forms of reentrancy
and a fine-grained static mechanism to reason about reentrancy security.
We presented \langname, a core calculus that
combines static and dynamic locking to provably enforce reentrancy security
in addition to providing standard information flow assurances.
We further showed that \langname is expressive enough to implement and analyze various challenging examples.
\langname's lightweight, inferable annotations support an independently-useful
verification process while complementing other verification methods.

We hope these foundational results will aid the development of practical
secure languages.
To ensure usability, languages will need to infer labels wherever possible
and use sensible defaults in many other areas.
They might further require polymorphic, finer-grained locks that we believe can fit into
the structure of a distributive lattice.
Finally, while we focused entirely on single-threaded reentrancy,
concurrency is common in real-world languages and applications.
The relationship between reentrancy and concurrency controls/consistency models
is unclear and, we believe, a promising area for future work.

\section*{Acknowledgments}

We would first like to thank our anonymous reviewers for their thoughtful comments and suggestions.
Additional thanks to Tom Magrino for help clarifying and explaining earlier versions of this work,
and to Rachit Nigam, Rolph Recto, and Drew Zagieboylo for help editing.

This work was funded in part by a National Defense Science and
Engineering Graduate (NDSEG) Fellowship, NSF grants 1704615 and 1704788, and a gift from Ripple.
Any opinions, findings, conclusions, or recommendations expressed here are those of the authors and may not reflect those of these sponsors.

\bibliographystyle{abbrvnat}
\bibliography{../../bibtex/pm-master,../blockchain}

\appendix
\section{Full \langname Rules}
\label{sec:full-lang}

The full operational semantics for \langname are given in Figure~\ref{fig:full-semantics} and
the full typing rules are given in
Figures~\ref{fig:type-system-expr} and~\ref{fig:type-system-class}.

\begin{figure}
  \centering
  \begin{ruleset}[0.8\textwidth]
    \EEvalRule* \\ \ELetRule* \\ \EIfTRule* \\ \EIfFRule* \\ \EAtPcRule*
    \\
    \ERefRule* \\ \EDerefRule* \\ \EAssignRule*
    \\
    \ECastRule* \\ \EFieldRule* \\ \ECallRule* \\ \ECallAtkRule* \\ \EReturnRule*
    \\
    \ELockRule* \\ \EUnlockRule* \\ \EEndorseRule*
    \\
    \EIgnoreLocksRule*
  \end{ruleset}
  \caption{Full small-step operational semantics for \langname.}
  \label{fig:full-semantics}
\end{figure}

\begingroup
\renewcommand{\rulefiguresize}{\fontsize{8pt}{10pt}\selectfont}
\newcommand{\groupsplit}{\vspace{0.75em}\hrule\vspace{0.75em}}
\begin{figure*}
  \begin{boxedminipage}{\textwidth}
    \small
    \vspace*{0.5em}
    \hspace*{\tabcolsep}\textbf{Value Typing}
    \begin{ruleset}
      \VarRule \and \UnitRule \and \TrueRule \and \FalseRule
      \and
      \NewRule \and \LocRule \and \NullRule \and \SubtypeVRule
    \end{ruleset}

    \groupsplit

    \hspace*{\tabcolsep}\textbf{Core Expression Typing}
    \begin{ruleset}
      \ValRule \and \EndorseRule \and \CastRule
      \and
      \FieldRule \and \parbox[b]{20.7em}{\parbox[b]{18em}{$\CallRule$}}
      \and
      \IfRule
      \and
      \RefRule \and \DerefRule \and \AssignRule
      \and
      \LockRule \and \LetRule \and \VarianceRule
    \end{ruleset}

    \groupsplit

    \hspace*{\tabcolsep}\textbf{Tracking Statement Typing}
    \begin{ruleset}
      \AtPcRule \and \WithLockRule \and \ReturnRule
    \end{ruleset}

    \groupsplit

    \hspace*{\tabcolsep}\textbf{Attacker-Model Expression Typing}
    \begin{ruleset}
      \IgnoreLocksRule
    \end{ruleset}

  \end{boxedminipage}
  \caption{Full typing rules for \langname values, expressions, and statements.}
  \label{fig:type-system-expr}
\end{figure*}

\begin{figure*}
  \begin{boxedminipage}{\textwidth}
    \small
    \vspace*{0.5em}
    \hspace*{\tabcolsep}\textbf{Class Typing}
    \begin{ruleset}
      \MethodOkRule \and \ClassOkRule \and \CtOkRule
    \end{ruleset}

    \groupsplit

    \hspace*{\tabcolsep}\textbf{Lookup Functions}
    \begin{ruleset}
      \FieldListRule \and \DefinedMethodRule \and \InheritedMethodRule \and \CanOverrideRule
    \end{ruleset}

    \groupsplit

    \hspace*{\tabcolsep}\textbf{Subtyping}
    \begin{ruleset}
      \infer{\ell \actsfor \ell'}{t^{\ell} \subtyp t^{\ell'}}
      \and
      \infer{\CT(C) = \cdefNoBody*}{C^\ell \subtyp D^\ell}
      \and
      \infer{
        \tau_1 \subtyp \tau_2 \\
        \tau_2 \subtyp \tau_3
      }{\tau_1 \subtyp \tau_3}
    \end{ruleset}

    \groupsplit

    \begin{tabular}{l|l}
      \textbf{Protection} & \textbf{Heap Typing} \\
      \begin{ruleset}[0.475\textwidth]
        \infer{\ell \actsfor \ell'}{\ell \prot t^{\ell'}}
      \end{ruleset}
      &
      \begin{ruleset}[0.475\textwidth]
        \infer{
          \sigma(\loc) = (v, \tau) ~\Longrightarrow~ \Sigma_\sigma \proves v : \tau
        }{\provesWt{\sigma}}
      \end{ruleset}
    \end{tabular}
    \vspace*{0.5em}
  \end{boxedminipage}
  \caption{Typing rules for \langname classes, auxiliary lookup functions, and relations.}
  \label{fig:type-system-class}
\end{figure*}
\endgroup

\section{Location--Name Isomorphism}
\label{sec:loc-name-iso}

The \ruleref{E-Ref} operational semantic rule allows for selection of any unmapped location name when creating a new location.
This makes the \langname operational semantics nondeterministic in its choice of location names.
However, this is the only source of nondeterminism in the semantics.
That is, for any pair of statement-heap pairs that are equivalent up to location names,
if one steps, then the other steps and the results are again equivalent up to location names.

To reason about these differences, we define an equivalence relation that relates statements and heaps
that differ only in their location names.
Formally, we define a location name permutation $\theta$ as an injective map from locations to locations.
We extend it to values by permuting location names, recursively permuting constructor arguments of objects, and leaving other values unmodified.
We further extend it to statements by recursively applying to each sub-statement
and to heaps as follows.
$$\theta(\sigma)(\loc) \triangleq (\theta(v), \tau) ~~\text{where}~~ \sigma(\theta^{-1}(\loc)) = (v, \tau)$$
This permutation supports the requisite equivalence relation.
\begin{definition}[Location--name isomorphism]
  Statements $s_1$ and $s_2$ are \emph{location--name isomorphic}, denoted $s_1 \locIso s_2$,
  if there exists some $\theta$ such that $s_1 = \theta(s_2)$.
  Similarly, for heaps $\sigma_1$ and $\sigma_2$, $\sigma_1 \locIso \sigma_2 \overset{\triangle}{\iff} \exists \theta.\, \sigma_1 = \theta(\sigma_2)$.

  We write $(s_1, \sigma_1) \locIso (s_2, \sigma_2)$ to mean there is a $\theta$ such that
  $(s_1, \sigma_1) = (\theta(s_2), \theta(\sigma_2))$
  and similarly for $(s_1, \config_1) \locIso (s_2, \config_2)$.
\end{definition}

This definition is sufficient to state and prove the important property that
the \langname semantics is deterministic up to location--name isomorphism.
\begin{theorem}
  \label{thm:loc-name-iso-semantics}
  For any $s_1$, $s_1'$, and $s_2$ and any $\config_1$, $\config_1'$ and $\config_2$,
  if $(s_1, \config_1) \locIso (s_2, \config_2)$ and $\conf{s_1}{\config_1} \stepsone \conf{s_1'}{\config_1'}$,
  then there exists $s_2'$ and $\config_2'$ such that $\conf{s_2}{\config_2} \stepsone \conf{s_2'}{\config_2'}$,
  and for all such $s_2'$ and $\config_2'$, $(s_1', \config_1') \locIso (s_2', \config_2')$.
\end{theorem}

\begin{proof}
  By induction on the operational semantics.
  We take the permutation to be defined only mapping location names between $\sigma_1$ and $\sigma_2$ and extend it on uses of \ruleref{E-Ref} (or inductively with \ruleref{E-Eval}).
\end{proof}

Finally, for use in the noninterference theorem (Theorem~\ref{thm:ni}),
we combine location--name isomorphism with $\ellHigh$\=/equivalence.
\begin{definition}[Location--name $\ellHigh$\=/isomorphism]
  Two states $\sigma_1$ and $\sigma_2$ are \emph{location--name $\ellHigh$\=/isomorphic},
  denoted $\sigma_1 \lequivLocIso \sigma_2$, if there exists a $\theta$ such that
  $\lerase{\sigma_1} = \lerase{\theta(\sigma_2)}$.
\end{definition}

\section{Preservation and Progress}
\label{sec:lang-prop-proofs}

We now prove preservation and progress theorems for \langname.

Because \langname is stateful, the type preservation theorem includes preservation of both the statement and the heap.
\begin{theorem}[Type Preservation]
  \label{thm:preservation}
  If
  \begin{itemize}[nosep]
    \item $\conf{s}{\confTuple*} \stepsone \conf{s'}{\confTuple{\sigma'}{\mlabList'}{\lockList'}}$,
    \item $\Sigma_\sigma \proves \CT~\mathsf{ok}$,
    \item $\Sigma_\sigma; \Gamma; \pc; \inLock \proves s : \tau \dashv \outLock$, and
    \item $\provesWt{\sigma}$,
  \end{itemize}
  then
  \begin{itemize}[nosep]
    \item $\Sigma_\sigma \subseteq \Sigma_{\sigma'}$,
    \item $\provesWt{\sigma'}$, and
    \item $\Sigma_{\sigma'}; \Gamma; \pc; \inLock \proves s' : \tau \dashv \outLock$.
  \end{itemize}
\end{theorem}

The proof of Theorem~\ref{thm:preservation} makes use of several simple lemmas.

\begin{lemma}[Closed Value Typing]
  \label{lem:val-typing}
  If $v \neq x$ and $\Sigma; \Gamma \proves v : t^\ell$, then $\Sigma; \Gamma' \proves v : t^{\ell'}$ for any $\Gamma'$ and $\ell'$.
\end{lemma}

\begin{proof}
  By inspection on the value typing rules.
\end{proof}

\begin{lemma}[Value Substitution]
  \label{lem:val-subst}
  The following rule is admissible
  $$\infer{
      \Sigma; \Gamma, x\ty \tau'; \pc; \inLock \proves s : \tau \dashv \outLock \\
      \Sigma; \Gamma \proves v : \tau'
  }{\Sigma; \Gamma; \pc; \inLock \proves \subst{s}{x}{v} : \tau \dashv \outLock}$$
\end{lemma}

\begin{proof}
  By simple structural induction on the proof that $\Sigma; \Gamma, x\ty\tau'; \pc; \inLock \proves s : \tau \dashv \outLock$.
\end{proof}

\begin{lemma}[Heap-type Extension]
  \label{lem:heap-type-ext}
  The following rules are admissible
  \begin{mathpar}
    \infer{
      \Sigma; \Gamma \proves v : \tau \\
      \Sigma \subseteq \Sigma'
    }{\Sigma'; \Gamma \proves v : \tau}
    \and
    \infer{
      \Sigma; \Gamma; \pc; \inLock \proves s : \tau \dashv \outLock \\
      \Sigma \subseteq \Sigma'
    }{\Sigma'; \Gamma; \pc; \inLock \proves s : \tau \dashv \outLock}
  \end{mathpar}
\end{lemma}

\begin{proof}
  By simple induction on the proofs of $\Sigma; \Gamma \proves v : \tau$ and $\Sigma; \Gamma; \pc; \inLock \proves s : \tau \dashv \outLock$.
\end{proof}

\begin{lemma}[Heap Extension]
  \label{lem:heap-ext}
  The following rule is admissible
  $$\infer{
    \provesWt{\sigma} \\
    \Sigma_\sigma \proves v : \tau \\
    \loc \notin \dom(\sigma)
  }{\proves \subst{\sigma}{\loc}{(v, \tau)}~\mathsf{wt}}$$
\end{lemma}

\begin{proof}
  For notational ease, let $\subst{\sigma}{\loc}{(v, \tau)} = \sigma'$.
  First we note that because $\loc \notin \dom(\sigma)$, we have that $\Sigma_{\sigma'} = \Sigma_\sigma \cup \{ \loc \mapsto \tau \}$
  with $\loc \notin \dom(\Sigma_\sigma) = \dom(\sigma)$.
  Now assume $\sigma'(\loc') = (v', \tau')$.
  If $\loc' = \loc$, then the premise of the rule gives us $\Sigma_\sigma \proves v' : \tau'$,
  otherwise inversion on $\provesWt{\sigma}$ gives us the same property.
  By Lemma~\ref{lem:heap-type-ext}, $\Sigma_{\sigma'} \proves v' : \tau'$, thereby proving $\provesWt{\sigma'}$.
\end{proof}

\begin{lemma}[Statement Substitution]
  \label{lem:expr-subst}
  If $\Sigma; \Gamma; \pc; \inLock \proves E[s_1] : \tau \dashv \outLock$
  then there is some $\Gamma'$, $\pc'$, $\inLock'$, $\tau'$, and $\outLock'$ such that $\Sigma; \Gamma'; \pc'; \inLock' \proves s_1 : \tau' \dashv \outLock'$
  and for any statement $s_2$ and heap-type $\Sigma' \supseteq \Sigma$, such that $\Sigma'; \Gamma'; \pc'; \inLock' \proves s_2 : \tau' \dashv \outLock'$,
  then $\Sigma'; \Gamma; \pc; \inLock \proves E[s_2] : \tau \dashv \outLock$.
\end{lemma}

\begin{proof}
  By simple induction on the proof of $\Sigma; \Gamma; \pc; \inLock \proves E[s_1] : \tau \dashv \outLock$.
\end{proof}

These lemmas are sufficient to prove type preservation.
\begin{proof}[Proof of Theorem~\ref{thm:preservation}]
  This will be a proof by induction on the typing rules and inversion on the operational semantics.
  \begin{description}
    \item[Case \ruleref{Val}:]
      Values cannot step, so this is impossible.

    \item[Case \ruleref{Endorse}:]
      Because $v$ must be a closed value, it type-checks with any label, so \ruleref{Val} proves the result.

    \item[Case \ruleref{Cast}:]
      Inversion on the operational semantics requires that $v = \new~C'(\overline{v})$ and $C' \subtyp C$.
      Therefore \ruleref{New}, \ruleref{SubtypeV}, and \ruleref{Val} prove the case.

    \item[Case \ruleref{Field}:]
      Inversion on the operational semantics says $v = \new~D(\overline{v})$ and
      the premise of \ruleref{Field} requires $\Sigma; \Gamma \proves \new~D(\overline{v}) : C^\ell$.
      By inversion on the value typing rules, $D^\ell \subtyp C^\ell$ and $\Sigma; \Gamma \proves v_i : \tau_i$.
      Therefore, \ruleref{SubtypeV} is sufficient to prove $\Sigma; \Gamma \proves v_i : \tau$,
      and \ruleref{Val} competes the case.

    \item[Case \ruleref{Call}:]
      Inversion on the operational semantics says $v = \new~D(\overline{v})$,
      and the premise of \ruleref{Call} requires $\Sigma; \Gamma \proves \new~D(\overline{v}) : C^\ell$.
      By inversion on the value typing rules, $D^\ell \subtyp C^\ell$.
      By the restrictions on overriding and the fact that $\mtype(C, m) = \mtypeExpr*$,
      we know that $\mbody(D, m) = \mbodyExpr*$.
      \ruleref{Method-Ok} further requires $\Sigma; \overline{x}\ty\overline{\tau_a}, \this\ty\tilde{D}^{\pc_2}; \pc_2; \inLock' \proves e : \tau \dashv \outLock'$
      where $\inLock' \join \outLock' \actsfor \outLock$ and $D \subtyp \tilde{D}$.
      Therefore, using the premise that $\Sigma; \Gamma \proves \overline{w} : \overline{\tau_a}$
      and Lemma~\ref{lem:val-subst}, it must be the case that $\Sigma; \cdot; \pc_2; \inLock' \proves e[\overline{x} \mapsto \overline{w}, \this \mapsto \new~D(\overline{v})] : \tau \dashv \outLock'$.
      This premise coupled with \ruleref{Return} and \ruleref{AtPc} prove the desired result.

    \item[Case \ruleref{If}:]
      Inversion on the operational semantics requires that the step must be \ruleref{E-IfT} or \ruleref{E-IfF}.
      The appropriate premise of \ruleref{If} requiring the branches to type-check in the same environment and \ruleref{AtPc} prove the case.

    \item[Case \ruleref{Ref}:]
      By construction $\Sigma_{\sigma'}(\loc) = \tau$, so \ruleref{Loc} and \ruleref{Val} prove the well-typed condition.
      Lemma~\ref{lem:heap-ext} ensures $\proves \sigma'~\mathsf{wt}$, and $\sigma \subset \sigma'$, so $\Sigma_\sigma \subset \Sigma_{\sigma'}$.

    \item[Case \ruleref{Deref}:]
      Inversion on the operational semantics shows the step uses \ruleref{E-Deref}, meaning $v = \loc$ and $\sigma(\loc) = (v', \tau)$.
      The assumption that $\proves \sigma~\mathsf{wt}$ means $\Sigma_\sigma \proves v' : \tau$,
      so that coupled with \ruleref{SubtypeV} and \ruleref{Val} proves the case.

    \item[Case \ruleref{Assign}:]
      Inversion on the operational semantics shows the step must be \ruleref{E-Assign},
      which means $v_1 = \loc$, so inversion on the premise $\Sigma; \Gamma \proves v_1 : (\RefType{\tau})^\ell$
      shows $\sigma(\loc) = (v, \tau)$, so \ruleref{E-Assign} requires $\Sigma_\sigma \proves v_2 : \tau$.
      Therefore, $\Sigma_\sigma = \Sigma_{\sigma'}$ and $\sigma' = \subst{\sigma}{\loc}{(v_2, \tau)}$ remains well-typed.
      Finally, \ruleref{Unit} and \ruleref{Val} prove $s'$ properly type-checks.

    \item[Case \ruleref{Lock}:]
      The semantic rule must be \ruleref{E-Lock}, so $s' = \withLock{e}$,
      and the premises of \ruleref{WithLock} are identical to \ruleref{Lock}, so \ruleref{WithLock} proves the case.

    \item[Case \ruleref{Let}:]
      Here we see $s = \letIn{s_1}{e_2}$.
      We consider two sub-cases, if $s_1 = v$ is a value, and if it is not.
      In the first sub-case, the operational semantic rule must be \ruleref{Let},
      and inversion on the typing rules proves that $\Sigma_\sigma; \Gamma \proves v : \tau_1$, so Lemma~\ref{lem:val-subst} proves the sub-case.

      In the second sub-case,
      inversion on the operational semantics proves that the step must be \ruleref{E-Eval}.
      The \ruleref{Let} rule's first premise is that $\Sigma_\sigma; \Gamma; \pc; \inLock \proves s_1 : \tau_1 \dashv \outLock'$.
      Coupled with the inductive step in \ruleref{E-Eval} that $\conf{s_1}{\confTuple*} \stepsone \conf{s_1'}{\confTuple{\sigma'}{\mlabList'}{\lockList'}}$,
      the inductive hypothesis proves that $\Sigma_{\sigma'}; \Gamma; \pc; \inLock \proves s_1' : \tau_1 \dashv \outLock'$
      with $\Sigma_\sigma \subseteq \Sigma_{\sigma'}$ and $\proves \sigma'~\mathsf{wt}$.
      Lemma~\ref{lem:heap-type-ext} therefore shows that $\Sigma_{\sigma'}; \Gamma, x\ty\tau_1; \pc; \inLock \proves e_2 : \tau_2 \dashv \outLock$,
      so \ruleref{Let} us sufficient to show $\Sigma_{\sigma'}; \Gamma; \pc; \inLock \proves \letIn{s_1'}{e_2} : \tau_2 \dashv \outLock$, finishing the case.

    \item[Case \ruleref{Variance}:]
      By induction on the typing rules.

    \item[Cases \ruleref{AtPc}, \ruleref{WithLock}, and \ruleref{Return}:]
      Each of these cases has two sub-cases: where the sub-statement is a value and where it is not.
      If the sub-statement is a value, the step must be \ruleref{E-AtPc}, \ruleref{E-Unlock}, or \ruleref{E-Return}, respectively.
      In each case \ruleref{Val} allows values to type-check with any $\pc$ and lock labels, proving the case.
      If the sub-statement is not a value, the only step possible is \ruleref{E-Eval}.
      Here the proof follows by induction on the typing rules in the same manner as the \ruleref{Let} case above.
      \qedhere
  \end{description}
\end{proof}

Several semantic steps (\ruleref{E-Ref}, \ruleref{E-Assign}, and \ruleref{E-Call})
include information-security checks to guarantee that the code performing the operation is sufficiently trusted.
The type system guarantees that these labels remain at least as trusted as the $\pc$~label of code executing.
We formally define this property as a relation between a label stack and a statement, denoted by $\mlabList \leftrightsquigarrow s$,
and then prove that the semantics maintains this relation.
The relation is formally defined on evaluation contexts and extended to statements $s = E[e]$ if $\mlabList \leftrightsquigarrow E$.
\begin{ruleset}
  \infer{ }{\mlabel \leftrightsquigarrow [\cdot]}
  \and
  \infer{\mlabList \leftrightsquigarrow E}{\mlabList \leftrightsquigarrow \letIn{E}{e}}
  \and
  \infer{\mlabList \leftrightsquigarrow E}{\mlabList \leftrightsquigarrow \withLock{E}}
  \and
  \infer{\mlabList \leftrightsquigarrow E}{\mlabList \leftrightsquigarrow \ignoreLocks{E}}
  \\
  \infer{\mlabList \leftrightsquigarrow E}{\ell,\mlabList \leftrightsquigarrow \funend{\tau}~E}
  \and
  \infer{
    \ell,\mlabList \leftrightsquigarrow E \\
    \ell \actsfor \pc
  }{\ell,\mlabList \leftrightsquigarrow \atpc{E}}
\end{ruleset}

\begin{prop}
  \label{prop:lab-list-step-correspond}
  For any statements $s$ and $s'$ and configurations $\config = \confTuple{\sigma}{(\mlabel, \mlabList)}{\lockList}$ and $\config' = \confTuple{\sigma'}{\mlabList'}{\lockList'}$,
  if \mbox{$\proves \CT~\mathsf{ok}$} and $(\mlabel, \mlabList) \leftrightsquigarrow s$ and $\Sigma_\sigma; \Gamma; \mlabel; \inLock \proves s : \tau \dashv \outLock$ and $\conf{s}{\config} \stepsone \conf{s'}{\config'}$,
  then $\mlabList' \leftrightsquigarrow s'$.
\end{prop}

The proof of Proposition~\ref{prop:lab-list-step-correspond} relies on two lemmas.

\begin{lemma}
  \label{lem:lab-list-ctx-split}
  For any label list $\mlabList$ and evaluation contexts $E_1$ and $E_2$,
  $\mlabList \leftrightsquigarrow E_1[E_2]$ if and only if there exist $\mlabList_1$, $\mlabList_2$, and $\mlabel$
  such that (1) $\mlabList_1, \mlabel, \mlabList_2 = \mlabList$,
  (2) $\mlabList_1, \mlabel \leftrightsquigarrow E_1$, and (3) $\mlabel, \mlabList_2 \leftrightsquigarrow E_2$.
\end{lemma}

\begin{proof}
  This is a proof by induction on $E_1$.
  \begin{description}
    \item[Case {$E_1 = [\cdot]$}:] \hfill
      \begin{itemize}[nosep]
        \item[$(\Rightarrow)$]
          Let $\mlabList_1$ be empty and note that $\mlabList$ cannot be empty, so $\mlabList = \mlabel, \mlabList_2$.
        \item[$(\Leftarrow)$]
          By inversion on the rules, $\mlabList_1$ must be empty, so $\mlabList = \mlabel, \mlabList_2$, proving the result.
      \end{itemize}

    \item[Case $E_1 = (\letIn{E_1'}{e})$, $\withLock{E_1'}$, or $\ignoreLocks{E_1'}$:] \hfill
      \begin{itemize}[nosep]
        \item[$(\Rightarrow)$]
          By induction, there exist $\mlabList_1$, $\mlabList_2$, and $\mlabel$ such that
          $\mlabList = \mlabList_1, \mlabel, \mlabList_2$, $\mlabList_1, \mlabel \leftrightsquigarrow E_1'$, and $\mlabel, \mlabList_2 \leftrightsquigarrow E_2$.
          Therefore, by the appropriate rule, $\mlabList_1, \mlabel \leftrightsquigarrow E_1$.
        \item[$(\Leftarrow)$]
          By induction, $\mlabList_1, \mlabel, \mlabList_2 \leftrightsquigarrow E_1'[E_2]$,
          so by the appropriate rule, $\mlabList_1, \mlabel, \mlabList_2 \leftrightsquigarrow E_1[E_2]$.
      \end{itemize}

    \item[Case $E_1 = \funend{\tau}~E_1'$:] \hfill
      \begin{itemize}[nosep]
        \item[$(\Rightarrow)$]
          Inversion on the correspondence proves $\mlabList = \ell, \mlabList'$
          and $\mlabList' \leftrightsquigarrow E_1'[E_2]$.
          By induction, there is some $\mlabList_1', \mlabel, \mlabList_2 = \mlabList'$ such that
          $\mlabList_1', \mlabel \leftrightsquigarrow E_1'$ and $\mlabel, \mlabList_2 \leftrightsquigarrow E_2$.
          Letting $\mlabList_1 = \ell, \mlabList_1'$ completes the case.
        \item[$(\Leftarrow)$]
          By inversion on the correspondence rules, if $\mlabList_1, \mlabel \leftrightsquigarrow E_1$,
          then $\mlabList_1 = \ell, \mlabList_1'$ for some $\ell$ and $\mlabList_1'$ and $\mlabList_1', \mlabel \leftrightsquigarrow E_1'$.
          By induction, $\mlabList_1', \mlabel, \mlabList_2 \leftrightsquigarrow E_1'[E_2]$,
          so therefore
          $$\mlabList_1, \mlabel, \mlabList_2 = \ell, \mlabList_1', \mlabel, \mlabList_2 \leftrightsquigarrow \funend{\tau}~E_1'[E_2] = E_1[E_2].$$
      \end{itemize}

    \item[Case $E_1 = \atpc{E_1'}$:] \hfill
      \begin{itemize}[nosep]
        \item[$(\Rightarrow)$]
          By inversion on the rules, $\mlabList \leftrightsquigarrow E_1'[E_2]$,
          so by induction $\mlabList = \mlabList_1, \mlabel, \mlabList_2$ with the desired properties.
          Moreover, $\mlabList = \ell, \mlabList'$ and $\ell \actsfor \pc$.
          Because $\mlabList_1, \mlabel$ is a non-empty prefix of $\mlabList$,
          it must be the case that $\mlabList_1, \mlabel = \ell, \mlabList_1'$,
          so therefore $\mlabList_1, \mlabel \leftrightsquigarrow \atpc{E_1'} = E_1$, as desired.
        \item[$(\Leftarrow)$]
          By inversion on the correspondence rules, $\mlabList_1, \mlabel \leftrightsquigarrow E_1'$,
          so by induction, $\mlabList = \mlabList_1, \mlabel, \mlabList_2 \leftrightsquigarrow E_1'[E_2]$.
          Moreover, $\mlabList_1, \mlabel = \ell, \mlabList_1'$ and $\ell \actsfor \pc$.
          Therefore $\mlabList = \ell, \mlabList_1', \mlabList_2$,
          satisfying the requirements to prove $\mlabList \leftrightsquigarrow \atpc{E_1'[E_2]} = E_1[E_2]$.
          \qedhere
      \end{itemize}
  \end{description}
\end{proof}

\begin{lemma}
  \label{lem:partial-lab-list-step}
  For statements $s$ and $s'$, configurations $\config = \confTuple*$ and $\config' = \confTuple{\sigma'}{\mlabList'}{\lockList'}$,
  and label lists $\mlabList_1$ and $\mlabList_2$,
  if $\mlabList = \mlabList_1, \mlabList_2$ and $\mlabList_2$ is not empty,
  then $\confExpr{s} \stepsone \conf{s'}{\config'}$ if and only if $\confExpr[\mlabList_2/\mlabList]{s} \stepsone \conf{s'}{\config'[\mlabList_2'/\mlabList]}$
  for some $\mlabList_2'$ where $\mlabList' = \mlabList_1, \mlabList_2'$.
\end{lemma}

\begin{proof}
  By simple induction on the operational semantics.
\end{proof}

\begin{proof}[Proof of Proposition~\ref{prop:lab-list-step-correspond}]
  This will be a proof by induction on the operational semantics.
  \begin{description}
    \item[Case \ruleref{E-Eval}:]
      In this case $s = E[\tilde{s}]$, and by definition, $\tilde{s} = \tilde{E}[e]$.
      By Lemma~\ref{lem:lab-list-ctx-split}, there exist $\mlabList_1$, $\mlabList_2$, and~$\ell$ such that
      $\mlabel, \mlabList = \mlabList_1, \ell, \mlabList_2$ where $\mlabList_1, \ell \leftrightsquigarrow E$ and $\ell, \mlabList_2 \leftrightsquigarrow \tilde{E}$.
      Therefore \ruleref{E-Eval} gives \mbox{$\conf{\tilde{s}}{\config} \stepsone \conf{\tilde{s}'}{\config'}$},
      and because $\ell, \mlabList_2$ is non-empty,
      Lemma~\ref{lem:partial-lab-list-step} proves
      $\confExpr[(\ell, \mlabList_2)/\mlabList]{\tilde{s}} \stepsone \conf{\tilde{s}'}{\config'[\mlabList_2'/\mlabList]}$,
      and moreover $\mlabList' = \mlabList_1, \mlabList_2'$.
      Induction on this step ensures that $\mlabList_2' \leftrightsquigarrow \tilde{s}'$,
      so therefore $\mlabList_2'$ must be non-empty.
      As a single step can only add or remove one element from $\mlabList$, that means $\mlabList_2' = \ell, \mlabList_2''$,
      so by Lemma~\ref{lem:lab-list-ctx-split}, $\mlabList' = \mlabList_1, \ell, \mlabList_2'' \leftrightsquigarrow E[\tilde{s}'] = s'$.

    \item[Case \ruleref{E-IfT} and \ruleref{E-IfF}:]
      Here $s = \IfThenElse{v}{e_1}{e_2}$.
      By inversion on the correspondence rules, $\mlabList = \cdot$,
      and by inversion on the typing rules $\mlabel \actsfor \pc$.
      Therefore $\mlabel \leftrightsquigarrow \atpc{[\cdot]}$, so by definition $\mlabList' = \mlabel \leftrightsquigarrow (\atpc{e_i}) = s'$.

    \item[Case \ruleref{E-AtPc}:]
      Here $s = \atpc{v}$ and $s' = v$.
      By inversion on the correspondence rules, $\mlabList = \cdot$ and $\mlabList' = \mlabel$.
      Because $\mlabel \leftrightsquigarrow v$ for any $v$, this completes the case.

    \item[Cases \ruleref{E-Call} and \ruleref{E-CallAtk}:]
      Here $s = \new~C(\overline{v}).m(\overline{w})$ and $\mbody(C, m) = \mbodyExpr*[\mlabel']$.
      By inversion on the correspondence rules, $\mlabList = \cdot$ and $\mlabList' = \mlabel, \mlabel'$.
      By \ruleref{Method-Ok}, $\mlabel' \actsfor \pc_2$.
      Therefore, letting $e' = e[\overline{x} \mapsto \overline{w}, \this \mapsto \new~C(\overline{v})]$,
      $$\infer*{
        \infer*{
          \infer*{ }{\mlabel' \leftrightsquigarrow e'} \\
          \mlabel' \actsfor \pc_2
        }{\mlabel' \leftrightsquigarrow \atpc[\pc_2]{e'}}
      }{\mlabel, \mlabel' \leftrightsquigarrow \funend{\tau}~(\atpc[\pc_2]{e'})}.$$

    \item[Case \ruleref{E-Return}:]
      Here $s = \funend{\tau}~v$, so inversion on the correspondence rules proves $\mlabList = \ell$.
      Therefore $\mlabList' = \mlabel \leftrightsquigarrow v = s'$ proves the case.
  \end{description}
  No other operational semantic rules modify $\mlabList$ or add or remove $\funend{}$ or $\atpcName$ terms.
  Therefore the same proofs apply before and after the step.
\end{proof}

The progress theorem is not without caveats.
\langname's type system intentionally leaves checking of explicit
casts, null dereferences, and dynamic reentrancy locks to run time.
As a result, the progress theorem states that these three are the \emph{only} ways a well-typed program can get stuck.
\begin{samepage}
\begin{theorem}[Progress]
  \label{thm:progress}
  For any statement $s$ and configuration $\config = \confTuple{\sigma}{(\mlabel, \mlabList)}{\lockList}$,
  if
  \begin{itemize}[nosep]
    \item $\Sigma_\sigma; \cdot; \pc; \inLock \proves s : \tau \dashv \outLock$,
    \item $\mlabel \actsfor \pc$, and
    \item $(\mlabel, \mlabList) \leftrightsquigarrow s$,
  \end{itemize}
  then one of the following holds:
  \begin{enumerate}[nosep]
    \item $s$ is a closed value,
    \item $\conf{s}{\config} \stepsone \conf{s'}{\config'}$ for some $s'$ and $\config'$,
    \item $s = E[(C)(\new~D(\overline{v}))]$ where $D \not\subtyp C$,
    \item $s = E[\deref{\Null}]$ or $s = E[\Null := v]$, or
    \item $s = E[\new~C(\overline{v}).m(\overline{w})]$ for a $C$ and $m$ such that $\mtype(C, m) = \mtypeExpr*$
      and there is some $\mlabel \in \lockList$ such that $\pc_1 \nactsfor \pc_2 \join \mlabel$.
  \end{enumerate}
\end{theorem}
\end{samepage}

\begin{proof}
  This is a proof by induction on the derivation that $\Sigma_\sigma; \cdot; \pc; \inLock \proves s : \tau \dashv \outLock$.
  \begin{description}
    \item[Case \ruleref{Val}:]
      Because $\Gamma = \cdot$, $s$ is a closed value.

    \item[Case \ruleref{Endorse}:]
      Here $s = \Endorse{v}{\ell}{\ell'}$.
      Since $\Gamma = \cdot$, $v$ is a closed value, so \ruleref{E-Endorse} applies.

    \item[Case \ruleref{Cast}:]
      Here $s = (C)v$.
      Inversion on the value typing rules coupled with the fact that $\Gamma = \cdot$ proves that $v = \new~D(\overline{v})$.
      If $D \subtyp C$, then \ruleref{E-Cast} applies with $\config' = \config$.
      Otherwise this is a bad cast.

    \item[Case \ruleref{Field}:]
      Here $s = v.f_i$.
      Again, inversion on the value typing rules with $\Gamma = \cdot$ proves $v = \new~C(\overline{v})$.
      Moreover \ruleref{Field} requires reference to a valid fields, so \ruleref{E-Field} steps $s$ with $\config' = \config$.

    \item[Case \ruleref{Call}:]
      Here $s = v.m(\overline{v})$.
      If a step can be taken, it must use \ruleref{E-Call} or \ruleref{E-CallAtk}.
      Because $\Gamma = \cdot$, inversion on the premise that $\Sigma_\sigma; \Gamma \proves v : C^\ell$
      proves $v = \new~C(\overline{w})$.
      The premise $\Sigma; \Gamma \proves \overline{v} : \overline{\tau_a}$ also directly proves the corresponding
      premise of \ruleref{E-Call}/\ruleref{E-CallAtk}.
      Inversion on the proof that $(\mlabel, \mlabList) \leftrightsquigarrow s$ proves that $\mlabList$ is empty,
      so therefore the premise of \ruleref{E-Call}/\ruleref{E-CallAtk} requiring the caller's integrity to act for $\pc_1$
      is satisfied by $\mlabel \actsfor \pc \actsfor \pc_1$.
      At this point, \ruleref{E-CallAtk} applies if $\ellAdv \actsfor \pc_2$
      and \ruleref{E-Call} applies if $\bigwedge_{\ell \in \lockList} (\pc_1 \actsfor \pc_2 \join \ell)$.
      Therefore, if the statement is stuck, neither is satisfied, and the second
      is precisely the condition of a dynamic reentrancy lock blocking a call.

    \item[Case \ruleref{If}:]
      Here $s = \IfThenElse[\pc']{v}{e_1}{e_2}$.
      Inversion on the value typing rules using $\Gamma = \cdot$ means $v = \True$ or $v = \False$.
      Therefore \ruleref{E-IfT} or \ruleref{E-IfF} apply.

    \item[Case \ruleref{Ref}:]
      Here $s = \RefOp{v}{\tau}$.
      This step will be with \ruleref{E-Ref}.
      Since $\Gamma = \cdot$, the requirement that $\Sigma_\sigma \proves v : \tau$ comes directly from \ruleref{Ref}.
      Moreover, inversion on the rules proving $(\mlabel, \mlabList) \leftrightsquigarrow s$ shows that $\mlabList = \cdot$,
      so the protection requirement of \ruleref{E-Ref} is $\mlabel \prot \tau$ and $\mlabel \actsfor \pc \prot \tau$,
      meaning the step applies with some fresh $\loc \notin \dom(\sigma)$.

    \item[Case \ruleref{Deref}:]
      Here $s = \deref{v}$.
      Since $\Gamma = \cdot$, inversion on the \ruleref{Deref} premise that $\Sigma \proves v : (\RefType{\tau'})^\ell$
      means $v = \loc$ with $\Sigma_\sigma(\loc) = \tau'$ or $v = \Null$.
      In the first case, by definition this means $\sigma(\loc) = (v', \tau')$ for some $v'$, so \ruleref{E-Deref} applies.
      In this second case, this is a null dereference.

    \item[Case \ruleref{Assign}:]
      Here $s = (v_1 := v_2)$.
      Again, $\Gamma = \cdot$ and inversion on the typing rules using the premise $\Sigma; \Gamma \proves v_1 : (\RefType{\tau})^\ell$
      proves that $v_1 = \loc$ or $v_1 = \Null$.
      If $v_1 = \Null$, then this is a null dereference.
      If $v_1 = \loc$, then the step must be \ruleref{E-Assign}.
      The requirement that $\Sigma_\sigma(\loc) = \tau$ and $\Sigma \proves v_2 : \tau$ stem from inversion on the typing derivation of $v_1$
      and the second premise of \ruleref{Assign}.
      Finally, inversion on the rules proving $(\mlabel, \mlabList) \leftrightsquigarrow s$ shows that $\mlabList = \cdot$,
      and \ruleref{Assign} requires $\pc \join \ell \prot \tau$, so the transitivity of~$\actsfor$ proves $\mlabel \prot \tau$, as needed.

    \item[Case \ruleref{Lock}:]
      \ruleref{E-Lock} always applies.

    \item[Case \ruleref{Let}:]
      Here $s = \letIn{\tilde{s}}{e}$.
      The first hypothesis of \ruleref{Let} proves $\Sigma_\sigma; \cdot; \pc; \inLock \proves \tilde{s} : \tau_1 \dashv \outLock'$,
      and $(\mlabel, \mlabList) \leftrightsquigarrow \tilde{s}$.
      Therefore, our inductive hypothesis applies to $\tilde{s}$.
      If $\tilde{s}$ is a closed value, then \ruleref{E-Let} applies to $s$, stepping to $s' = \subst{e}{x}{\tilde{s}}$ letting $\config' = \config$.
      If $\conf{\tilde{s}}{\config} \stepsone \conf{\tilde{s}'}{\config'}$, then by \ruleref{E-Eval}, $\conf{s}{\config} \stepsone \conf{\letIn{\tilde{s}'}{e}}{\config'}$.
      For the other three cases where $\tilde{s} = E[e']$ where $e'$ is a failure condition,
      we note that $\letIn{E}{e}$ is an evaluation context, so $s$ falls into the same failure case.

    \item[Case \ruleref{Variance}:]
      Because $\mlabel \actsfor \pc \actsfor \pc'$, this case follows directly by induction.

    \item[Case \ruleref{AtPc}:]
      Here $s = \atpc[\pc']{\tilde{s}}$.
      Inversion on the proof that $(\mlabel, \mlabList) \leftrightsquigarrow s$ proves $\mlabel \actsfor \pc'$.
      The hypothesis of \ruleref{AtPc} is $\Sigma_\sigma; \cdot; \pc'; \inLock \proves \tilde{s} : \tau \dashv \outLock$,
      so the inductive hypothesis applies to $\tilde{s}$.

      If $\tilde{s}$ is a closed value, then \ruleref{E-AtPc} applies letting $\config' = \config$.
      If $\tilde{s}$ steps to $\tilde{s}'$, then \ruleref{E-Eval} proves $\conf{s}{\config} \stepsone \conf{\atpc[\pc']{\tilde{s}'}}{\config'}$.
      For the other three cases, as with \ruleref{Let}, $\tilde{s} = E[e]$ where $e$ is a failure condition,
      so $\atpc[\pc']{E}$ is an evaluation context proving that $s$ falls into the same failure case as $\tilde{s}$.

    \item[Cases \ruleref{WithLock} and \ruleref{IgnoreLocks}:]
      The logic of these cases is the same as the logic of the \ruleref{AtPc} case, but using $\pc$ instead of $\pc'$.

    \item[Case \ruleref{Return}:]
      Here $s = \funend{\tau}~\tilde{s}$.
      Inversion on the proof that $(\mlabel, \mlabList) \leftrightsquigarrow s$ shows that $\mlabList$ is not empty
      and $\mlabList \leftrightsquigarrow \tilde{s}$.
      Additionally, a premise of \ruleref{Return} is $\Sigma_\sigma; \cdot; \pc; \inLock' \proves \tilde{s} : \tau \dashv \outLock'$.
      Therefore, the inductive hypothesis applies using Lemma~\ref{lem:partial-lab-list-step} to replace $(\mlabel, \mlabList)$ with $\mlabList$ in $\config$.

      If $\tilde{s}$ is a closed value, the well-typed premise of \ruleref{Return} proves $\Sigma_\sigma \proves v : \tau$,
      and since $(\mlabel, \mlabList)$ is non-empty, \ruleref{E-Return} applies.
      If $\tilde{s}$ steps to $\tilde{s}'$, then \ruleref{E-Eval} allows $s$ to step as well.
      Again, for the three failure cases where $\tilde{s} = E[e]$, simply replacing $E$ with $\funend{\tau}~E$ creates the expected form.
      \qedhere
  \end{description}
\end{proof}

Note that, for any invocation $I = (\ell, \loc, m(\overline{v}))$, $\ell \leftrightsquigarrow \deref{\loc}.m(\overline{v})$.
Therefore, if the invocation and class table are well-typed in $\Sigma_\sigma$ for a well-typed heap $\sigma$,
Theorems~\ref{thm:preservation} and~\ref{thm:progress} combine with Proposition~\ref{prop:lab-list-step-correspond} to prove that
the invocation either steps to a closed value with a well-typed heap
or gets stuck on one of the three run-time error checks.

\input{ni-proof}

\input{security-proofs}

%% file: abstract.tex
The disastrous vulnerabilities in smart contracts 
sharply remind us of our ignorance: we do not
know how to write code that is secure in composition with
malicious code.
Information flow control has long been proposed as a way to achieve
compositional security, offering strong guarantees
even when combining software from different trust domains.
Unfortunately, this appealing story breaks down in the
presence of reentrancy attacks.
We formalize a general definition of reentrancy
and introduce a security condition
that allows software modules like smart contracts to protect their key invariants
while retaining the expressive power of safe forms of reentrancy.
We present a security type system that
provably enforces secure information flow;
in conjunction with run-time mechanisms, it
enforces secure reentrancy even in the presence of unknown code;
and it helps locate and correct recent high-profile vulnerabilities.

%% file: ni-proof.tex
\section{Proof of Noninterference}
\label{sec:ni-proof}

\begingroup
\renewcommand{\lequiv}[1][\ellHigh*]{\approx_{#1}}
\renewcommand{\lequivLocIso}[1][\ellHigh*]{\locIso_{#1}}

We now provide a proof of Theorem~\ref{thm:ni} presented in Section~\ref{sec:proof}.
We prove Theorem~\ref{thm:ni} using an erasure-based construction.
Specifically, we will erase low-integrity values in the heap and then execute the same program using a modified semantics
that continues to omit low-integrity values from the state and uses a special value, $\erased$, when one would be read.
We prove that, if the original execution terminated and the code is endorsement-free, this modified execution must terminate and,
critically, the high-integrity components of the state must match.
The theorem then follows by noting that if $\sigma_1 \lequiv \sigma_2$,
then both executions must produce heaps that's high-integrity components are the same as the modified execution on a partially-erased heap.

Formally, we introduce a new value to denote erased data.
$$\begin{array}{rcl}
  v & ::= & \cdots \alt \erased
\end{array}$$
The typing and semantic rules that handle~$\erased$ are parameterized on a label~$\ellHigh$ defining high-integrity values.
For notational ease, we omit that label in our syntax.
However, as our theorems are all parameterized over~$\ellHigh$, they remain true for any possible choice of~$\ellHigh$.

The type system allows $\erased$ to be any type, as long as that type is low-integrity.
To simplify notation, we define $\getLab(t^\ell) = \ell$.
$$\Rule{Bullet}{
  \getLab(\tau) \nactsfor \ellHigh
}{\Sigma; \Gamma \proves \erased : \tau}$$

We introduce an expanded operational semantics to deal with these terms.
To separate executions with and without bullets, we define a new step function denoted $\bulletstepsone$ when working with erased terms.
We also define a context $B$ defining syntactic forms that normally require a decision based on the value.
The \ruleref{B-BulletCtx} rule simply erases the entire expression when the given value is $\erased$.
$$\begin{array}{rcl}
  B & ::= & \IfThenElse{[\cdot]}{e}{e} \alt \deref{[\cdot]} \alt (C)[\cdot] \alt [\cdot].f \alt [\cdot].m(\overline{v})
\end{array}$$
\begin{center}
\begin{ruleset}
  \Rule{B-PureStep}{
    \conf{s}{\confTuple*} \stepsone \conf{s'}{\confTuple{\sigma}{\mlabList'}{\lockList'}}
  }{\conf{s}{\confTuple*} \bulletstepsone \conf{s'}{\confTuple{\sigma}{\mlabList'}{\lockList'}}}
  \and
  \Rule{B-Eval}{
    \confExpr{s} \bulletstepsone \conf{s'}{\config'}
  }{\confExpr{E[s]} \bulletstepsone \conf{E[s']}{\config'}}
  \and
  \Rule{B-BulletCtx}{ }{\confExpr{B[\erased]} \bulletstepsone \confExpr{\erased}}
  \\
  \Rule{B-TRef}{
    \loc \notin \dom(\sigma) \\
    \Sigma_\sigma \proves v : \tau \\\\
    \mlabList = \mlabList', \mlabel \\
    \mlabel \prot \tau \\
    \getLab(\tau) \actsfor \ellHigh
  }{\confExpr{\RefOp{v}{\tau}} \bulletstepsone \confExpr[\subst{\sigma}{\loc}{(v, \tau)}/\sigma]{\loc}}
  \and
  \Rule{B-URef}{
    \loc \notin \dom(\sigma) \\
    \Sigma_\sigma \proves v : \tau \\\\
    \mlabList = \mlabList', \mlabel \\
    \mlabel \prot \tau \\
    \getLab(\tau) \nactsfor \ellHigh
  }{\confExpr{\RefOp{v}{\tau}} \bulletstepsone \confExpr[\subst{\sigma}{\loc}{(\erased, \tau)}/\sigma]{\loc}}
  \and
  \Rule{B-BAssign}{ }{\confExpr{\erased := v} \bulletstepsone \confExpr{()}}
  \and
  \Rule{B-TAssign}{
    \Sigma_\sigma(\loc) = \tau \\
    \Sigma_\sigma \proves v : \tau \\\\
    \mlabList = \mlabList', \mlabel \\
    \mlabel \prot \tau \\
    \getLab(\tau) \actsfor \ellHigh
  }{\confExpr{\loc := v} \bulletstepsone \confExpr[\subst{\sigma}{\loc}{(v, \tau)}/\sigma]{()}}
  \and
  \Rule{B-UAssign}{
    \Sigma_\sigma(\loc) = \tau \\
    \Sigma_\sigma \proves v : \tau \\\\
    \mlabList = \mlabList', \mlabel \\
    \mlabel \prot \tau \\
    \getLab(\tau) \nactsfor \ellHigh
  }{\confExpr{\loc := v} \bulletstepsone \confExpr[\subst{\sigma}{\loc}{(\erased, \tau)}/\sigma]{()}}
\end{ruleset}
\end{center}

These semantics inherit from our original operation semantics whenever the step does not modify the heap.
When modifying the heap, however, $\bulletstepsone$ omits any values that are in low-integrity memory locations,
while treating high-integrity memory locations normally.
When reading from the heap, it produces $\erased$ whenever it tries to read from a location that has a type but not a value.
In our construction for our proof, these will be precisely the low-integrity locations.

We now claim that, if $\CT$ is endorsement-free at $\ellHigh$,
then any invocation with input state $\sigma_1$ will, under normal semantics, produce
a state $\sigma_2$ that is $\ellHigh$-equivalent to executing the same invocation under bullet semantics with input state $\lerase{\sigma_1}$.

\begin{lemma}[Label Stack Maintenance]
  \label{lem:lab-stack-maintain}
  For any expression $e$, if
  $$\conf{e}{\confTuple*} \stepsmany \conf{v}{\confTuple{\sigma'}{\mlabList'}{\lockList'}},$$
  then $\mlabList' = \mlabList$ and $\lockList' = \lockList$.
\end{lemma}

\begin{proof}
  This will be a proof by induction on the number of steps from $e$ to $v$ and on the operational semantics.
  In the base case, there are zero steps, so the result trivially holds.

  We now assume $\conf{e}{\confTuple*} \stepsmany \conf{v}{\confTuple{\sigma'}{\mlabList'}{\lockList'}}$ takes $n \geq 1$ steps
  and the result holds for all executions of $k < n$ steps.
  We consider the following cases.
  \begin{description}
    \item[Case \ruleref{E-Eval}:]
      If $E = [\cdot]$, we can replace this step with another, so without loss of generality, we assume $E \neq [\cdot]$.
      Since $e = E[\tilde{e}]$ is an expression, $\tilde{e}$ is also an expression and $E = \letIn{x}{E'}{e''}$.
      By inversion on the operational semantics, \ruleref{E-Eval} is the only rule that can apply until $\tilde{e}$ reaches some value $\tilde{v}$.
      Moreover, $E[\tilde{v}]$ is not a value, so $\confExpr{\tilde{e}} \stepsmany \conf{\tilde{v}}{\tilde{\config}}$ in fewer steps.
      By induction, we therefore have that $\tilde{\mlabList} = \mlabList$ and $\tilde{\lockList} = \lockList$.
      Moreover, since $E[\tilde{e}]$ was surface syntax, $E[\tilde{v}]$ must be as well.
      Therefore, another application of our inductive hypothesis proves
      $$\conf{E[\tilde{e}]}{\confTuple*} \stepsmany \conf{E[\tilde{v}]}{\confTuple{\tilde{\sigma}}{\mlabList}{\lockList}}
        \stepsmany \conf{v}{\confTuple{\sigma'}{\mlabList}{\lockList}}$$

    \item[Cases \ruleref{E-IfT} and \ruleref{E-IfF}:]
      Because $e = \IfThenElse{v'}{e_1}{e_2}$ was surface-syntax, $e_i$ must also be surface syntax.
      Inspection on the semantic rules says that any expression of the form $\atpc{\tilde{e}}$ can only step using \ruleref{E-Eval} if $\tilde{e}$ steps
      or \ruleref{E-AtPc} if $\tilde{e}$ is a value.
      Therefore, we know that $\confExpr{\tilde{e}} \stepsmany \conf{v}{\config'}$, and then $\atpc{v}$ steps once using \ruleref{E-AtPc}.
      By induction on the number of steps, we therefore have that $\mlabList' = \mlabList$ and $\lockList' = \lockList$.

    \item[Case \ruleref{E-Lock}:]
      This case is similar to the previous case.
      Again, we know that $e = \lockIn{\tilde{e}}{\ell}$ and $\tilde{e}$ is surface-syntax.
      We also know that $\confExpr{e} \stepsone \confExpr[\lockList, \ell/\lockList]{\withLock{\tilde{e}}}$.
      By the same argument as above, $\tilde{e}$ must step to a value in fewer steps, so by induction
      $$\conf{\tilde{e}}{\confTuple{\sigma}{\mlabList}{(\lockList, \ell)}} \stepsmany \conf{v}{\confTuple{\sigma'}{\mlabList}{(\lockList, \ell)}}$$
      A single application of \ruleref{E-Unlock} then gives us the desired result.

    \item[Cases \ruleref{E-Call} and \ruleref{E-CallAtk}:]
      These cases are identical to the previous one, but modifying $\mlabList$ instead of $\lockList$ and using \ruleref{E-Return} instead of \ruleref{E-Unlock}.

    \item[Cases \ruleref{E-Unlock} and \ruleref{E-Return}:]
      These are impossible because $e$ is surface-syntax.
  \end{description}
  In all other cases, stepping $e$ once continues to be surface syntax and leaves $\mlabList$ and $\lockList$ unmodified.
  We can therefore remove a single step and apply our inductive hypothesis.
\end{proof}

\begin{lemma}[Step Confinement]
  \label{lem:step-confinement}
  For a state $\sigma_1$ where $\Sigma \subseteq \Sigma_{\sigma_1}$ and a statement $s_1$, if
  \begin{enumerate}[nosep]
    \item $\Sigma \proves \CT~\mathsf{ok}$ is endorsement-free at $\ellHigh$,
    \item $\provesWt{\sigma_1}$,
    \item\label{lem:low-pc-heap:li:well-typed} $\Sigma; \Gamma; \pc; \inLock \proves s_1 : \tau \dashv \outLock$,
    \item\label{lem:low-pc-heap:li:low-pc} $\pc \nactsfor \ellHigh$,
    \item\label{lem:low-pc-heap:li:all-pcs-low} for all sub-statements $\atpc[\pc']{s}$ of $s_1$, $\pc' \nactsfor \ellHigh$, and
    \item $\conf{s_1}{\confTuple{\sigma_1}{\mlabList_1}{\lockList_1}} \stepsone \conf{s_2}{\confTuple{\sigma_2}{\mlabList_2}{\lockList_2}}$,
  \end{enumerate}
  then
  \begin{itemize}[nosep]
    \item $\sigma_1 \lequiv \sigma_2$ and
    \item for all sub-statements $\atpc[\pc']{s}$ of $s_2$, $\pc' \nactsfor \ellHigh$.
  \end{itemize}
\end{lemma}

\begin{proof}
  This will be a proof by induction on the semantic rule used to take a step.
  The following are the nontrivial cases.
  \begin{description}
    \item[Case \ruleref{E-Eval}:]
      Here we have $s_1 = E[\tilde{s}_1]$.
      We claim by induction on $E$ that $\Sigma; \Gamma; \pc'; \inLock' \proves \tilde{s}_1 : \tau' \dashv \outLock'$ for some $\pc'$, $\inLock'$, $\tau'$, and $\outLock'$ where $\pc' \nactsfor \ellHigh$.
      If $E = [\cdot]$, this follows directly from our assumptions.
      If $E = \letIn{x}{E'}{s'}$, $\funend{\tau}~E'$, or $\withLock{E'}$, we note that $\Sigma; \Gamma; \pc; \inLock' \proves E'[\tilde{s}_1] : \tau' \dashv \outLock'$ for some $\inLock'$, $\tau'$, and $\outLock'$,
      so by induction on $E$, we have the desired result.
      If $E = \atpc[\pc'']{E'}$, we note that $\Sigma; \Gamma; \pc''; \inLock' \proves E'[\tilde{s}_1] : \tau \dashv \outLock$ and, by assumption, $\pc'' \nactsfor \ellHigh$.
      Thus induction on $E$ again gets us the desired typing judgment.

      \ruleref{E-Eval} tells us $\conf{\tilde{s}_1}{\confTuple{\sigma_1}{\mlabList_1}{\lockList_1}} \stepsone \conf{\tilde{s}_2}{\confTuple{\sigma_2}{\mlabList_2}{\lockList_2}}$ and $s_2 = E[\tilde{s}_2]$.
      Since $\tilde{s_1}$ is a sub-statement of $s_1$, it must satisfy hypothesis~\ref{lem:low-pc-heap:li:all-pcs-low},
      and the typing judgment above gives us hypotheses~\ref{lem:low-pc-heap:li:well-typed} and~\ref{lem:low-pc-heap:li:low-pc}.
      Induction on the operational semantics therefore gives us $\sigma_1 \lequiv \sigma_2$
      and, for all sub-statements $\atpc[\pc']{e}$ of $\tilde{s}_2$, $\pc' \nactsfor \ellHigh$.
      By hypothesis~\ref{lem:low-pc-heap:li:all-pcs-low}, the same must be true of $E$, so therefore $E[\tilde{s}_2] = s_2$ satisfies the required condition.

    \item[Cases \ruleref{E-IfT} and \ruleref{E-IfF}:]
      Here $s_1 = \IfThenElse[\pc']{v}{s_1'}{s_2'}$.
      We know that $\sigma_1 = \sigma_2$, so that condition is trivially true.
      Both $s_1'$ and $s_2'$ are surface-syntax, so they contain no sub-statement of the form $\atpc[\pc'']{e}$,
      meaning the only such sub-statement in $s_2 = \atpc[\pc']{s_i'}$ is the outer one.
      By inversion on the typing rules, we know that $\pc \actsfor \pc'$, so by transitivity, $\pc' \nactsfor \ellHigh$.

    \item[Case \ruleref{E-Ref}:]
      Here $s_1 = \RefOp{v}{\tau'}$.
      By inversion on the typing rules, we know that $\pc \prot \tau'$, and by assumption, $\pc \nactsfor \ellHigh$.
      Therefore, since $\loc \notin \dom(\sigma_1)$ and $\sigma_2 = \subst{\sigma_1}{\loc}{(v, \tau')}$,
      we know that $\lerase{\sigma_1} = \lerase{\sigma_2}$, which is exactly the definition of $\sigma_1 \lequiv \sigma_2$.
      There are no sub-statements of the form $\atpc[\pc']{e}$, so that result is trivially true.

    \item[Case \ruleref{E-Assign}:]
      Here $s_1 = \loc := v$.
      By inversion on the typing rules, we know $\Sigma(\loc) = \tau'$ and $\pc \prot \tau'$.
      By assumption, $\pc \nactsfor \ellHigh$, so therefore $\loc \notin \dom(\lerase(\sigma_1))$.
      Given this and the fact that $\Sigma_{\sigma_2} = \Sigma_{\sigma_1}$, again $\sigma_1 \lequiv \sigma_2$, as desired.
      As in the previous case, there are no sub-statements of the form $\atpc[\pc']{e}$.

    \item[Cases \ruleref{E-Call} and \ruleref{E-CallAtk}:]
      Here $s_1 = \new~C(\overline{v}).m(\overline{w})$ with $\mbody(C, m) = \mbodyExpr*$.
      Inversion on the typing rules proves that $\pc \actsfor \pc_1$, and by assumption, $\pc \nactsfor \ellHigh$.
      Therefore, by transitivity, $\pc_1 \nactsfor \ellHigh$, so, by the definition of $\CT$ being endorsement-free at $\ellHigh$,
      it must be the case that $\pc_2 \nactsfor \ellHigh$.
      Moreover, $e[\overline{x} \mapsto \overline{w}, \this \mapsto \new~C(\overline{v})]$ is surface-syntax,
      so the only sub-statement of the form $\atpc[\pc']{s'}$ on $s_2$ is the outer one where $\pc' = \pc_2$, and we just proved $\pc_2 \nactsfor \ellHigh$.
      Finally, the step leaves the heap and heap type unmodified, finishing the case.
  \end{description}
  All other cases leave the heap and heap type unmodified and do not add sub-statement of the form $\atpc[\pc']{e}$,
  making the result trivial in those cases.
\end{proof}

\begin{corollary}[Confinement]
  \label{cor:confinement}
  Given a class table $\CT$ and an expression (not statement) $e$, if
  \begin{itemize}[nosep]
    \item $\CT$ and $e$ are both endorsement-free at $\ellHigh$,
    \item $\Sigma_\sigma \proves \CT~\mathsf{ok}$,
    \item $\Sigma_\sigma; \Gamma; \pc; \inLock \proves e : \tau \dashv \outLock$ for some $\pc \nactsfor \ellHigh$, and
    \item $\conf{e}{\confTuple*} \stepsmany \conf{v}{\confTuple{\sigma'}{\mlabList'}{\lockList'}}$,
  \end{itemize}
  then $\sigma \lequiv \sigma'$, $\mlabList = \mlabList'$, and $\lockList = \lockList'$.
\end{corollary}

\begin{proof}
  We apply Lemma~\ref{lem:lab-stack-maintain} and inductively apply Lemma~\ref{lem:step-confinement} using the fact that expressions
  cannot contain any subexpressions of the form $\atpc[\pc']{e'}$.
\end{proof}

We aim to prove something about execution in our regular semantics through execution in our semantics with bullets,
so we need a way to relate terms with and without bullets.
We do this using a syntactic relation denoted $e_1 \geqErased e_2$ to indicate that $e_2$ is just $e_1$ but possibly with some information erased.
On values, the relation is defined as follows.
\begin{mathpar}
  \infer{v \neq \bullet}{v \geqErased v}
  \and
  \infer{v \notin \{x, \bullet\}}{v \geqErased \bullet}
  \and
  \infer{\overline{v} \geqErased \overline{w}}{\new~C(\overline{v}) \geqErased \new~C(\overline{w})}
\end{mathpar}

We extend this relation to typing proofs.
We first relate typing proofs of closed values (so proofs that do not use \ruleref{Var})
to value typing proofs using \ruleref{Bullet}.
Note that we do not mandate that the heap types be the same at every location
so long as they are the same at the locations used in the typing proof.
That is
$$\textsc{\rulefiguresize[Loc]~}\infer{\Sigma_1(\loc) = \tau}{\Sigma_1; \Gamma \proves \loc : (\RefType{\tau})^\ell}
\geqErased \infer{\Sigma_2(\loc) = \tau}{\Sigma_2; \Gamma \proves \loc : (\RefType{\tau})^\ell}\textsc{\rulefiguresize~[Loc]}$$
whenever $\Sigma_1(\loc) = \Sigma_2(\loc)$, even if $\Sigma_1$ and $\Sigma_2$ differ on other locations.
Notably, if $\Sigma_1(\loc) = \tau \neq \Sigma_2(\loc)$ (possibly because $\loc \notin \dom(\Sigma_2)$) and $\ell \nactsfor \ellHigh$, then,
$$\textsc{\rulefiguresize[Loc]~}\infer{\Sigma_1(\loc) = \tau}{\Sigma_1; \Gamma \proves \loc : (\RefType{\tau})^\ell}
\geqErased \infer{\ell \nactsfor \ellHigh}{\Sigma_2; \Gamma \proves \erased : (\RefType{\tau})^\ell}\textsc{\rulefiguresize~[Bullet]}.$$

We finally extend the relation to typing proofs of expressions and statements by extending it structurally.
That is, if the typing proofs of each sub-statement is related, then the typing proof of the whole statement is related.
For example,
$$\infer{
  \infer{\pi_1}{\Sigma_1; \Gamma \proves v_1 : \tau} \geqErased \infer{\pi_2}{\Sigma_2; \Gamma \proves v_2 : \tau}
}{
  \infer{
    \infer*{\pi_1}{\Sigma_1; \Gamma \proves v_1 : \tau} \\
    \pc \prot \tau
  }{\Sigma_1; \Gamma; \pc; \inLock \proves \RefOp{v}{\tau} : (\RefType{\tau})^\ell \dashv \outLock}
  \geqErased
  \infer{
    \infer*{\pi_2}{\Sigma_2; \Gamma \proves v_2 : \tau} \\
    \pc \prot \tau
  }{\Sigma_2; \Gamma; \pc; \inLock \proves \RefOp{v}{\tau} : (\RefType{\tau})^\ell \dashv \outLock}
}$$
We usually denote this relation $\Sigma_1; \Gamma; \pc; \inLock \proves s_1 : \tau \dashv \outLock \geqErased \Sigma_2; \Gamma; \pc; \inLock \proves s_2 : \tau \dashv \outLock$.

We now use this relation to relate executions in the regular semantics and the erasure semantics.
For this we use a slightly modified erasure procedure on heaps, $\lerase{\sigma}^\erased$.
Instead of simply removing all low-integrity mappings, it instead replaces the values with $\erased$.
$$\lerase{\sigma}^\erased(\loc) \triangleq \begin{cases}
  (v, t^\ell) & \text{if } \sigma(\loc) = (v, t^\ell) \text{ and } \ell \actsfor \ellHigh \\
  (\erased, t^\ell) & \text{if } \sigma(\loc) = (v, t^\ell) \text{ and } \ell \nactsfor \ellHigh
\end{cases}$$

\begin{lemma}[Bullet Semantics Completeness]
  \label{lem:bullet-complete}
  Let $\config_i = \confTuple{\sigma_i}{\mlabList}{\lockList}$.
  If
  \begin{itemize}[nosep]
    \item $\Sigma_{\sigma_1}; \Gamma; \pc; \inLock \proves s_1 : \tau \dashv \outLock \geqErased \Sigma_{\sigma_2}; \Gamma; \pc; \inLock \proves s_2 : \tau \dashv \outLock$,
    \item $\conf{s_1}{\config_1} \stepsone \conf{s_1'}{\config_1'}$,
  \end{itemize}
  then $\conf{s_2}{\config_2} \bulletstepsone \conf{s_2'}{\config_2'}$.
\end{lemma}

\begin{proof}
  This is a proof by induction on the operational semantics of $s_1 \stepsone s_1'$.
  \begin{description}
    \item[Case \ruleref{E-Eval}:]
      By induction on $E$, if $s_1 = E[\tilde{s}_1]$, then $s_2 = E'[\tilde{s}_2]$ where $\tilde{s}_1 \geqErased \tilde{s}_2$.
      By induction on the operational semantics, $\conf{\tilde{s}_2}{\config_2} \bulletstepsone \conf{\tilde{s}_2'}{\config_2'}$,
      and \ruleref{B-Eval} applies to complete the case.

    \item[Cases \ruleref{E-IfT} and \ruleref{E-IfF}:]
      Here we have $s_1 = \IfThenElse{\tilde{v}}{\tilde{e}_1}{\tilde{e}_2}$,
      and $s_2 = \IfThenElse{v^\bullet}{e^\bullet_1}{e^\bullet_2}$.
      We consider two sub-cases.
      First, if $v^\bullet = \bullet$, we see that $\conf{s_2}{\config_2} \bulletstepsone \conf{\bullet}{\config_2}$ by \ruleref{B-BulletCtx}.
      If $v^\bullet \neq \bullet$, then $v^\bullet = \tilde{v}$, and therefore \ruleref{B-PureStep} allows $s_2$ to step, as desired.

    \item[Cases \ruleref{E-Cast}, \ruleref{E-Field}, \ruleref{E-Call}, and \ruleref{E-CallAtk}:]
      These cases follow the same logic as the previous case, with their corresponding syntax.

    \item[Case \ruleref{E-Ref}:]
      Here we have $s_1 = \RefOp{v_1}{\tau}$ so therefore $s_2 = \RefOp{v_2}{\tau}$ where $v_1 \geqErased v_2$.
      Inversion on \ruleref{E-Ref} proves $\mlabList = \mlabList', \mlabel$ where $\mlabel \prot \tau$.
      Since $\mlabList$ is the same in $\config_1$ and $\config_2$,
      if $\getLab(\tau) \actsfor \ellHigh$, then \ruleref{B-TRef} applies, and if not, \ruleref{B-URef} applies.

    \item[Case \ruleref{E-Deref}:]
      Here $s_1 = \deref{\loc}$ and $s_2 = \deref{v}$ where either $v = \erased$ or $v = \loc$.
      If $v = \erased$, then \ruleref{B-BulletCtx} applies.
      Otherwise, we know that $\Sigma_{\sigma_2}; \Gamma; \pc, \inLock \proves \deref{\loc} : \tau \dashv \outLock$.
      By inversion on the expression typing rules, we know that $\Sigma_{\sigma_2}; \Gamma \proves \loc : (\RefType{\tau})^\ell$,
      and by inversion on the value typing rules, we therefore have $\Sigma_{\sigma_2}(\loc) = \tau$.
      In other words, $\loc \in \dom(\sigma_2)$, so \ruleref{B-PureStep} applies with \ruleref{E-Deref}.

    \item[Case \ruleref{E-Assign}:]
      In this case $s_1 = \loc := v$ and $s_2 = v_1 := v_2$ where $v_1 = \erased$ or $v_1 = \loc$.
      If $v_1 = \erased$, then \ruleref{B-BAssign} applies.
      Otherwise, because $s_1$ is well-typed with $\Sigma_{\sigma_1}$, inversion on the typing rules proves $\Sigma_{\sigma_1}(\loc) = \tau'$.
      Because $s_1$ steps with \ruleref{E-Assign}, inversion on \ruleref{E-Assign} proves $\mlabList = \mlabList', \mlabel$ where $\mlabel \prot \tau'$.
      By inversion on the $\geqErased$ relation, it must be the case that $\Sigma_{\sigma_2}(\loc) = \tau'$ and $\Sigma_{\sigma_2}; \Gamma \proves v_2 : \tau'$.
      Since $\mlabList$ is the same in $\config_1$ and $\config_2$,
      this is sufficient to apply one of \ruleref{B-TAssign} or \ruleref{B-UAssign}, depending on $\getLab(\tau')$.
  \end{description}
  For all other cases, the heap remains unmodified and no decisions are made based on a value that may be $\erased$,
  so \ruleref{B-PureStep} applies to $s_2$ using the same step that applied to $s_1$.
\end{proof}

\begin{lemma}[Bullet Step Correspondence]
  \label{lem:bullet-step-correspond}
  For any class table $\CT$, statements $s_1$ and $s_2$, heaps $\sigma_1$ and $\sigma_2$,
  and heap-type $\Sigma$, if
  \begin{itemize}[nosep]
    \item $\Sigma \proves \CT~\mathsf{ok}$ is endorsement-free at $\ellHigh$,
    \item $s_1$ and $s_2$ are endorsement-free at $\ellHigh$,
    \item $\provesWt{\sigma_i}$ for both $i = 1, 2$,
    \item $\sigma_1 \lequiv \sigma_2$ with $\sigma_2 \subseteq \lerase{\sigma_1}^\erased$,
    \item $\Sigma \subseteq \Sigma_{\sigma_2}$,
    \item $\Sigma_{\sigma_1}; \Gamma; \pc; \inLock \proves s_1 : \tau \dashv \outLock \geqErased \Sigma_{\sigma_2}; \Gamma; \pc; \inLock \proves s_2 : \tau \dashv \outLock$, and
    \item $\conf{s_1}{\confTuple{\sigma_1}{\mlabList}{\lockList}} \stepsone^+ \confExpr{v}$,
  \end{itemize}
  then there exists statements $s_1'$ and $s_2'$,
  heaps $\sigma_1'$ and $\sigma_2'$,
  and label stacks $\mlabList'$ and $\lockList'$ such that
  \begin{itemize}[nosep]
    \item $\conf{s_1}{\confTuple{\sigma_1}{\mlabList}{\lockList}} \stepsone^+ \conf{s_1'}{\confTuple{\sigma_1'}{\mlabList'}{\lockList'}}$,
    \item $\conf{s_2}{\confTuple{\sigma_2}{\mlabList}{\lockList}} \bulletstepsone \conf{s_2'}{\confTuple{\sigma_2'}{\mlabList'}{\lockList'}}$,
    \item $s_1'$ and $s_2'$ are endorsement-free at $\ellHigh$,
    \item $\provesWt{\sigma_i'}$ for both $i = 1, 2$,
    \item $\sigma_1' \lequiv \sigma_2'$ with $\sigma_2' \subseteq \lerase{\sigma_1'}^\erased$, and
    \item $\Sigma_{\sigma_1'}; \Gamma; \pc; \inLock \proves s_1' : \tau \dashv \outLock \geqErased \Sigma_{\sigma_2'}; \Gamma; \pc; \inLock \proves s_2' : \tau \dashv \outLock$.
  \end{itemize}
\end{lemma}

\begin{proof}
  By Lemma~\ref{lem:bullet-complete}, the fact that $s_1 \stepsone^+ v$ means that $s_2 \bulletstepsone s_2'$.
  This will be a proof by induction on the rule used to prove $s_2 \bulletstepsone s_2'$,
  though in the case of \ruleref{B-TRef}, we may need to construct a new, different $s_2'$.
  \begin{description}
    \item[Case \ruleref{B-PureStep}:]
      We have that $s_1 \stepsone s_1'$ by whatever step was used in the hypothesis of \ruleref{B-PureStep}.
      To prove the typing proofs correspond, we note that, for most possible steps,
      both $s_1'$ and $s_2'$ type-check by the same logic as in the proof of Theorem~\ref{thm:preservation}, meaning the typing proofs transform in the same way.
      The exception is \ruleref{E-Endorse}.
      Here let $s_2 = \Endorse{v_2}{\ell'}{\ell}$ and consider two cases: if $v_2 = \erased$ and if $v_2 \neq \erased$.
      When $v_2 \neq \erased$, the same argument as in Theorem~\ref{thm:preservation} applies, and $s_1 = \Endorse{v_1}{\ell'}{\ell}$ follows the same step by the same argument.
      When $v_2 = \erased$, inversion on the typing rules gives us that $\ell' \nactsfor \ellHigh$.
      Because we know $s_2$ is endorsement-free at $\ellHigh$, this means $\ell \nactsfor \ellHigh$, so therefore $\Sigma; \Gamma \proves \erased : t^\ell$.
      Again, $s_1$ follows \ruleref{E-Endorse} and the typing proofs correspond.

      For the heap correspondence and well-typed conditions, we note that the heaps and their types remain unchanged for both executions.
      To maintain endorsement-freedom at $\ellHigh$, most possible steps cannot add new terms, so they cannot add new $\EndorseName$ terms.
      \ruleref{E-Call} and \ruleref{E-CallAtk}, however, can introduce new terms into $s_1'$ and $s_2'$ that may not have been present in $s_1$ and $s_2$.
      Because $\CT$ is endorsement-free at $\ellHigh$, any new sub-statements of the form $\Endorse{v}{\ell'}{\ell}$ must have the required property.

    \item[Case \ruleref{B-Eval}:]
      In this case $s_2 = E_2[\tilde{s}_2]$ and $\tilde{s}_2 \bulletstepsone \tilde{s}_2'$.
      By inversion on $s_1 \geqErased s_2$, it must be the case that $s_1 = E_1[\tilde{s}_1]$ where $\tilde{s}_1 \geqErased \tilde{s}_2$.
      By inversion on the set of evaluation contexts, $s_1$ can only step through \ruleref{E-Eval} and no other steps.
      Therefore, by induction, on the $\bulletstepsone$ relation, $\tilde{s}_1 \stepsone \tilde{s}_1'$ with the required properties,
      so \ruleref{E-Eval} gives us everything except correspondence of the typing proof.
      We get that by noting that we can apply Lemma~\ref{lem:expr-subst} in exactly the same way to both proofs.

    \item[Case \ruleref{B-BulletCtx} with {$B = \deref{[\cdot]}$, $(C)[\cdot]$, or $[\cdot].f$}:]
      Here we have that $s_2 = B[\erased]$, so by inversion on $s_1 \geqErased s_2$, we know that $s_1 = B[v_1]$ for some non-bullet value $v_1$.
      By the fact that $s_1 \stepsone^+ v$, we know that $s_1$ must step, so by inspection on the operational semantics,
      it must step with \ruleref{E-Deref}, \ruleref{E-Cast}, or \ruleref{E-Field}, depending on the syntactic form.
      In each case the result is a non-variable value $v_1'$, so therefore $s_1' = v_1' \geqErased \erased = v_2'$ with typing proofs
      using \ruleref{Val} to get to a value typing judgment that allows them to differ on $\erased$.
      The heap does not change.

    \item[Case \ruleref{B-BulletCtx} with {$B = \IfThenElse[\pc']{[\cdot]}{e^2_1}{e^2_2}$}:]
      First we note that $s_2' = \erased$ and $\sigma_2' = \sigma_2$.
      We also note that inversion on the typing rules shows $\Sigma_{\sigma_2}; \Gamma \proves \erased : \bool^\ell$ for some $\ell \nactsfor \ellHigh$
      and $\ell \prot \tau$, meaning \ruleref{Bullet} gives us $\Sigma_{\sigma_2}; \Gamma \proves \bullet : \tau$.
      We now examine $s_1$ and the corresponding steps.

      Because $\Sigma_{\sigma_1}; \Gamma; \pc; \inLock \proves s_1 : \tau \dashv \outLock \geqErased \Sigma_{\sigma_2}; \Gamma; \pc; \inLock \proves s_2 : \tau \dashv \outLock$,
      we know $s_1 = \IfThenElse[\pc']{v_1}{e^1_1}{e^1_2}$.
      This syntactic structure means $s_1$ must step with one of \ruleref{E-IfT} or \ruleref{E-IfF}.
      Because we have assumed that $\conf{s_1}{\confTuple{\sigma_1}{\mlabList}{\lockList}} \stepsone^+ \confExpr{v}$,
      we further know that
      $$\conf{s_1}{\confTuple{\sigma_1}{\mlabList}{\lockList}} \stepsone \conf{\atpc[\pc']{e^1_i}}{\confTuple{\sigma_1}{\mlabList}{\lockList}} \stepsmany \confExpr{\atpc[\pc']{v}} \stepsone \confExpr{v}.$$
      By inspection on the semantic rules, we know that \ruleref{E-Eval} must apply in each of the steps in the middle segment,
      meaning $\conf{e^1_i}{\confTuple{\sigma_1}{\mlabList}{\lockList}} \stepsmany \confExpr{v}$.

      The correspondence of the typing proof with $s_1$ proves that $\Sigma_{\sigma_1}; \Gamma \proves v_1 : \bool^\ell$ for some $\ell \nactsfor \ellHigh$.
      Inversion on that typing rules therefore tells us $\ell \actsfor \pc'$ and $\Sigma_{\sigma_1}; \Gamma; \pc'; \inLock \proves e^1_i : \tau \dashv \outLock$.
      By Corollary~\ref{cor:confinement}, $\mlabList' = \mlabList$ and $\lockList' = \lockList$,
      and $\sigma_1' \lequiv \sigma_1 \lequiv \sigma_2 = \sigma_2'$.
      By inductively applying Theorem~\ref{thm:preservation}, we get $\sigma_1' \supseteq \sigma_1$, so
      $$\sigma_2' = \sigma_2 \subseteq \lerase{\sigma_1}^\erased \subseteq \lerase{\sigma_1'}^\erased.$$
      By letting $s_1' = v$ and noting that all values are endorsement-free at~$\ellHigh$, we complete the case.

    \item[Case \ruleref{B-BulletCtx} with {$[\cdot].m(\overline{v})$}:]
      This case is very similar to the previous case.
      Again, $s_2' = \erased$ and $\sigma_2' = \sigma_2$.
      Also, inversion on the typing rules gives us $\Sigma_{\sigma_2}; \Gamma \proves \erased : C^\ell$ for some $\ell \nactsfor \ellHigh$, and $\ell \prot \tau$,
      again allowing \ruleref{Bullet} to prove $\Sigma_{\sigma_2}; \Gamma \proves \erased : \tau$.
      We again turn to $s_1$.

      The typing correspondence now means $s_1 = v_1.m(\overline{w})$, so it must step using \ruleref{E-Call} or \ruleref{E-CallAtk}.
      Therefore $v_1 = \new~C(\overline{w'})$ and $\mbody(C, m) = \mbodyExpr*$.
      Again, we know that it steps to a value, so now
      \begin{align*}
        \conf{s_1}{\confTuple{\sigma_1}{\mlabList}{\lockList}} & \stepsone \conf{\atpc[\pc_2]{(\funend{\tau}~e')}}{\confTuple{\sigma_1}{(\mlabList, \mlabel)}{\lockList}} \\
        & \stepsmany \conf{\atpc[\pc_2]{(\funend{\tau}~v)}}{\confTuple{\sigma_1'}{\mlabList_1'}{\lockList_1'}}
      \end{align*}
      where $e' = e[\overline{x} \mapsto \overline{w}, \this \mapsto \new~C(\overline{w'})]$ is an expression.
      Additionally, $\Sigma_{\sigma_1}; \Gamma; \pc_2; \inLock' \proves e' : \tau \dashv \outLock'$.
      By the correspondence of the typing proofs of $s_1$ and $s_2$, we know that $\Sigma_{\sigma_2}; \Gamma; \pc; \inLock \proves \erased.m(\overline{v}) : \tau \dashv \outLock$
      interpreting $\erased$ as $\Sigma_{\sigma_2}; \Gamma \proves \erased : C^\ell$.
      Inversion on the typing rules therefore gives us that $\ell \nactsfor \ellHigh$ and $\ell \actsfor \pc_1$.
      By transitivity, we know that $\pc_1 \nactsfor \ellHigh$, so by the fact that $\CT$ is endorsement-free at $\ellHigh$, we have that $\pc_2 \nactsfor \ellHigh$.
      Therefore, we can apply Corollary~\ref{cor:confinement} to our above semantic steps, giving:
      \begin{itemize}[nosep]
        \item $\sigma_1' \lequiv \sigma_1 \lequiv \sigma_2 = \sigma_2'$,
        \item $\mlabList_1' = \mlabList, \mlabel$, and
        \item $\lockList_1' = \lockList$.
      \end{itemize}
      Again, Theorem~\ref{thm:preservation}'s result tells us $\sigma_1' \supseteq \sigma_1$,
      meaning $\sigma_2' \subseteq \lerase{\sigma_1'}^\erased$.
      Applying \ruleref{E-AtPc} and \ruleref{E-Return} while letting $s_1' = v$ completes the case.

    \item[Case \ruleref{B-TRef}:]
      Here we take the $\tilde{s}_2'$ from Lemma~\ref{lem:bullet-complete} as a candidate, which we may modify.
      In particular, we note that $s_2 = \RefOp{v_2}{\tau'}$, so by the typing correspondence, $s_1 = \RefOp{v_1}{\tau'}$.
      Therefore, $s_1$ must step using \ruleref{E-Ref}, giving $s_1' = \loc$ for some $\loc \notin \dom(\sigma_1)$ and
      $\sigma_1' = \subst{\sigma_1}{\loc}{(v_1, \tau')}$.

      For $s_2'$, we know that $\dom(\sigma_2) \subseteq \dom(\sigma_1)$, so $\loc \notin \dom(\sigma_2)$.
      We also know that $s_2$ could step using \ruleref{B-TRef}, so we can use the same step, setting the location to $\loc$.
      Therefore, $\sigma_2' = \subst{\sigma_2}{\loc}{(v_2, \tau')}$.
      The fact that $s_2'$ is well-typed in $\Sigma_{\sigma_2'}$ follows directly from this extension.

      Endorsement-freedom of $s_1'$ and $s_2'$, typing correspondence, and that $\sigma_1' \lequiv \sigma_2'$ are now straightforward.
      To show that $\sigma_2' \subseteq \lerase{\sigma_1'}^\erased$,
      we note that the typing correspondence between $v_1$ and $v_2$ means that either $v_2 = v_1$ or $v_2 = \erased$.
      The second case is impossible because $\getLab(\tau') \actsfor \ellHigh$, so inversion on the typing rules demonstrates $\Sigma_{\sigma_2}; \Gamma \nproves \erased : \tau'$.
      With $v_2 = v_1$, the relation between $\sigma_1'$ and $\sigma_2'$ follows directly from their definitions and the corresponding relation between $\sigma_1$ and $\sigma_2$.

    \item[Case \ruleref{B-URef}:]
      Using the same logic as the previous case, $s_1$ must step using \ruleref{E-Ref},
      and we can make $s_2' = s_1' = \loc$ for some $\loc \notin \dom(\sigma_1) \supseteq \dom(\sigma_2)$.
      We again have that $s_1'$ and $s_2'$ correspond and are well-typed and that $\sigma_1' \lequiv \sigma_2'$.
      Finally, we note that, by assumption from \ruleref{B-URef}, $s_2 = \RefOp{v_2}{\tau'}$ where $\getLab(\tau') \nactsfor \ellHigh$.
      Therefore, $\lerase{\sigma_1'}^\erased = \lerase{\subst{\sigma_1}{\loc}{\tau'}}^\erased = \subst{\lerase{\sigma_1}^\erased}{\loc}{(\erased, \tau')}$,
      and correspondingly, $\sigma_2' = \subst{\sigma_2}{\loc}{(\erased, \tau')}$.
      The correspondence follows from the correspondence between $\sigma_1$ and $\sigma_2$.

    \item[Case \ruleref{B-BAssign}:]
      Here $s_2 = \erased := v_2$, so $s_1 = \loc := v_1$ where $v_1 \geqErased v_2$.
      By inversion on the typing rules and the $\geqErased$ relation, we know that $\Sigma_{\sigma_1}; \Gamma \proves \loc : \RefType{\tau'}^\ell$
      and $\Sigma_{\sigma_2}; \Gamma \proves \erased : \RefType{\tau'}^\ell$.
      Moreover, we know that $\ell \nactsfor \ellHigh$ and $\ell \prot \tau$.
      Since $s_1$ steps, it must step with \ruleref{E-Assign}, meaning $\sigma_1' = \subst{\sigma_1}{\loc}{(v_1, \tau')}$.
      Therefore $\lerase{\sigma_1'}^\erased = \lerase{\sigma_1}^\erased \supseteq \sigma_2 = \sigma_2'$.
      Letting $s_1' = s_2' = ()$ completes the case.

    \item[Case \ruleref{B-TAssign}:]
      This is similar to the \ruleref{B-TRef} case, but we do not need to construct a new location, as $s_2 = \loc := v$.
      We also know by the same logic as in that case that $v \neq \erased$, so $s_1 = s_2$.
      \ruleref{B-TAssign} and \ruleref{E-Assign} produce precisely the same output on the same input, proving the case.

    \item[Case \ruleref{B-UAssign}:]
      Here we note that $s_2 = \loc := v_2$, meaning $s_1 = \loc := v_1$ where $v_1 \geqErased v_2$ and $i \in \dom(\sigma_2)$.
      By inversion on the typing rules, we know that $\Sigma_{\sigma_i}(\loc) = \tau'$ for both $i = 1, 2$ and some $\tau'$ where $\getLab(\tau') \nactsfor \ellHigh$.
      Therefore, $s_1$ steps using \ruleref{E-Assign}, so $\sigma_1' = \subst{\sigma_1}{\loc}{(v_1, \tau')}$ where $\sigma_1(\loc) = (v, \tau')$ for some $v$.
      As a result, $\sigma_2' = \sigma_2 \subseteq \lerase{\sigma_1}^\erased = \lerase{\sigma_1'}^\erased$.
      The two steps result in $s_1' = s_2' = ()$, so the statements type-check with corresponding rules.
    \qedhere
  \end{description}
\end{proof}

\begin{corollary}
  \label{cor:erased-full-eval}
  For any class table $\CT$, heap type $\Sigma$, and expression (not statement) $e$, if
  \begin{itemize}[nosep]
    \item $\Sigma \proves \CT~\mathsf{ok}$ is endorsement-free at $\ellHigh$,
    \item $e$ is endorsement-free at $\ellHigh$,
    \item $\Sigma \subseteq \Sigma_{\sigma_1}$,
    \item $\Sigma; \Gamma; \pc; \inLock \proves e : \tau \dashv \outLock$,
    \item $\provesWt{\sigma_1}$, and
    \item $\conf{e}{\confTuple{\sigma_1}{\mlabList}{\lockList}} \stepsmany \conf{v}{\confTuple{\sigma_1'}{\mlabList'}{\lockList'}}$,
  \end{itemize}
  then there is some value $v'$, heap $\sigma_2'$, and heap type $\Sigma_2'$ such that
  \begin{itemize}[nosep]
    \item $\conf{e}{\confTuple{\lerase{\sigma_1}^\erased}{\mlabList}{\lockList}} \bulletstepsmany \conf{v'}{\confTuple{\sigma_2'}{\mlabList'}{\lockList'}}$ and
    \item $\sigma_1' \lequivLocIso \sigma_2'$.
  \end{itemize}
\end{corollary}

\begin{proof}
  This proof follows from Lemma~\ref{lem:bullet-step-correspond} and induction on the number of steps,
  letting $s_1 = s_2 = e$ and $\sigma_2 = \lerase{\sigma_1}^\bullet$ to start.
  If there are zero steps---that is $e = s_1 = v$---then we are done.
  Otherwise Lemma~\ref{lem:bullet-step-correspond} allows us to step $s_2$ once using $\bulletstepsone$ and provides a corresponding set of
  steps using $\stepsone$ for $s_1$.
  The result may have differently-named locations from the original,
  but Theorem~\ref{thm:loc-name-iso-semantics} allows us to continue stepping a location-name isomorphic expression.
  The steps therefore maintain all requirements to apply Lemma~\ref{lem:bullet-step-correspond} again until $s_1$ reaches a value.
  At that point, we are assured $\sigma_2' \subseteq \lerase{\sigma_1''}^\erased$ and $\sigma_1' \lequiv \sigma_1''$ for some $\sigma_1'' \locIso \sigma_1'$.
  Therefore $\sigma_1' \lequivLocIso \sigma_2'$.
\end{proof}

\begin{retheorem}{thm:ni}[Noninterference]
  Let $\CT$ be a class table where $\Sigma \proves \CT~\mathsf{ok}$ is endorsement-free at $\ellHigh$.
  For any well-typed heaps $\sigma_1$ and $\sigma_2$ such that $\Sigma \subseteq \Sigma_{\sigma_i}$
  and any invocation $I$ such that $\Sigma \proves I$ and $(I, \CT, \sigma_i) \Downarrow \sigma_i'$,
  if $\sigma_1 \lequivLocIso \sigma_2$, then $\sigma_1' \lequivLocIso \sigma_2'$.
\end{retheorem}

\begin{proof}
  First we note that since $\Sigma \subseteq \Sigma_{\sigma_i}$, Lemma~\ref{lem:heap-ext} means $\Sigma_{\sigma_i} \proves \CT~\mathsf{ok}$ for both $i = 1, 2$,
  meaning our various lemmas apply in both cases.
  Without loss of generality, we assume $\sigma_1 \lequiv \sigma_2$, since we can permute the location names in one to match the other and permute the results back later.
  There exists a unique
  $$\tilde{\sigma} = \lerase{\sigma_1}^\erased = \lerase{\sigma_2}^\erased$$

  Let $I = (\loc, m(\overline{v}), \ell)$.
  Note that $\deref{\loc}.m(\overline{v})$ is an expression with no $\EndorseName$ statements and $\Sigma \proves \deref{\loc}.m(\overline{v}) : \tau$ for some $\tau$.
  Therefore, by Corollary~\ref{cor:erased-full-eval},
  $$\conf{\deref{\loc}.m(\overline{v})}{\confTuple{\tilde{\sigma}}{\ell}{\cdot}} \bulletstepsmany \conf{v}{\confTuple{\tilde{\sigma}'}{\ell}{\cdot}}$$
  where $\tilde{\sigma}' \lequivLocIso \sigma_i'$ for both $i = 1, 2$.
  Transitivity of $\lequivLocIso$ then proves $\sigma_1' \lequivLocIso \sigma_2'$.
\end{proof}

\endgroup

%% file: security-proofs.tex
\section{Proof of Reentrancy Security}
\label{sec:security-proofs-full}

We now prove Theorem~\ref{thm:security}.
As discussed in Section~\ref{sec:reentrancy-defn-formal},
we do this by first proving Theorem~\ref{thm:all-tail-reenter} saying all reentrancy is tail-reentrancy
and Theorem~\ref{thm:tail-reenter-safe} that says tail reentrancy is secure according on Definition~\ref{defn:reentrancy-security}.

\subsection{\langname Allows Only Tail Reentrancy}
\label{sec:only-tail-reenter-proof}

We start by proving Theorem~\ref{thm:all-tail-reenter}.
We prove this theorem using the general formulation of ``trusted'' and ``untrusted'' labels.
In particular, we partition $\Labs$ into a downward-closed sublattice~$\T$ and the attacker-controlled labels~$\adv = \overline{\T}$.
Notationally, we will use $\ellHigh$ to denote some trusted label ($\ellHigh \in \T$), rather than a distinguished one.
We refer to code complying with locks in $\T$\=/code, to mean it complies with locks in $\ellHigh$\=/code for all $\ellHigh \in \T$.

Finally, we will prove the result for two adversarial models:
one where \ruleref{E-CallAtk} is admissible and $\adv$ is a sublattice,
and the other where \ruleref{E-CallAtk} is \emph{not} admissible, but $\adv$ has no restrictions beyond $\adv = \overline{\T}$.
These two proofs are extremely similar.
Indeed, they differ only in a single case of Lemma~\ref{lem:lock-respect} and Lemma~\ref{lem:method-call-lock-list} on which it relies.
We will specifically call out the differences when they arise.

The proof follows the following general structure.
First we show that high-integrity code maintains all of the input locks $\inLock$ it claims to and the operational semantics maintain all dynamic locks.
Second, we will show that, if a statement that complies with locks steps to an auto-endorse call,
it cannot comply with a lock on any label that call endorses through (i.e., one that does not trust $\pc_1$ but does trust $\pc_2$).
Finally, we connect these to show that, for low-integrity call from a high-integrity context that proceeds to make a reentrant call,
the original low-integrity call must have been in tail position form the original high-integrity execution.

To discuss the security of an invocation mid-evaluation, we need to discuss the security of a statement $s$ with respect to locks.
We do this using several different tools.
First, we extend our notion of lock compliance in $\T$\=/code to statements.
We do this with a judgment $\provesCwl{\pc}{s}$.
The nontrivial rules are as follows.
\begin{mathpar}
  \infer{
    \provesCwl{\pc}{e_1} \\
    \provesCwl{\pc}{e_2}
  }{\provesCwl{\pc}{(\IfThenElse[\pc']{v}{e_1}{e_2})}}
  \and
  \infer{
    \provesCwl{\pc}{e}
  }{\provesCwl{\pc}{(\lockIn{\ell}{e})}}
  \and
  \infer{
    \provesCwl{\pc}{s}
  }{\provesCwl{\pc}{(\withLock{s})}}
  \and
  \infer{
    \provesCwl{\pc}{s} \\
    \provesCwl{\pc}{e}
  }{\provesCwl{\pc}{(\letIn{s}{e})}}
  \and
  \infer{
    \provesCwl{\pc}{s}
  }{\provesCwl{\pc}{(\funend{\tau}~s)}}
  \\
  \infer{
    \provesCwl{\pc'}{s}
  }{\provesCwl{\pc}{(\atpc[\pc']{s})}}
  \and
  \infer{
    \provesCwl{\pc}{s} \\
    \pc \notin \T
  }{\provesCwl{\pc}{(\ignoreLocks{s})}}
\end{mathpar}
If $s$ has none of the syntactic forms in the rules defined above, then $\provesCwl{\pc}{s}$ for any $\pc$ and $\T$.
Note that, because $\atpcName$ terms are statements but not expressions, for any expression $e$,
$\provesCwl{\pc}{e}$ if one of two conditions holds:
either $\pc \notin \T$ or $e$ has no subexpressions of the form $\ignoreLocks{e'}$.
As a result, we can also specify our definition of lock compliance from class tables using this judgment.
Specifically,
\begin{minipage}{\textwidth}
\small
$$\CT~\text{complies with locks in $\T$\=/code} \iff
\infer{
  \CT(C) = \cdef* \\\\
  \mdef* \in \overline{M}
}{\provesCwl{\ell_C}{e}}~\text{is admissible for $\CT$}$$
\end{minipage}

\begin{lemma}
  \label{lem:statement-cwl-variance}
  If $\provesCwl{\pc}{s}$ and $\pc \actsfor \pc'$, then $\provesCwl{\pc'}{s}$.
\end{lemma}

\begin{proof}
  By simple induction on the definition of $\provesCwl{\pc}{s}$.
\end{proof}

We will also be considering statements in the middle of evaluation,
so we need a way to extract the $\pc$ label that we expect sub-statements to type-check with,
and similarly we need to extract the list of dynamic locks that will be present when a sub-statement completes executing.
We do that using the following two recursive functions defined on evaluation contexts.
\begin{align*}
  \getLocks(\lockList, E) & = \begin{cases}
    \lockList & \text{if } E = [\cdot] \\
    \getLocks((\lockList, \ell), E') & \text{if } E = \withLock{E'} \\
    \getLocks(\lockList, E') & \text{if } E = \letIn{E'}{e},~\funend{\tau}~E',~\text{or}~\atpc{E'}
  \end{cases}
  \\
  \innerPc(\pc, E) & = \begin{cases}
    \pc & \text{if } E = [\cdot] \\
    \innerPc(\pc', E') & \text{if } E = \atpc[\pc']{E'} \\
    \innerPc(\pc, E') & \text{if } E = \letIn{E'}{e},~\funend{\tau}~E',~\text{or}~\withLock{E'}
  \end{cases}
\end{align*}
We extend both of these to statements by $\getLocks(\lockList, E[e]) = \getLocks(\lockList, E)$, and similarly for $\innerPc$.

\begin{definition}[Configuration Safety]
  \label{defn:config-safety}
  A statement-configuration pair $\conf{s}{\confTuple*}$ is \emph{$\T$\=/safe with $\pc$ and $\hat{\lockList}$} if
  \begin{enumerate}[nosep]
    \item\label{safety:li:ct-ok} $\Sigma_\sigma \proves \CT~\mathsf{ok}$ complies with locks in $\T$\=/code,
    \item\label{safety:li:heap-ok} $\proves \sigma~\mathsf{wt}$,
    \item\label{safety:li:well-typed} $\Sigma_\sigma; \Gamma; \pc; \inLock \proves s : \tau \dashv \outLock$,
    \item\label{safety:li:statement-rl} $\provesCwl{\pc}{s}$,
    \item\label{safety:li:lock-list} $\lockList = \getLocks(\hat{\lockList}, s)$, and
    \item\label{safety:li:safe-locks}
      for any $E$ and $s'$ where $E \neq E'[\funend{\tau}~[\cdot]]$ and $\pc' = \innerPc(\pc, E) \in \T$,
      if $s = E[s']$ then there is some $\inLock'$ such that
      $\Sigma_\sigma; \Gamma'; \pc'; \inLock' \proves s' : \tau' \dashv \outLock'$ and $(\bigmeet \getLocks(\hat{\lockList}, E)) \meet \inLock' \actsfor \pc'$.
  \end{enumerate}
\end{definition}

\begin{lemma}
  \label{lem:safe-in-ctx}
  If $\confExpr{E[s]}$ is $\T$\=/safe at $\pc$ and $\lockList$,
  then $\confExpr{s}$ is $\T$\=/safe at $\innerPc(\pc, E)$ and $\getLocks(\lockList, E)$.
\end{lemma}

\begin{proof}
  By induction on $E$ and the definitions of $\innerPc$ and $\getLocks$.
\end{proof}

\begin{lemma}[Preservation of $\T$\=/Safety]
  \label{lem:config-safety}
  If $\confExpr{s}$ is $\T$\=/safe with $\pc$ and $\hat{\lockList}$, and $\confExpr{s} \stepsone \conf{s'}{\config'}$,
  then $\conf{s'}{\config'}$ is $\T$\=/safe with $\pc$ and $\hat{\lockList}$.
\end{lemma}

\begin{proof}
  Condition~\ref{safety:li:ct-ok} follows from Lemma~\ref{lem:heap-type-ext} and the fact that $\CT$ must remain unchanged.
  Conditions~\ref{safety:li:heap-ok} and~\ref{safety:li:well-typed} follow directly from Theorem~\ref{thm:preservation}.
  We prove the other three conditions by induction on the operational semantics.
  Notationally, we let $\config = \confTuple*$ and $\config' = \confTuple{\sigma'}{\mlabList'}{\lockList'}$.
  Also, by assumption, there is some $\inLock'$ such that $\Sigma_\sigma; \Gamma; \pc; \inLock' \proves s : \tau \dashv \outLock$ and $(\bigmeet \hat{\lockList}) \meet \inLock' \actsfor \pc$.
  We assume without loss of generality that $\inLock$ has this property.

  \begin{description}
    \item[Case \ruleref{E-Eval}:]
      In this case $s = E[\tilde{s}]$, $\confExpr{\tilde{s}} \stepsone \conf{\tilde{s}'}{\config'}$, and $s' = E[\tilde{s}']$.
      Let $\pc' = \innerPc(\pc, E)$ and $\hat{\lockList}' = \getLocks(\hat{\lockList}, E)$.
      By Lemma~\ref{lem:safe-in-ctx}, we know that $\tilde{s}$ is $\T$\=/safe at $\pc'$ and $\hat{\lockList}'$,
      so by induction on the operational semantics, $\conf{\tilde{s}'}{\config'}$ is as well.
      We also note that $\Sigma_\sigma; \Gamma'; \pc'; \inLock' \proves \tilde{s} : \tau' \dashv \outLock'$.

      By the safety of $\conf{s}{\config}$, for every pair of sub-contexts $E_1$ and $E_2$ such that $E = E_1[E_2]$,
      either $\tilde{\pc} = \innerPc(\pc, E_1) \notin T$, $E_1 = E_1'[\funend{\tau}~[\cdot]]$, or $\Sigma_\sigma; \tilde{\Gamma}; \tilde{\pc}; \tilde{\inLock} \proves E_2[\tilde{s}] : \tau_1 \dashv \tilde{\outLock}$
      for some $\tilde{\inLock}$ where $(\bigmeet \getLocks(\hat{\lockList}, E_1)) \meet \tilde{\inLock} \actsfor \tilde{\pc}$.
      By Theorem~\ref{thm:preservation}, $\Sigma_{\sigma'}; \Gamma'; \pc'; \inLock' \proves \tilde{s}' : \tau' \dashv \outLock'$,
      so by Lemma~\ref{lem:expr-subst}, we also have that $\Sigma_{\sigma'}; \tilde{\Gamma}; \tilde{\pc}; \tilde{\inLock} \proves E_2[\tilde{s}'] : \tau_1 \dashv \tilde{\outLock}$.
      As this holds for every choice of $E_1$ and $E_2$, this proves the case.

    \item[Cases \ruleref{E-IfT} and \ruleref{E-IfF}:]
      In both cases we have $s = \IfThenElse[\pc']{v}{e_1}{e_2}$ and $s' = \atpc[\pc']{e_i}$ for either $i = 1$ or $2$.
      By inversion on the typing rules, $\pc \actsfor \pc'$ and $\Sigma_\sigma; \Gamma; \pc'; \inLock \proves e_i : \tau \dashv \outLock$ for both $i = 1, 2$.
      Moreover, by Lemma~\ref{lem:statement-cwl-variance}, $\provesCwl{\pc'}{e_i}$, so Condition~\ref{safety:li:statement-rl} holds for $s'$.
      Because $e_1$ and $e_2$ are expressions, we know that if $e_i = E[\tilde{s}]$, then $E$ consists entirely of $\Let$ and $\ignoreLocksName$ statements and $\tilde{s}$ is an expression.
      Therefore, if $\pc' \notin T$, then Condition~\ref{safety:li:safe-locks} is trivial.

      If $\pc' \in T$, then because $\T$ is downward-closed, $\pc \in T$.
      Because $\provesCwl{\pc}{e_i}$, $\ignoreLocksName$ cannot appear in $e$ in this sub-case, so $E$ consists entirely of $\Let$ statements.
      As a result, $\Sigma_\sigma; \Gamma'; \pc'; \inLock \proves \tilde{s} : \tau' \dashv \outLock$ for some $\Gamma' \supseteq \Gamma$ and $\tau'$.
      Because $(\bigmeet \hat{\lockList}) \meet \inLock \actsfor \pc \actsfor \pc'$,
      this proves that $\conf{e_i}{\config'}$ is $\T$\=/safe at $\pc'$ and $\hat{\lockList}$.
      Since $s' = \atpc[\pc']{e_i}$, the $\T$\=/safety transfers to $\conf{s'}{\config'}$.

    \item[Case \ruleref{E-Let}:]
      Here $s = (\letIn{v}{e})$ and $s' = \subst{e}{x}{v}$.
      Theorem~\ref{thm:preservation} proves $\Sigma_\sigma; \Gamma; \pc; \inLock \proves s' : \tau \dashv \outLock$.
      Moreover, because $s'$ is an expression, by the same logic as in the previous case, $\conf{s'}{\config'}$ must be $\T$\=/safe at $\pc$ and $\hat{\lockList}$.

    \item[Case \ruleref{E-Lock}:]
      Here $s = \lockIn{\ell}{e}$.
      First we note that $\lockList' = (\lockList, \ell) = (\hat{\lockList}, \ell) = \getLocks(\hat{\lockList}, \withLock{e})$, as is required by Condition~\ref{safety:li:lock-list}.

      Next, inversion on the typing rules tells that $\Sigma_\sigma; \Gamma; \pc; \inLock' \proves e : \tau \dashv \outLock'$
      where $\inLock' \meet \ell \actsfor \inLock$ and $\outLock' \meet \ell \actsfor \outLock$.
      Further, we know that $s' = \withLock{e}$, $\lockList = \hat{\lockList}$, and $\lockList' = (\lockList, \ell)$
      By Condition~\ref{safety:li:statement-rl} on $s$,
      first Condition~\ref{safety:li:statement-rl} holds trivially on $s'$,
      and second, either $\pc \in T$ or $e$ contains no $\ignoreLocksName$ terms, as $e$ is an expression.
      Therefore, by the same logic as in the previous two cases, it suffices to show Condition~\ref{safety:li:safe-locks} holds when $\pc \in T$ and $E = \withLock{[\cdot]}$.
      Here we know that $\Sigma_\sigma; \Gamma; \pc; \inLock' \proves e : \tau \dashv \outLock'$ with $\inLock'$ defined as above.
      As a result,
      \begin{align*}
        \left(\bigmeet \getLocks(\hat{\lockList}, E)\right) \meet \inLock' & = \left(\bigmeet (\hat{\lockList}, \ell)\right) \meet \inLock' \\
        & = \left(\bigmeet \hat{\lockList}\right) \meet \ell \meet \inLock' \\
        & \actsfor \left(\bigmeet \hat{\lockList}\right) \meet \inLock \\
        & \actsfor \pc.
      \end{align*}

    \item[Case \ruleref{E-Unlock}:]
      Here $s = \withLock{v}$ and $s' = v$, so Condition~\ref{safety:li:statement-rl} is trivial.
      Condition~\ref{safety:li:lock-list} follows from the semantic rule that requires $\lockList = (\lockList', \ell)$,
      so if $\getLocks(\hat{\lockList}, s) = \lockList$, then $\hat{\lockList} = \lockList' = \getLocks(\hat{\lockList}, v)$.
      Condition~\ref{safety:li:safe-locks} follows from the fact that values type-check with any $\inLock$, including $\pc$.

    \item[Cases \ruleref{E-Call} and \ruleref{E-CallAtk}:]
      Here $s = \new~C(\overline{v}).m(\overline{w})$
      If we let $\mbody(C, m) = \mbodyExpr*$, then $s' = \atpc[\pc_2]{(\funend{\tau}~e')}$ where $e' = e[\overline{x} \mapsto \overline{w}, \this \mapsto \new~C(\overline{v})]$.
      By Condition~\ref{safety:li:ct-ok} on $s$, we know that $\provesCwl{\ell_C}{e'}$.
      The \ruleref{Method-Ok} rule requires that $\ell_C \actsfor \pc_2$,
      so therefore by Lemma~\ref{lem:statement-cwl-variance} proves $\provesCwl{\pc_2}{e'}$,
      proving Condition~\ref{safety:li:statement-rl}.

      Since the body of the method is an expression and $\lockList = \lockList'$, Condition~\ref{safety:li:lock-list} holds trivially.

      For Condition~\ref{safety:li:safe-locks}, we consider multiple possible evaluation contexts $E$.
      If $E = [\cdot]$, note that $\funend{\tau}~s''$ type-checks with any $\inLock$.
      If $E = \funend{\tau}~[\cdot]$, then this is precisely the caveat that Condition~\ref{safety:li:safe-locks} does not restrict.
      If $E = \atpc[\pc_2]{\funend{\tau}~[\cdot]}$, \ruleref{Method-Ok} ensures that
      $\Sigma_\sigma; \cdot; \pc_2; \inLock' \proves e' : \tau \dashv \outLock'$ for some $\inLock' \actsfor \pc_2$.
      In particular, this means $(\bigmeet \hat{\lockList}) \meet \inLock' \actsfor \pc_2$ regardless of the contents of $\hat{\lockList}$.
      Moreover, because we know that $\provesCwl{\pc_2}{e'}$ and $e'$ is an expression, either $\pc_2 \notin \T$,
      in which case Condition~\ref{safety:li:safe-locks} is trivial in $e'$, or $\pc_2 \in \T$ and $e'$ does not contain $\ignoreLocksName$ terms.
      In the second case, the same logic as in several previous cases completes the proof that Condition~\ref{safety:li:safe-locks} holds, and thus the case.
  \end{description}
  In all other cases the step leaves $\lockList$ unchanged and produces a value.
  All well-typed value type check with any $\inLock$ and $\provesCwl{\pc}{v}$ for any label $\pc$ and value $v$, so all conditions hold.
\end{proof}

\begin{lemma}
  \label{lem:lock-respect}
  For any label $\ellHigh \in \T$, statement $s$, configuration $\config = \confTuple*$, lock list $\hat{\lockList}$,
  if
  \begin{enumerate}[nosep]
    \item\label{lem:locks:li:safety} $\conf{s}{\config}$ is $\T$\=/safe with $\pc$ and $\hat{\lockList}$ for some label $\pc$,
    \item\label{lem:locks:li:no-ignore-locks} $s$ contains no sub-statements of the form $\ignoreLocks{s'}$,
    \item\label{lem:locks:li:steps} $\conf{s}{\config} \stepsmany \conf{E[\new~C(\overline{v}).m(\overline{w})]}{\config'} \stepsone \conf{s'}{\config''}$, and
    \item\label{lem:locks:li:endorsement} $\mtype(C, m) = \mtypeExpr{\overline{\tau_a}}{\pc_1}{\pc_2}{\hat{\outLock}}{\hat{\tau}}$ with $\pc_1 \nactsfor \ellHigh$ and $\pc_2 \actsfor \ellHigh$,
  \end{enumerate}
  then for any $\inLock$ and $\outLock$ such that $\Sigma_\sigma; \Gamma; \pc; \inLock \proves s : \tau \dashv \outLock$, then $(\bigmeet \hat{\lockList}) \meet (\inLock \join \outLock) \nactsfor \ellHigh$.
\end{lemma}

\begin{proof}
  This is a proof by induction on the number of steps in premise~\ref{lem:locks:li:steps}.
  For the base case of zero steps, $s = E[\new~C(\overline{v}).m(\overline{w})]$.
  We prove this case by induction on $E$.
  For these cases, we will use the notational short-hand $s' = E'[\new~C(\overline{v}).m(\overline{w})]$
  where $E'$ will be defined in each inductive case.
  \begin{description}
    \item[Case {$E = [\cdot]$}:]
      The $\T$\=/safety of $\conf{s}{\config}$ with $\pc$ and $\hat{\lockList}$
      ensures $\Sigma_\sigma; \Gamma; \pc; \inLock \proves \new~C(\overline{v}).m(\overline{w}) : \tau \dashv \outLock$.
      By inversion on the typing rules, we know that $\pc_1 \actsfor \pc_2 \join \inLock$.
      We also know that this expression steps again, so by inversion on the operational semantics,
      it must step using \ruleref{E-Call}, meaning $\bigwedge_{\ell \in \lockList} (\pc_1 \actsfor \pc_2 \join \ell)$.
      Therefore, by the fact that $\meet$ produces the greatest lower bound and the distributive property of the lattice,
      $$\pc_1 \actsfor \bigmeet_{\ell \in \lockList} (\pc_2 \join \ell) \meet (\pc_2 \join \inLock) = \pc_2 \join \left(\left(\bigmeet \lockList\right) \meet \inLock\right).$$
      Moreover, because $\pc_1 \nactsfor \ellHigh$, transitivity of $\actsfor$ tells us that this label does not act for $\ellHigh$.
      Yet $\pc_2 \actsfor \ellHigh$, so by the definition of join, it must be the case that $((\bigmeet \lockList) \meet \inLock) \nactsfor \ellHigh$.
      Because $\inLock \actsfor \inLock \join \outLock$, and $\lockList = \hat{\lockList}$ in this case,
      transitivity of $\actsfor$ and equality substitution proves $(\bigmeet \hat{\lockList}) \meet (\inLock \join \outLock) \nactsfor \ellHigh$, as desired.

    \item[Case $E = \letIn{E'}{e}$:]
      By inversion on the typing rules, we note that $\Sigma_\sigma; \Gamma; \pc; \inLock \proves s' : \tau' \dashv \outLock'$ where $\outLock' \actsfor \inLock$.
      Premises~\ref{lem:locks:li:safety} and~\ref{lem:locks:li:no-ignore-locks} are clearly true for $\confExpr{s'}$,
      so by induction on $E$, $(\bigmeet \hat{\lockList}) \meet (\inLock \join \outLock') \nactsfor \ellHigh$.
      Since $\outLock' \actsfor \inLock$, we know that $\inLock \join \outLock' = \inLock \actsfor \inLock \join \outLock$.
      Transitivity of $\actsfor$ then proves the desired result.

    \item[Case $E = \funend{\tau}~E'$:]
      Here inversion on the typing rules tells us $\Sigma_\sigma; \cdot; \pc; \inLock' \proves s' : \tau \dashv \outLock'$
      for some $\inLock'$ and $\outLock'$ where $\inLock' \join \outLock' \actsfor \outLock$.
      As with the previous case, our inductive hypothesis on $E$ applies, giving us $(\bigmeet \hat{\lockList}) \meet (\inLock' \join \outLock') \nactsfor \ellHigh$.
      Since $\inLock' \join \outLock' \actsfor \outLock \actsfor \inLock \join \outLock$, transitivity of $\actsfor$ again gives us the desired result.

    \item[Case $E = \withLock{E'}$:]
      Here inversion on the typing rules tells us $\Sigma_\sigma; \Gamma; \pc; \inLock' \proves s' : \tau \dashv \outLock'$
      where $\inLock' \meet \ell \actsfor \inLock$ and $\outLock' \meet \ell \actsfor \outLock$.
      By the definition of $\getLocks$, we know that $\confExpr{s'}$ must be $\T$\=/safe with $\pc$ and $(\hat{\lockList}, \ell)$.
      Premise~\ref{lem:locks:li:no-ignore-locks} is clearly true of $s'$ since we have not added new syntax,
      so induction on $E$ tells us $(\bigmeet (\hat{\lockList}, \ell)) \meet (\inLock' \join \outLock') \nactsfor \ellHigh$.
      Using the above facts and the distributive property of the lattice,
      \begin{align*}
        \left(\bigmeet (\hat{\lockList}, \ell)\right) \meet (\inLock' \join \outLock') & = \left(\bigmeet \hat{\lockList}\right) \meet \ell \meet (\inLock' \join \outLock') \\
        & = \left(\bigmeet \hat{\lockList}\right) \meet \left((\inLock' \meet \ell) \join (\outLock' \meet \ell)\right) \\
        & \actsfor \left(\bigmeet \hat{\lockList}\right) \meet (\inLock \join \outLock).
      \end{align*}
      Transitivity of $\actsfor$ finishes the case.

    \item[Case {$E = \atpc[\pc']{E'}$}:]
      Here $\confExpr{s'}$ is $\T$\=/safe at $\pc'$ and $\hat{\lockList}$ and premise~\ref{lem:locks:li:no-ignore-locks} clearly holds,
      so induction on $E$ proves the case.

    \item[Case $E = \ignoreLocks{E'}$:] This case is impossible by assumption~\ref{lem:locks:li:no-ignore-locks}.
  \end{description}

  We now move on to the inductive case on the number of steps.
  For all cases, Lemma~\ref{lem:config-safety} ensures that premise~\ref{lem:locks:li:safety} remains true after a single step.
  By inspection on the operational semantics, we can introduce $\ignoreLocksName$ terms in only two ways:
  directly through \ruleref{E-Call} and \ruleref{E-CallAtk} and indirectly through \ruleref{E-Eval}.
  Thus premise~\ref{lem:locks:li:no-ignore-locks} inductively holds for all other steps.
  Similarly, premises~\ref{lem:locks:li:steps} and~\ref{lem:locks:li:endorsement} remain true by assumption at top-level.
  We can therefore directly apply our inductive hypothesis for all steps except \ruleref{E-Eval}, \ruleref{E-Call}, and \ruleref{E-CallAtk}.
  We handle those cases explicitly.

  For the case of \ruleref{E-Eval} where $s = \tilde{E}[\tilde{s}]$, we induct on $\tilde{E}$ and the operational semantics.
  \begin{description}
    \item[Case {$\tilde{E} = [\cdot]$}:]
      Here induction on the operational semantic rule proves the case.

    \item[Case {$\tilde{E} = \letIn{\tilde{E}'}{e}$}:]
      We now consider two sub-cases: if $\confExpr{\tilde{E}'[\tilde{s}]} \stepsmany \conf{\tilde{E}''[\new~C(\overline{v}).m(\overline{w})]}{\config'}$ or not.
      If there is such an evaluation, then all of the inductive hypotheses hold for $\tilde{E}'[\tilde{s}]$, so induction on $\tilde{E}$ prove the case.
      If there is no such evaluation, inspection on the operational semantics tells us that we can only step $s$ using \ruleref{E-Eval} stepping $\tilde{E}'[\tilde{s}]$ until it steps to a value.
      Therefore, premise~\ref{lem:locks:li:steps}, ensures that there is some value $v$ and context $\config_v$ such that $\confExpr{\tilde{E}'[\tilde{s}]} \stepsone^+ \conf{v}{\config_v}$.
      Using \ruleref{E-Eval} on each step gives us
      $$\confExpr{s} \stepsone^+ \conf{\letIn{v}{e}}{\config_v} \stepsmany \conf{E[\new~C(\overline{v}).m(\overline{w})]}{\config'} \stepsone \conf{\tilde{s}}{\config''}.$$
      Therefore, $\conf{\letIn{v}{e}}{\config_v}$ satisfies our inductive hypothesis, so induction completes the case.
  \end{description}
  For the other three possible cases of $\tilde{E}$, the same logic as in the base-case proof above applies.

  We now turn to when the step is \ruleref{E-Call} or \ruleref{E-CallAtk}.
  In both cases $s = \new~D(\overline{v'}).m'(\overline{w'})$ and $\mbody(D, m') = \mbodyExpr{\mlabel}{\overline{x}}{\overline{\tau_a'}}{\pc_1'}{\pc_2'}{e}{\tau}$.
  If we let $e' = e[\overline{x} \mapsto \overline{w'}, \this \mapsto \new~D(\overline{v'})]$, this steps to $\atpc[\pc_2']{(\funend{\tau}~e')}$.
  We handle this in two sub-cases: if $\pc_2' \in \T$ and if $\pc_2' \notin \T$.

  If $\pc_2' \in \T$, then \ruleref{Method-Ok} proves that $\mlabel \actsfor \pc_2'$ and therefore $\mlabel \in \T$.
  Because $\confExpr{s}$ is $\T$\=/safe, the method body $e$, and hence $e'$, cannot have any subexpressions of the form $\ignoreLocks{e''}$.
  Therefore the new statement satisfies premise~\ref{lem:locks:li:no-ignore-locks} of this lemma, allowing us to apply the inductive hypothesis.

  If $\pc_2' \notin \T$, we claim that $(\bigmeet \hat{\lockList}) \meet \pc_2' \nactsfor \ellHigh$,
  which we prove differently based on the security assumptions of the system:
  either $\adv = \overline{\T}$ is a sublattice, or \ruleref{E-CallAtk} is not admissible.
  In both cases we will apply Lemma~\ref{lem:method-call-lock-list} to the configuration after taking this step.
  To meet the requirement of hte lemma that there is no sub-statement of the form $\atpc[\pc']{s'}$,
  we use $\conf{e'}{\confTuple{\sigma}{(\mlabList, \mlabel)}{\lockList}}$, noting that this
  configuration is $\T$-safe with $\pc_2'$ and $\hat{\lockList}$.

  Because $\pc_2' \notin \T$, Lemma~\ref{lem:method-call-lock-list} proves that $(\bigmeet \hat{\lockList}) \in \adv$.
  When $\adv$ is a sublattice, it is closed under join, so $(\bigmeet \hat{\lockList}) \meet \pc_2' \in \adv$.
  By the downward-closed property of $\T$, that means $(\bigmeet \hat{\lockList}) \meet \pc_2' \nactsfor \ellHigh$.

  If \ruleref{E-CallAtk} is not admissible, Lemma~\ref{lem:method-call-lock-list}
  proves that $\pc_2' \actsfor \pc_2 \join (\bigmeet \hat{\lockList})$.
  By the definition of meet and the distributive property of the lattice,
  $$\pc_2' = \left(\pc_2 \join \left(\bigmeet \hat{\lockList}\right)\right) \meet \pc_2' = (\pc_2 \meet \pc_2') \join \left(\left(\bigmeet \hat{\lockList}\right) \meet \pc_2'\right).$$
  By assumption on this sub-case, $\pc_2' \notin \T$ and therefore $\pc_2' \nactsfor \ellHigh$,
  so at least one of the two sides of the join cannot act for $\ellHigh$.
  However, the definition of meet gives $\pc_2 \meet \pc_2' \actsfor \pc_2 \actsfor \ellHigh$.
  Therefore $(\bigmeet \hat{\lockList}) \meet \pc_2' \nactsfor \ellHigh$.

  By inversion on the typing rules, if $\Sigma_\sigma; \Gamma; \pc; \inLock \proves s : \tau \dashv \outLock$,
  then $\outLock' \join \pc_2' \actsfor \outLock$ where $\outLock'$ is the lock label on $D.m$.
  In particular, $\pc_2' \actsfor \outLock \actsfor \inLock \join \outLock$.
  As a result, $(\bigmeet \hat{\lockList}) \meet \pc_2' \actsfor (\bigmeet \hat{\lockList}) \meet (\inLock \join \outLock)$,
  so transitivity of $\actsfor$ proves $(\bigmeet \hat{\lockList}) \meet (\inLock \join \outLock) \nactsfor \ellHigh$.
\end{proof}

\begin{lemma}
  \label{lem:method-call-lock-list}
  For any statement $s$, configuration $\config = \confTuple*$, label $\pc$, and lock list $\hat{\lockList}$, if
  \begin{itemize}[nosep]
    \item $\conf{s}{\config}$ is $\T$\=/safe with $\pc$ and $\hat{\lockList}$,
    \item $s$ contains no sub-statements of the form $\atpc[\pc']{s'}$,
    \item $\conf{s}{\config} \stepsmany \conf{E[\new~C(\overline{v}).m(\overline{w})]}{\config'} \stepsone \conf{s'}{\config''}$, and
    \item $\mtype(C, m) = \mtypeExpr{\overline{\tau_a}}{\pc_1}{\pc_2}{\outLock}{\tau}$ with $\pc_2 \in \T$,
  \end{itemize}
  then $\pc \notin \T$ implies $(\bigmeet \hat{\lockList}) \notin \T$,
  and $\pc \actsfor \pc_2 \join (\bigmeet \hat{\lockList})$ if no step uses \ruleref{E-CallAtk}.
\end{lemma}

\begin{proof}
  This proof follows by induction on the number of steps.
  For the base case where $s = E[\new~C(\overline{v}).m(\overline{w})]$, we induct on $E$ to prove $\pc \actsfor \pc_2 \join (\bigmeet \hat{\lockList})$.
  \begin{description}
    \item[Case {$E = [\cdot]$}:]
      Inversion on the typing rules tells us $\pc \actsfor \pc_1$ and inversion on the operational semantics tell us $\pc_1 \actsfor \pc_2 \join (\bigmeet \lockList)$.
      By the definition of $\getLocks$, $\hat{\lockList} = \lockList$, so transitivity proves the case.

    \item[Case {$E = \atpc[\pc']{E}$}:]
      This case is impossible by assumption.

    \item[Case $E = \withLock{E'}$:]
      In this case we note that if $s = E[\tilde{s}]$, then $\confExpr{E'[\tilde{s}]}$ must be $\T$\=/safe with $\pc$ and $(\hat{\lockList}, \ell)$.
      Therefore, by induction on $E$, $\pc \actsfor \pc_2 \join (\bigmeet (\hat{\lockList}, \ell))$.
      However,
      $$\bigmeet (\hat{\lockList}, \ell) = \left(\bigmeet \hat{\lockList}\right) \meet \ell \actsfor \bigmeet \hat{\lockList}.$$
      Therefore, by transitivity, $\pc \actsfor \pc_2 \join (\bigmeet \hat{\lockList})$.

    \item[All other cases:]
      The $\pc$ remains unmodified and $\getLocks(\hat{\lockList}, E'[\tilde{s}]) = \getLocks(\hat{\lockList}, E[\tilde{s}]) = \lockList$,
      so a simple inductive application completes the case.
  \end{description}
  This directly proves the second conclusion in this case.
  When $\pc \notin \T$, the fact that $\T$ is downward-closed means $\pc_2 \join (\bigmeet \hat{\lockList}) \notin \T$.
  However, $\pc_2 \in \T$ and $\T$ is a sublattice, so therefore it must be the case that $(\bigmeet \hat{\lockList}) \notin \T$.

  We now move to the inductive step.
  Lemma~\ref{lem:config-safety} ensures that $\T$\=/safety is retained.
  By inspection on the operational semantic rules, we can introduce new syntax only with \ruleref{E-Eval}, \ruleref{E-IfT}, \ruleref{E-IfF}, \ruleref{E-Call}, and \ruleref{E-CallAtk}.
  For all other steps, a direct application of the inductive hypothesis proves the lemma.
  We now prove those cases.
  \begin{description}
    \item[Case \ruleref{E-Eval}:]
      This case is by induction on $\tilde{E}$ where $s = \tilde{E}[\tilde{s}]$.
      If $\tilde{E} = [\cdot]$, induction on the operational semantics completes the case.
      When $\tilde{E} = \letIn{\tilde{E}'}{e}$, we must consider whether $\tilde{E}'[\tilde{s}]$ steps to the relevant method call or not.
      If it does, a direct inductive application proves the case.
      If it does not, we note that $\confExpr{\tilde{E}'[\tilde{s}]} \stepsone^+ \conf{v}{\config_v}$ for some value $v$ and configuration $\config_v$.
      This new expression satisfies the premises of our top-level inductive hypothesis, so we can apply that.

      By assumption, $\tilde{E} \neq \atpc[\pc']{\tilde{E}'}$, and the other possible options are the same as in the base case.

    \item[Cases \ruleref{E-IfT} and \ruleref{E-IfF}:]
      In this case we note that $s = \IfThenElse[\pc']{v}{e_1}{e_2}$.
      Inversion on the typing rules proves that $\Sigma_\sigma; \Gamma; \pc'; \inLock \proves e_i : \tau \dashv \outLock$ for both $i = 1, 2$
      and $\pc \actsfor \pc'$.
      Therefore, $\confExpr{e_i}$ is $\T$\=/safe with $\pc'$ and $\hat{\lockList}$.
      Moreover, $e_1$ and $e_2$ are expressions, so they contain no sub-statements of the form $\atpc[\pc'']{s''}$, allowing us to apply our inductive hypothesis.
      If $\pc \notin \T$, then because $\T$ is downward-closed, $\pc' \notin \T$,
      so induction proves that $(\bigmeet \hat{\lockList}) \notin \T$.
      If \ruleref{E-CallAtk} is not admissible,
      induction proves $\pc' \actsfor \pc_2 \join (\bigmeet \hat{\lockList})$, so transitivity gets us the desired result.

    \item[Case \ruleref{E-Call}:]
      In this case $s = \new~D(\overline{v'}).m'(\overline{w'})$ with $\mbody(D, m') = \mbodyExpr{\ell_{m'}}{\overline{x}}{\overline{\tau_a'}}{\pc_1'}{\pc_2'}{e}{\tau'}$,
      and $\pc_1' \actsfor \pc_2' \join (\bigmeet \lockList)$.
      Inversion on the typing rules proves that $\pc \actsfor \pc_1'$.
      Additionally, the statement after the step is $\funend{\tau'}~(\atpc[\pc_2']{e'})$ for some expression $e'$.

      By Lemma~\ref{lem:config-safety} the new configuration is $\T$\=/safe at $\pc$ and $\hat{\lockList}$,
      so inductively, replacing the statement with $e'$ is $\T$\=/safe at $\pc_2'$ and $\hat{\lockList}$.
      If $\pc_2' \notin \T$, then by induction $(\bigmeet \hat{\lockList}) \notin \T$.
      If $\pc \notin \T$ but $\pc_2' \in \T$, then the same logic as in the base case proves $(\bigmeet \hat{\lockList}) \notin \T$.

      If \ruleref{E-CallAtk} is never used, induction on the number of steps proves $\pc_2' \actsfor \pc_2 \join (\bigmeet \hat{\lockList})$.
      Combining this with the flow above, we get
      \[
        \pc \actsfor \pc_2' \join \left(\bigmeet \hat{\lockList}\right)
            \actsfor \left(\pc_2 \join \left(\bigmeet \hat{\lockList}\right)\right) \join \left(\bigmeet \hat{\lockList}\right)
            = \pc_2 \join \left(\bigmeet \hat{\lockList}\right).%
        \qedhere
      \]

    \item[Case \ruleref{E-CallAtk}:]
      Inversion on the semantic rules proves $\pc_2' \in \adv = \overline{\T}$.
      Using the same argument as in the \ruleref{E-Call} case to apply the inductive hypothesis,
      induction proves that $(\bigmeet \hat{\lockList}) \notin \T$ regardless of the value of $\pc$.
      This case is impossible by assumption when \ruleref{E-CallAtk} is not taken.
  \end{description}
\end{proof}

We formalize the concept of a tail call, which is a call initiated in a tail position of some expression,
by defining a \textit{tail context} $T$ which, by construction, does nothing after the call returns.

\begin{definition}[Tail Context]
  \label{defn:tail-context}
  $$\begin{array}{rcl}
    T & ::= & [\cdot] \alt \funend{\tau}~T \alt \withLock{T} \alt \atpc{T}
  \end{array}$$
\end{definition}

The following lemma captures our intuition that a tail context ``does nothing''.

\begin{lemma}
  \label{lem:tail-ctx-noop}
  If $\conf{T[v]}{\confTuple*} \stepsone \conf{s}{\confTuple{\sigma'}{\mlabList'}{\lockList'}}$,
  then for some tail context $T'$, $s = T'[v]$ and $\sigma = \sigma'$.
\end{lemma}

\begin{proof}
  By simple induction on the operational semantics, noting for \ruleref{E-Eval} that, if $T[v] = E[s']$, then $E = T_1$ and $s' = T_2[v]$ for some tail contexts $T_1$ and $T_2$.
\end{proof}

\begin{definition}[Tail Reentrancy]
  \label{defn:tail-reentrancy}
  We say a statement $s$ is in an \emph{$\ellHigh$-tail-reentrant} state if $s$ is $\ellHigh$-reentrant---that is, %
  $s = E_0[\atpc[\pc_1]{E_1[\atpc[\pc_2]{E_2[\atpc[\pc_3]{s'}]}]}]$
  where $\pc_1, \pc_3 \actsfor \ellHigh$ and $\pc_2 \nactsfor \ellHigh$---%
  and there is some tail context $T$, evaluation context $\tilde{E}_2$, and label $\pc_2'$ such that $\pc_2' \nactsfor \ellHigh$ and
  $$E_1[\atpc[\pc_2]{[\cdot]}] = T[\atpc[\pc_2']{\tilde{E}_2}]$$
\end{definition}

\begin{retheorem}{thm:all-tail-reenter}
  For any label~$\ellHigh \in \T$, class table~$\CT$, and well-typed heap~$\sigma_1$,
  if $\Sigma_{\sigma_1} \proves \CT~\mathsf{ok}$ complies with locks in $\T$\=/code,
  then for any invocation~$I$ and heap $\sigma_2$ where $\Sigma_{\sigma_1} \proves I$ and $(I, \CT, \sigma_1) \Downarrow \sigma_2$,
  all $\ellHigh$\=/reentrant states in the execution are $\ellHigh$\=/tail-reentrant.
\end{retheorem}

\begin{proof}
  By Definition~\ref{defn:reentrancy}, if $I = (\loc, m(\overline{v}), \ell)$ is an $\ellHigh$\=/reentrant invocation in $\sigma_1$,
  there must exists a statement $s$ such that
  $$s = E_0\Big[\atpc[\pc_1]{E_1\big[\atpc[\pc_2]{E_2[\atpc[\pc_3]{s'}]}\big]}\Big]$$
  where $\pc_1, \pc_3 \actsfor \ellHigh$ but $\pc_2 \nactsfor \ellHigh$, and
  $\conf{\deref{\loc}.m(\overline{v})}{\confTuple{\sigma_1}{\ell}{\cdot}} \stepsmany \confExpr{s}$.
  We prove by induction on~$E_1$ that~$s$ is $\ellHigh$-tail-reentrant according to Definition~\ref{defn:tail-reentrancy}.
  Specifically, we claim the following.

  \begin{claim}
    If $s = E_0'[E_1[\atpc[\pc_2]{s''}]]$ where $\innerPc(\ell, E_0') \actsfor \ellHigh$,
    then there is some $\tilde{E}_2$, $T$, and $\pc_2'$ such that $E_1[\atpc[\pc_2]{[\cdot]}] = T[\atpc[\pc_2']{\tilde{E}_2}]$ and $\pc_2' \nactsfor \ellHigh$.
  \end{claim}

  \begin{proof}[Proof of claim]
    This is a proof by induction on $E_1$.
    \begin{description}
      \item[Case {$E_1 = [\cdot]$}:]
        Because $\pc_2 \nactsfor \ellHigh$ by assumption, letting $\pc_2' = \pc_2$, $\tilde{E}_2 = [\cdot]$, and $T = [\cdot]$ proves the case.

      \item[Case {$E_1 = \atpc[\pc']{E_1'}$}]:
        There are two sub-cases to consider.
        If $\pc' \actsfor \ellHigh$, then the inductive hypothesis applies by
        replacing $E_0'$ with $E_0'[\atpc[\pc']{[\cdot]}]$ and $E_1$ by $E_1'$.
        It then proves that $E_1' = T'[\atpc[\pc_2']{\tilde{E}_2}]$ for some $\pc_2' \nactsfor \ellHigh$.
        Letting $T = \atpc[\pc']{T'}$ completes the sub-case.

        If $\pc' \nactsfor \ellHigh$, then letting $\tilde{E}_2 = E_1'$, $\pc_2' = \pc'$, and $T = [\cdot]$ proves the case.

      \item[Case $E_1 = \withLock{E_1'}$:]
        Replacing $E_0'$ with $E_0'[\withLock{[\cdot]}]$ and $E_1$ with $E_1'$,
        the inductive hypothesis proves $E_1' = T'[\atpc[\pc_2']{\tilde{E}_2}]$ for some $\pc_2' \nactsfor \ellHigh$.
        Letting $T = \withLock{T'}$ completes the case.

      \item[Case $E_1 = \funend{\tau}~E_1'$:]
        This case follows from the same logic as the previous case.

      \item[Case $E_1 = (\letIn{E_1'}{e})$:]
        Let $\pc = \innerPc(\ell, E_0')$.
        Lemma~\ref{lem:config-safety} and induction on the number of steps to get to $s$,
        proves that if $\conf{\deref{\loc}.m(\overline{v})}{\confTuple{\sigma_1}{\ell}{\cdot}} \stepsmany \conf{s}{\confTuple{\sigma}{\mlabList}{\lockList}}$
        then it must be the case that each configuration encountered along the way is $\ellHigh$\=/safe with $\ell$ and $\cdot$.

        To step to $s$, there must be some expression $e_1$ such that
        \[\begin{array}{c}
          \conf{\deref{\loc}.m(\overline{v})}{\confTuple{\sigma_1}{\ell}{\cdot}} \stepsmany \conf{E_0'[\letIn{e_1}{e}]}{\confTuple{\sigma'}{\mlabList}{\lockList}} \\
          \text{and} \\
          \conf{e_1}{\confTuple{\sigma'}{\mlabList}{\lockList}} \stepsmany \conf{E[\new~D(\overline{v'}).m'(\overline{w})]}{\config'}
        \end{array}\]
        where $\mtype(D, m') = \mtypeExpr{\overline{\tau_a}}{\tilde{\pc}_1}{\tilde{\pc}_2}{\tilde{\outLock}}{\tilde{\tau}}$
        such that $\tilde{\pc}_1 \nactsfor \ellHigh$ and $\tilde{\pc}_2 \actsfor \ellHigh$.
        Inversion on the typing rules and the safety of $\conf{E_0'[\letIn{e_1}{e}]}{\confTuple{\sigma'}{\mlabList}{\lockList}}$
        prove that $\Sigma; \Gamma; \pc; \inLock \proves e_1 : \tau_1 \dashv \outLock$ for some $\Sigma$, $\Gamma$, $\inLock$, $\tau_1$, and $\outLock$,
        where $\outLock \actsfor \inLock$ and $(\bigmeet \getLocks(\cdot, E_0')) \meet \inLock \actsfor \pc \actsfor \ellHigh$.
        Moreover, the safety of the configuration guarantees that $\getLocks(\cdot, E_0')$ is a prefix of $\lockList$,
        so in particular, $\bigmeet \lockList \actsfor \bigmeet \getLocks(\cdot, E_0')$.

        However, Lemma~\ref{lem:lock-respect} mandates that, since $\Sigma; \Gamma; \pc; \inLock \proves e_1 : \tau_1 \dashv \outLock$,
        $(\bigmeet \lockList) \meet (\inLock \join \outLock) \nactsfor \ellHigh$.
        Yet we know already that $\outLock \actsfor \inLock$, meaning $\inLock \join \outLock = \inLock$, and $\bigmeet \lockList \actsfor \bigmeet \getLocks(\cdot, E_0')$.
        Therefore, this proves that $(\bigmeet \getLocks(\cdot, E_0')) \meet \inLock \nactsfor \ellHigh$.
        This contradicts the safety result, so this case is impossible.

      \item[Case $E_1 = \ignoreLocks{E_1'}$:]
        Safety of the configuration, as argued in the previous case, proves that $\provesCwl{\ell}{s}$.
        Because, by assumption, $\innerPc(\ell, E_0') \actsfor \ellHigh$, inversion on the proof rules for $\provesCwl{\ell}{s}$
        demonstrates that this case is impossible.
        \qedhere
    \end{description}
  \end{proof}

  Letting $E_0' = E_0[\atpc[\pc_1]{[\cdot]}]$ clearly satisfies the assumptions of the claim.
  Therefore,
  $$s = E_0[\atpc[\pc_1]{T[\atpc[\pc_2']{\tilde{E}_2[s'']}]}]$$
  for some $\pc_2' \nactsfor \ellHigh$.
  This form satisfies Definition~\ref{defn:tail-reentrancy} and proves the theorem.
\end{proof}

\subsection{All Tail Reentrancy is Secure}
\label{sec:tail-reenter-secure-proof}

We now present a proof for Theorem~\ref{thm:tail-reenter-safe}, proving that all tail reentrancy is secure.
The proof follows the structure outlined in the proof sketch in Section~\ref{sec:reentrancy-defn-formal}.
It requires one simple lemma and follows essentially as a corollary from a more complicated statement.

\begin{lemma}
  \label{lem:all-types-inhabited}
  For any type $\tau$ and heap-type $\Sigma$, there exists a value $v$ such that $\Sigma \proves v : \tau$.
\end{lemma}

\begin{proof}
  This proof is by induction on the structure of $\tau$.
  If $\tau = \unit^\ell$, $v = ()$.
  If $\tau = \bool^\ell$, $v = \True$.
  If $\tau = (\RefType{\tau'})^\ell$, $v = \Null$.
  If $\tau = C^\ell$, let $\fields(C) = \overline{x}\ty\overline{\tau}$.
  For each $\tau_i$, by induction, there is some $v_i$ such that $\Sigma \proves v_i : \tau_i$.
  Therefore, by \ruleref{New}, $\Sigma \proves \new~C(\overline{v}) : C^\ell$.
\end{proof}

For the main proof, we assume the existence of an $\nat$ type and constant $\nat$ values.
This assumption is without loss of generality as natural numbers are simple to encode using objects.
The class simply has \programfont{isZero} and \programfont{previous} methods.
There are two implementations: zero returns $\True$ and $\this$, respectively,
while non-zero values have a single field pointing to the previous $\nat$ and return $\False$ and the value of their one field.
We will only use $\nat$ to increment and check the value, each of which is simple with this implementation.

\begin{lemma}
  \label{lem:separate-tail-reenter}
  For any class table $\CT$, invocation $I$, and heaps $\sigma_1$ and $\sigma_2$, if
  \begin{itemize}[nosep]
    \item $\Sigma_{\sigma_1} \proves \CT~\mathsf{ok}$ complies with locks in $\ell$\=/code,
    \item $\proves \sigma_1~\mathsf{wt}$,
    \item $\Sigma_{\sigma_1} \proves I$, and
    \item $(I, \CT, \sigma_1) \Downarrow \sigma_2$ where all $\ell$\=/reentrant states are $\ell$\=/tail-reentrant,
  \end{itemize}
  then there exist $\CT'$, $\overline{I}$, $\sigma_1'$, and $\sigma_2'$ such that
  \begin{enumerate}[nosep]
    \item $\Sigma_{\sigma_1'} \proves \CT'~\mathsf{ok}$ complies with locks in $\ell$\=/code,
    \item $\CT \lequiv \CT'$,
    \item $\proves \sigma_1'~\mathsf{wt}$,
    \item $\Sigma_{\sigma_1'} \proves \overline{I}$,
    \item $(\overline{I}, \CT', \sigma_1') \Downarrow \sigma_2'$ are all non-$\ell$\=/reentrant, and
    \item\label{sep-tail:li:state-equiv} $\sigma_i \lequiv \sigma_i'$ with $\sigma_i \subseteq \sigma_i'$ for both $i = 1, 2$.
  \end{enumerate}
\end{lemma}

\begin{proof}
  For notation, let $I = (\ell_I, \loc_I, m_I(\overline{v_I}))$.

  Step through the execution of $(I, \CT, \sigma_1) \Downarrow \sigma_2$ and create a log of the following relevant events:
  \begin{enumerate}
      \item Calls from low-integrity environments into high-integrity environments.
      \item\label{event:li:hi-to-low-call} Calls from high-integrity environments into low-integrity environments.
      \item Returns from low-integrity environments into high-integrity environments.
      \item State modifications from low-integrity environments.
  \end{enumerate}
  For most events, we will only need to reply the event later, so logging the type of event and the statement that is evaluated is sufficient.
  For event~\ref{event:li:hi-to-low-call}, however, $\CT'$ will need to have different code than $\CT$,
  so there must be a link to the original piece of code.
  Method calls already have a name and clear location in the code,
  but $\If$ statements can also move from high-integrity to low-integrity and have no names.
  To allow for unique tracking, we attach a unique name $a$ to each branch of each conditional statement in $\CT$.
  They have the same typing and semantic rules as before, but syntactically include this new annotation, denoted $\If\{\pc\}~v~\programfont{then}_{a_1}~e_1~\programfont{else}_{a_2}~e_2$.

  For a semantic step $\conf{s}{\confTuple{\sigma}{\mlabList}{\lockList}} \stepsone \conf{s'}{\confTuple{\sigma'}{\mlabList'}{\lockList'}}$,
  let $\pc_s = \innerPc(\ell_I, s)$ and $\pc_{s'} = \innerPc(\ell_I, s')$.
  The following formally defines when each type of event is emitted.
  \begin{enumerate}
    \item When $s = E[\new~C(\overline{v}).m(\overline{w})]$ and $\mtype(C, m) = \mtypeExpr*$,
      if $\pc_s \nactsfor \ell$ and $\pc_2 \actsfor \ell$,
      emit $\UpEvent(\pc_s, \new~C(\overline{v}).m(\overline{w}), \sigma)$.
    \item
      \begin{itemize}[leftmargin=*,nosep,label=-]
        \item When $s = E[\new~C(\overline{v}).m(\overline{w})]$ and $\mtype(C, m) = \mtypeExpr*$,
          if $\pc_s \actsfor \ell$ but $\pc_2 \nactsfor \ell$,
          emit $\DownEvent(\pc_2, C.m)$.
        \item When $s = E[\If\{\pc\}~v~\programfont{then}_{a_1}~e_1~\programfont{else}_{a_2}~e_2]$,
          if $\pc_s \actsfor \ell$ but $\pc \nactsfor \ell$,
          emit $\DownEvent(\pc, a_1)$ if $v = \True$ and $\DownEvent(\pc, a_2)$ if $v = \False$.
      \end{itemize}
    \item When $\sigma' = \subst{\sigma}{\loc}{(v, \tau)} \neq \sigma$, emit $\SetEvent(\loc \mapsto (v, \tau))$.
    \item When $s = E[\atpc[\pc_s]{v}]$, if $\pc_s \nactsfor \ell$ and $\pc_{s'} \actsfor \ell$, emit $\RetEvent(v)$.
  \end{enumerate}
  By inspection on the operational semantics, each step will emit at most one of the above events.

  There are several important properties to note about the log.
  First, the only semantic steps that can change the value of $\innerPc(\ell_I, s)$ are \ruleref{E-Call}, \ruleref{E-CallAtk}, \ruleref{E-IfT}, \ruleref{E-IfF}, and \ruleref{E-AtPc}.
  Type preservation (Theorem~\ref{thm:preservation}) ensures that each statement is well-typed,
  so $\If$ statements can only lower the integrity of the $\pc$, not raise it.
  Therefore, whenever $\pc_s \actsfor \ell$ and $\pc_{s'} \nactsfor \ell$, the log will contain a \DownEvent event,
  and whenever $\pc_s \nactsfor \ell$ and $\pc_{s'} \actsfor \ell$, the log will contain either a \UpEvent event or \RetEvent event.
  As a result, any two \DownEvent events must be separated by either an \UpEvent event or a \RetEvent event.

  Additionally, the \DownEvent and \RetEvent events must follow a stack discipline as the represent calls and returns.
  This stack discipline creates a correspondence between each \DownEvent and exactly one \RetEvent,
  which we will refer to as the ``corresponding \RetEvent'' event.

  We now use the log constructed from the execution of $(I, \CT, \sigma_1) \Downarrow \sigma_2$
  to construct $\CT'$, $\overline{I}$, and $\sigma_1'$.
  We will ensure by construction that all conditions hold aside from Condition~\ref{sep-tail:li:state-equiv} with $\sigma_2$ and $\sigma_2'$.
  We will then argue Condition~\ref{sep-tail:li:state-equiv} on $\sigma_2$ holds.

  \paragraph{Constructing $\CT'$, $\overline{I}$, and $\sigma_1'$}
  Initialize $\overline{I} = I$ if $\ell_I \actsfor \ell$ and empty otherwise,
  and initialize $\sigma_1' = \sigma$ and $\CT' = \CT$.
  We will add to $\overline{I}$ and $\sigma_1'$ and modify $\CT'$ as the construction progresses.

  Step through the log.
  When a $\DownEvent(\pc, a)$ event appears, where $a$ can either be $C.m$ or a unique name for the branch of an \If statement,
  note that the log must be of the form $\dotsc, \DownEvent, \overline{\SetEvent}, \ev, \dotsc$ where $\ev$ is either $\UpEvent$ or $\RetEvent$.
  If this is the first $\DownEvent$ event at location $a$, add a new mapping $\loc_a \mapsto (0, \nat^\pc)$ to $\sigma_1'$
  where $\loc_a$ is fresh, meaning $\loc_a \notin \dom(\sigma_1') \cup \dom(\sigma_2)$.
  Also modify the code at location $a$ in $\CT'$.
  If this is the first time encountering $a$, replace the existing code with code that increments $\loc_a$ and conditions on it.
  If this is the $n$th $\DownEvent(\pc, a)$ event in the log for $n > 1$,
  add a new branch to the code in $\CT'$ for if $\loc_a \mapsto n$.

  The code in the conditional branch for $\loc_a \mapsto n$ will do different things depending on $\ev$.
  If $\ev = \RetEvent(v)$, the code in $\CT'$ performs all state modification in $\overline{\SetEvent}$ and then returns~$v$.
  Making these state modification may require constructing new low-integrity methods if $\pc$ does not have sufficient integrity for each.
  Since we know that none of the modified cells are trusted by~$\ell$, however,
  making the modifications is always possible using low-integrity code.
  Moreover, because the state modifications were possible in the original execution without violating locks or entering high-integrity code
  (there was no $\UpEvent$ prior to $\ev = \RetEvent(v)$),
  a call graph with the same $\pc$ labels where $\pc \nactsfor \ell$ for each label must be possible.
  This guarantees that $\CT'$ continues to type-check.

  If $\ev$ is an \UpEvent event and this is the $n$th $\DownEvent(\pc, a)$ event for location $a$,
  then the $n$th entry into $a$ in $\CT'$ simply returns some value $v$ of the appropriate type.
  By Lemma~\ref{lem:all-types-inhabited}, some such well-typed $v$ must exist.

  When a $\UpEvent(\pc, \new~C(\overline{v}).m(\overline{w}), \sigma)$ event appears in the log, modify both $\sigma_1'$ and $\overline{I}$.
  For $\sigma_1'$, add a mapping $\loc \mapsto (\new~C(\overline{v}), C^\pc)$ for a fresh location $\loc \notin \dom(\sigma_1') \cup \dom(\sigma_2)$.
  For $\overline{I}$, add two new invocations.
  The first performs all state modifications from all \SetEvent events in the log prior to this \UpEvent
  that have not already been performed by a previous invocation.
  As before, constructing such an invocation may require adding new low-integrity code to $\CT'$.
  The second invocation added to $\overline{I}$ is $(\pc, \loc, m(\overline{w}))$ where $\loc$ is the new location added to $\sigma_1'$.

  Finally, after completing all \UpEvent and \DownEvent events in the log,
  include one final invocation with associated new code to apply any \SetEvent events not included in any previous invocations.

  \paragraph{The construction satisfies all requirements}
  By construction, the resulting invocations $\overline{I}$ are non-reentrant in $\CT'$ with initial state $\sigma_1'$.
  All code changes in $\CT'$ were low-integrity and remained well-typed,
  so $\Sigma_{\sigma_1'} \proves \CT'~\mathsf{ok}$ complies with locks in $\ell$\=/code and $\CT \lequiv \CT'$.
  We constructed $\sigma_1'$ by adding new well-typed low-integrity mappings to $\sigma_1$,
  meaning $\proves \sigma_1'~\mathsf{wt}$, $\sigma_1 \lequiv \sigma_1'$, and $\sigma_1 \subseteq \sigma_1'$, as desired.
  It remains to show that there is a $\sigma_2'$ such that $(\overline{I}, \CT', \sigma_1') \Downarrow \sigma_2'$
  with $\sigma_2 \lequiv \sigma_2'$ and $\sigma_2 \subseteq \sigma_2'$.

  Let $\tilde{\sigma}_1, \dotsc, \tilde{\sigma}_n$ be the sequence of heaps appearing in the \UpEvent events in the log.
  Let $I_1, \dotsc, I_n$ be the elements of $\overline{I}$ that call into high-integrity code
  (note that these are every other element of $\overline{I}$),
  and let $\tilde{\sigma}_k'$ be the heap provided as input to $I_k$ when executing $(\overline{I}, \CT', \sigma_1') \Downarrow \sigma_2'$.
  We now argue by induction on $k$ that $\tilde{\sigma}_k \lequiv \tilde{\sigma}_k'$ and $\tilde{\sigma}_k \subseteq \tilde{\sigma}_k'$.

  For the base case let $k = 1$.
  There are two sub-cases to consider: if $\ell_I \actsfor \ell$ and if it does not.
  If $\ell_I \actsfor \ell$, then $I_1 = I$ and there are no elements of $\overline{I}$ before it,
  so $\tilde{\sigma}_1 = \sigma_1$ and $\tilde{\sigma}_1' = \sigma_1'$, meaning the conditions on $\sigma_1$ and $\sigma_1'$ proved above are precisely the goal.
  If $\ell_I \nactsfor \ell$, there is one invocation $I_0$ in $\overline{I}$ before $I_1$, and it executes only low-integrity code to set mappings.
  By construction, the code invoked by $I_0$ performs exactly the modifications to $\sigma_1'$ that occurred to $\sigma_1$ prior to the \UpEvent event
  in the original invocation.
  Note that some of these modifications may be adding new mappings through using \ruleref{E-Ref}, which is non-deterministic.
  Because all mappings in $\sigma_1'$ not in $\sigma_1$ were taken to be fresh with respect to $\sigma_2$ as well,
  the names used in the original invocation must be free, so we can pick the same names when evaluating to $\tilde{\sigma}_1'$.
  Therefore, for some set of mappings $\overline{\loc} \mapsto (\overline{v}, \overline{\tau})$,
  $\tilde{\sigma}_1 = \subst{\sigma_1}{\overline{\loc}}{(\overline{v}, \overline{\tau})}$ and
  $\tilde{\sigma}_1' = \subst{\sigma_1'}{\overline{\loc}}{(\overline{v}, \overline{\tau})}$.
  Since $\sigma_1 \lequiv \sigma_1'$ and $\sigma_1 \subseteq \sigma_1'$, the same must therefore be true of $\tilde{\sigma}_1$ and $\tilde{\sigma}_1'$, as desired.

  Now assume $k > 1$ and, by induction, that $\tilde{\sigma}_{k-1} \lequiv \tilde{\sigma}_{k-1}'$ with $\tilde{\sigma}_{k-1} \subseteq \tilde{\sigma}_{k-1}'$.
  There are two sub-cases to consider depending on whether or not $k$th \UpEvent event stems from a $\ell$\=/reentrant call
  inside the call resulting in the ($k-1$)st \UpEvent event.

  If $I_k$ does \emph{not} correspond to a reentrant call,
  then $I_{k-1}$ corresponds to a high-integrity call that executed to completion without reentrancy in the original execution.
  By construction of $\CT'$, any part of that execution that operated at low-integrity corresponds to a \DownEvent in the log,
  and since none of those produced any high-integrity calls (that would cause reentrancy),
  they modified the state by incrementing new low-integrity counters
  and otherwise making the same modifications and returning the same values as the original execution.
  In particular, the changes to $\tilde{\sigma}_{k-1}'$ needed to achieve the state $\hat{\sigma}'$ after completing $I_{k-1}$,
  are updates to new low-integrity counters and
  the changes to $\tilde{\sigma}_{k-1}$ to achieve the state $\hat{\sigma}$ after completing the original high-integrity call.
  Because $\tilde{\sigma}_{k-1} \lequiv \tilde{\sigma}_{k-1}'$ and $\tilde{\sigma}_{k-1} \subseteq \tilde{\sigma}_{k-1}'$,
  it must be that $\hat{\sigma} \lequiv \hat{\sigma}'$ and $\hat{\sigma} \subseteq \hat{\sigma}'$.

  Further, any state modifications made after the high-integrity call returns (and thus after $I_{k-1}$ completes)
  but before the $k$th \UpEvent event (the beginning of $I_k$) must be made in a low-integrity environment.
  By the same logic as Lemma~\ref{lem:step-confinement} from the proof of Noninterference,
  they must be updates to low-integrity state.
  As a result, each has a corresponding \SetEvent event in the log, which we denote $\SetEvent(\overline{\loc} \mapsto (\overline{v}, \overline{\tau}))$.
  The extra low-integrity invocation added to $\overline{I}$ before $I_k$ makes exactly these modifications to the state.
  Therefore, $\tilde{\sigma}_k = \subst{\hat{\sigma}}{\overline{\loc}}{(\overline{v}, \overline{\tau})}$
  and $\tilde{\sigma}_k' = \subst{\hat{\sigma}'}{\overline{\loc}}{(\overline{v}, \overline{\tau})}$.
  The desired result follows from the above-proved correspondence of $\hat{\sigma}$ and $\hat{\sigma}'$.

  Lastly, consider the case where $I_k$ corresponds to a reentrant call inside the call that $I_{k-1}$ corresponds to.
  That is, the log must have the form $\dotsc, \UpEvent_{k-1}, \overline{\ev}, \DownEvent(\pc, a), \overline{\SetEvent}, \UpEvent_k, \dotsc$
  where $\overline{\ev}$ contains no \UpEvent events.
  In this case, the code created to replace the $\DownEvent(\pc, a)$ event in $\CT'$
  simply returns an arbitrary value of the correct type without modifying the state.
  Because we assumed all reentrancy was tail-reentrancy, this means $\UpEvent_k$ occurred when stepping a term of the form
  $$E_0[\atpc[\pc_1]{T[\atpc[\pc_2]{E_2[\new~C(\overline{v}).m(\overline{w})]}]}]$$
  where $\pc_1 \actsfor \ell$, $\pc_2 \nactsfor \ell$, and $\mtype(C, m) = \mtypeExpr{\overline{\tau_a}}{\pc_2'}{\pc_3}{\outLock}{\tau}$ with $\pc_3 \actsfor \ell$.

  In $\CT'$, we replaced the code corresponding to $E_2[\new~C(\overline{v}).m(\overline{w})]$
  with code that returns an arbitrary value of the correct type,
  and splitting the invocations means inside $I_{k-1}$, $E_0$ will be empty.
  Therefore, by Lemma~\ref{lem:tail-ctx-noop}, once $E_2[\new~C(\overline{v}).m(\overline{w})]$ evaluates to some value $v$,
  $T[v]$ will evaluate to $v$ with no changes to the state.
  Similarly, $I_{k-1}$ will return the arbitrary value returned in $\CT'$ without examining it
  or modifying the state at all.
  That means that the change from $\tilde{\sigma}_{k-1}'$ to $\hat{\sigma}'$, the heap when $I_{k-1}$ returns,
  is, as before, updates to new low-integrity counters coupled with
  exactly the change from $\tilde{\sigma}_{k-1}$ to the heap $\hat{\sigma}$ when the $\DownEvent(\pc, a)$ event occurred.
  The low-integrity state modifications in the extra invocation before $I_k$
  are again those made by the low-integrity code in $\CT$ before the call corresponding to $\UpEvent(\pc', \new~C(\overline{v}).m(\overline{w}), \tilde{\sigma}_k)$.
  By the same argument as before, $\tilde{\sigma}_k \lequiv \tilde{\sigma}_k'$ and $\tilde{\sigma}_k \subseteq \tilde{\sigma}_k'$, as desired.

  We have now shown that the state before each $I_k$ is a $\ell$\=/equivalent superset of
  the state before the corresponding call in the original execution.
  To see that this result extends to $\sigma_2$ and $\sigma_2'$,
  note that the logic above for non-reentrant calls applies to show that
  the state after completing $I_n$ is a $\ell$\=/equivalent superset
  of the state after completing the call that generated the final \UpEvent event in the original execution.
  There may be further low-integrity code in the original execution that modifies the state,
  but all such modifications generate \SetEvent events and are updated by the final invocation in $\overline{I}$ as described above.
  Therefore, again, $\sigma_2$ and $\sigma_2'$ are acquired by making identical modifications
  to the heap after the return of the final high-integrity call, thereby proving $\sigma_2 \lequiv \sigma_2'$ and $\sigma_2 \subseteq \sigma_2'$.
\end{proof}

\begin{retheorem}{thm:tail-reenter-safe}
  Let $\CT$ be a class table, $\sigma_1$ and $\sigma_2$ be well-typed heaps, and $I$ be an invocation
  such that \mbox{$(I, \CT, \sigma_1) \Downarrow \sigma_2$}
  where all $\ell$\=/reentrant states are $\ell$\=/tail-reentrant.
  For any $\ell$\=/integrity predicates $P$ and $Q$,
  if $\singleEntSatPQ*[\Sigma_{\sigma_1}]{\CT}$ and $P(\sigma_1)$, then $Q(\sigma_2)$.
\end{retheorem}

\begin{proof}
  Lemma~\ref{lem:separate-tail-reenter} proves that there exists $\CT'$, $\overline{I}$, $\sigma_1'$, and $\sigma_2'$
  with the properties stated in the lemma.
  Because $P$ is a $\ell$\=/integrity predicate and $\sigma_1 \lequiv \sigma_1'$, the assumption that $P(\sigma_1)$ means $P(\sigma_1')$.
  The definition of $\singleEntSatPQ*[\Sigma_{\sigma_1}]{\CT}$,
  coupled with $\CT \lequiv \CT'$ and $\Sigma_{\sigma_1} \subseteq \Sigma_{\sigma_1'}$
  mean that since $P(\sigma_1')$ holds, $Q(\sigma_2')$ must hold.
  Finally, since $\sigma_2 \lequiv \sigma_2'$,
  the fact that $Q$ is also a $\ell$\=/integrity predicate proves $Q(\sigma_2)$, as desired.
\end{proof}